\documentclass[10pt,a4paper]{article}

\usepackage[spanish,english]{babel} 
\usepackage[utf8]{inputenc}	

\usepackage[T1]{fontenc}
\usepackage[utf8]{inputenc}
\usepackage{authblk}

\usepackage{cite}

\usepackage{color}			

\usepackage{courier} 

\usepackage{multicol}		
\usepackage{graphicx}
\DeclareGraphicsExtensions{.png,.jpg,.pdf,.mps,.gif,.bmp,.png}
\usepackage{float}
\usepackage{subfig}
\usepackage[margin=0cm,font=footnotesize]{caption} 

\usepackage{amsmath}
\usepackage{eurosym}

\usepackage{indentfirst}	
\usepackage{anysize}	
\marginsize{2.6cm}{2.6cm}{2.2cm}{2.2cm}	

\usepackage{enumerate}

\usepackage{fancyhdr}
\fancypagestyle{plain}{
\fancyhead[L]{}
\fancyhead[C]{}
\fancyhead[R]{}
\fancyfoot[L]{}
\fancyfoot[C]{}
\fancyfoot[R]{}

}
\usepackage{psfrag}
\usepackage{pstool}
\usepackage{epsfig}
\usepackage{graphics}

\usepackage[x11names]{xcolor}
\usepackage[linktocpage]{hyperref}
\usepackage{color}
\hypersetup{
	colorlinks = true,
	linkcolor=blue,
	citecolor=SpringGreen4,
}

\usepackage{amsmath}
\usepackage{amsmath,amssymb,amsfonts,latexsym}

\usepackage{listings}
\definecolor{mygreen}{RGB}{28,172,0} 
\definecolor{mylilas}{RGB}{170,55,241}

\title{\textbf{Closed SPARSE\textemdash a predictive particle cloud tracer}}

\author[1]{Daniel Dom\'inguez-V\'azquez\thanks{ddominguezvazquez@sdsu.edu}}

\author[1]{Bjoern F. Klose\thanks{bklose@sdsu.edu}}
\author[1]{Gustaaf B. Jacobs\thanks{gjacobs@sdsu.edu, Corresponding Author}}

\affil[1]{Department of Aerospace Engineering, \\
San Diego State University, San Diego, CA 92182, USA}

\begin{document}

\maketitle

\begin{abstract}
A closed and predictive particle cloud tracer method is presented. The tracer builds upon the Subgrid Particle-Averaged Reynolds Stress Equivalent (SPARSE) formulation first introduced in [Davis \textit{et al.}, Proceedings of the Royal Society A, 473(2199), 2017] for the tracing of particle clouds. 
It was  later extended to a Cloud-In-Cell (CIC) formulation in [Taverniers \textit{et al.}, Journal of Computational Physics, 390, 2019] using a Gaussian distribution of a cloud's influence over a mesh-based, velocity field solution.
SPARSE corrects the cloud's trace to second order by combining a Taylor series expansion of the drag coefficient and Nusselt number correction factors around the mean relative velocity of a cloud of particles with a Reynolds decomposition of the particle equations to obtain a governing system for the first two statistical moments of the cloud's position, velocity and temperature.
Here, we close the thus far unclosed SPARSE formulation by determining
the velocity field in the vicinity of the mean cloud location using a truncated Taylor series velocity representation and by combining that with averaging.
The resulting tracer is predictive. 
It enables the tracing of a cloud of particles through a single point 
and so reduces the required degrees of freedom in the accurate tracing of groups of particles. 
We demonstrate the accuracy and convergence of the method
in several one-, two- and three-dimensional test cases. 
\end{abstract}

\begin{center}
\rule{0.9\textwidth}{0.2mm}
\end{center}


\section{Introduction} \label{sec: introduction}

Particle-laden flows are encountered in many natural and technological  environments ranging from  
spray flows in combustion engines and medical devices to dispersion of pollutants in the atmosphere,  and geological sedimentation among others. The Eulerian-Lagrangian or Euler-Lagrange (EL) approach combined with a volumeless, point-particle assumption, i.e., the so-called Particle-Source-In-Cell (PSIC) method as first introduced by Crowe \textit{et al.}~\cite{crowe1977particle}, is often used for the simulation of such flows at a process-scale.
In PSIC the carrier phase equations are solved in the Eulerian frame on a grid and
particles are traced along their path in the Lagrangian frame.
The reduced point-particle method permits the simulation of a large number of particles up to billions that are commonly encountered in large scale problems. 
The model reduction comes with limitations both in terms of numerical approximation and physics omissions, that are well documented in  literature~\cite{CST98,balachandar2019self,balachandar2009scaling,balachandar2010turbulent,mashayek2003analytical,sen2018role,capecelatro2021modeling,capecelatro2015fluid,gualtieri2015exact,battista2018application},
including convergence issues and approximation of the singular point distribution over a grid, considerable empiricism in particle forcing, omission of physics such as the finite size particle effects and wake effects, subgrid turbulence-particle interactions and particle-particle interactions,  etc. 

These issues continue to inspire and drive improvements to the PSIC method. 
Some of the recent research for example includes the tackling of convergence issues through a volume averaged method~\cite{capecelatro2013euler,ireland2017improving,shallcross2020volume} extending the original work by Anderson and Jackson~\cite{anderson1967fluid},
and the modeling of interparticle forcing by the pairwise interaction extended point-particle (PIEP) model developed by Balachandar \textit{et al}~\cite{akiki2017pairwiseFJM,akiki2017pairwiseJCF,balachandar2020toward}.
In ongoing work,  we are developing high-order EL methods and high-fidelity models and combination thereof, including EL methods
with high-order time integration \cite{jacobs2009implicit} and interpolations \cite{jacobs2006high,jacobs2009high,suarez2014high}, and machine-learned, multi-scale models with a quantified uncertainty ~\cite{sen2015evaluation,sen2017evaluation,sen2018evaluation,jacobs2019uncertainty,dominguez2021lagrangian}.  

Even with the point-particle approach, the degrees of freedom in a given problem may exceed the capacity of current-day computational resources. A further problem size reduction is hence desirable, especially
is one is interested  computing on desktop computers for design purposes.
The Cloud-In-Cell (CIC) method~\cite{birdsall1969clouds} addresses the computational cost problem by amalgamating groups of particles.  
Unfortunately, the CIC is usually not used well as it does not account the cloud scale dynamics but rather just scales a computational point or parcel with the number of particles within the physical cloud, i.e. a zeroth order model.
Doing so, means that one does not account for velocity distribution of both the carrier-phase and the particle phase within the cloud  \cite{davis2017sparse,mehrabadi2015pseudo,sun2016pseudo,baker2020reynolds}.
These effects can yield rather inaccurate predictions with the zeroth cloud method as was shown in \cite{sen2015evaluation,sen2017evaluation}.

Another method for reducing computational degrees of freedom is to use high-order approximation in the form of smooth macro-particles that distribute the particle influence
over a mesh using a Gaussian distribution or a polynomial distribution function. This approach was first introduced for discontinuous Galerkin based particle-mesh methods in \cite{jacobs2006high} and later for finite difference based methods in \cite{jacobs2009high}. While much effort
has gone into the high-order distribution and a naturally accompanying macro-particle approximation \cite{suarez2014high,CHING2021110266}, to the best of our knowledge no effort  has gone towards ensuring high-order corrections to the point tracer  that should naturally accompany a high-order macro-particle's distribution function. 

To address the shortcomings of CIC and high-order PSIC methods, we coined the Subgrid Particle-Averaged Reynolds Stress-Equivalent (SPARSE) formulation in Davis \textit{et al.}~\cite{davis2017sparse}.
SPARSE is based on a method of moments to capture the effect of subcloud scales in one-way coupled simulations. 
By combining a Reynolds averaging with a truncated Taylor expansion of the forcing correction within 
a cloud, SPARSE
augments the CIC method in two ways. Firstly,
 it provides a second order correction to the forcing. 
 Secondly, it accounts for interphase, drift, kinetic energy and stresses. 
In Taverniers \textit{et al.}~\cite{taverniers2019two}, SPARSE was extended for the simulation of two-way coupled and non-isothermal particle-laden flows by modeling the cloud deformation with a 
bivariate Gaussian function whose principle strains
depend on the subgrid scale strain tensor.
In tests of a shock interaction with a particle cloud it was shown that SPARSE
captures the same physics as the point-particle model, but requires two orders of magnitude fewer degrees of freedom. 
So far, the SPARSE tracer has been closed \textit{a priori} with data from PSIC simulations. 


Here, we propose a closed SPARSE algorithm that makes the tracer predictive. 
Following the SPARSE approach, covariance terms are closed using a combination of averaging and Taylor expansion of the carrier phase variables around the mean cloud location.  
To enable the closure, the SPARSE tracer presented in \cite{davis2017sparse} is first extended to account for position, velocity and temperature covariances. 
The resulting particle cloud tracer has a second order correction to the motion and deformation caused by the carrier-phase velocity field along its mean trajectory. 
We perform a variety of computations to verify accuracy and  convergence including one-dimensional tests with prescribed velocity fields, the two dimensional stagnation flow, and the three dimensional ABC flow. 
We also validate SPARSE with a simulation of an isotropic turbulence flow, where the gas is simulated with a Direct Numerical Simulation (DNS) solver and the SPARSE particles
are traced in the DNS flow field.

In the Section~\ref{sec: formulation}, we first summarize the one-way coupled SPARSE derivation.
Then we present the extended second moment formulation and closure. 
Verification  tests are discussed in Section~\ref{sec: tests}, followed
by the simulation results of two- and three-dimensional, particle-laden flow test cases in Section~\ref{sec: 2and3Dtests}. Concluding remarks and future work are reserved for the last Section.

\newpage
\section{Closed SPARSE: Governing Equations} \label{sec: formulation}


\subsection{Point-Particle Method}

For completeness, we start the derivation of SPARSE from the dimensional point-particle equations that describe the kinematics, dynamics and heat transfer of a small spherical particle immersed in a carrier flow as follows \cite{crowe1977particle,michaelides2016multiphase,jacobs2009high}
\begin{subequations}\label{eq: xp_up_Tp_dimensional}
\begin{align}
    \frac{d\tilde{\boldsymbol x}_p}{d\tilde{t}} &= \tilde{\boldsymbol u}_p ,
    \label{eq: dxpdt_dimensional} \\
    \tilde{m}_p\frac{d\tilde{\boldsymbol u}_p}{d\tilde{t}} &= \frac{1}{2} C_D \frac{\pi {\tilde{d}_p}^2}{4} \tilde{\rho} \left|\boldsymbol{\tilde{u}} - \boldsymbol{\tilde{u}}_p \right|\left(\boldsymbol{\tilde{u}} - \boldsymbol{\tilde{u}}_p \right) ,
    \label{eq: dupdt_dimensional} \\
    \tilde{m}_p \tilde{c}_p \frac{d\tilde{\boldsymbol u}_p}{d\tilde{t}} &= Nu \tilde{k} \pi \tilde{d}_p \left(\tilde{T}-\tilde{T}_p \right),
    \label{eq: dTpdt_dimensional} 
\end{align}
\end{subequations}
where $\boldsymbol{\tilde{x}}_p$, $\boldsymbol{\tilde{u}}_p$, $\tilde{T}_p$, and, $\tilde{m}_p$ and $\tilde{c}_p $ are the particle's position vector, velocity vector, temperature, mass and specific heat (at constant pressure), respectively.
The carrier flow at the particle position is described by the velocity vector $\boldsymbol{\tilde{u}}$, density $\tilde{\rho}$, temperature $\tilde{T}$, conductivity $\tilde{k}$ and dynamic viscosity $\tilde{\mu}$. 
The mass of a spherical particle is related
to its diameter $\tilde{d}_p$, and density $\tilde{\rho}_p$, as 
$\tilde{m}_p=\tilde{\rho}_p \pi \tilde{d}_p^3/6$.
The $C_D$ is the drag coefficient and $Nu$ the Nusselt number.
For a small particle Reynolds number, $Re_p=\tilde{\rho} \left|\boldsymbol{\tilde{u}}-\boldsymbol{\tilde{u}}_p \right|\tilde{d}_p/\tilde{\mu}$ and incompressible flow, the drag coefficient and Nusselt number are described analytically as $C_D=24/Re_p$ and $Nu=2$.
For higher particle Reynolds numbers and/or other flow parameters, these can be empirically corrected \cite{boiko2005drag,loth2008compressibility,tedeschi1999motion,feng2012drag} with the functions $f_1$ and $f_2$ as

\begin{subequations}\label{eq: f1_f2_definition}
\begin{align}
    C_D&=\frac{24}{Re_p}f_1 , 
    \label{eq: f1_definition} \\
    Nu&=2f_2 .
    \label{eq: f2_definition}
\end{align}
\end{subequations}
 
\noindent Using the non-dimensional variables 
     $t ={\tilde{t}}/{\tau_f}$, \ $\boldsymbol u = {\boldsymbol{\tilde{u}}}/{U_\infty}$, \ $\rho = {\tilde{\rho}}/{\rho_\infty}$, \ $T = {\tilde{T}}/{T_\infty}$, \ $\mu = {\tilde{\mu}}/{\mu_\infty}$, \ $k = {\tilde{k}}/{k_\infty}, \ c = {\tilde{c}}/{c_\infty}$,
    \ $\boldsymbol x_p = {\boldsymbol{\tilde{x}}_p}/{L_\infty}$, \ $\boldsymbol u_p = {\boldsymbol{\tilde{u}}_p}/{U_\infty}$, \ $T_p = {\tilde{T}_p} /{T_\infty}$, \  $\rho_p={\tilde{\rho}_p}/{\rho_\infty}$, \ $c_p={\tilde{c}_p}/{c_\infty} , \ d_p={\tilde{d}_p}/{L_\infty}$, 
where the $\infty$ subscript identifies reference scales, we arrive
at the non-dimensional formulation
\begin{subequations}\label{eq: xp_up_Tp_nondimensional}
\begin{align}
    \frac{d\boldsymbol{x}_p}{dt} &= \boldsymbol{u}_p ,
    \label{eq: dxpdt_nondimensional} \\
    \frac{d\boldsymbol{u}_p}{dt} &= \frac{f_1}{St}  \left( \boldsymbol{u} -\boldsymbol{u}_{p}  \right) , 
    \label{eq: dupdt_nondimensional} \\
    \frac { d T_p }{ dt } &= \frac{2c_r}{3Pr}\frac{f_2}{St}  \left( T -T_p  \right),
    \label{eq: dTpdt_nondimensional} 
\end{align}
\end{subequations}
where  $St=\tau_p/\tau_f$ is the Stokes number, i.e. the
ratio of the characteristic particle time scale,
$\tau_p={\tilde{\rho}_p \tilde{d}}_p^2 / \left(18\tilde{\mu} \right)$
to the convective carrier phase time scale, $\tau_f={L_\infty}/{U_\infty}$.
The Prandtl number is denoted by $Pr=\tilde{\mu} \tilde{c} / \tilde{k}$, 
and the specific heat ratio of the particle to the carrier phase with $c_r=c_p/c$. 
Defining a carrier phase, reference Prandtl number $Pr_\infty=\mu_\infty c_\infty/k_\infty$ and Reynolds number $Re_\infty=\rho_\infty U_\infty L_\infty/\mu_\infty$, we express the 
particle Reynolds number, Stokes number and Prandtl number in terms
of non-dimensional variables as follows
\begin{align}
    Re_p = Re_\infty\frac{\rho}{\mu}\left|\boldsymbol{u} -\boldsymbol{u}_p \right| d_p , \ \ \ \ St   = Re_\infty\frac{\rho_p d_p^2}{18\mu} , \ \ \ \ Pr   = Pr_\infty \frac{\mu c}{k}.
    \label{eq: Rep_St_Pr}
\end{align}
For the sake of simplicity,  we take the dynamic viscosity, conductivity and specific heat ratio of the carrier phase to be constant so that $\mu=k=c=1$.


\newpage
\subsection{SPARSE Particle Cloud Tracer}

Following Davis \textit{et al.}~\cite{davis2017sparse} and Taverniers \textit{et al.}~\cite{taverniers2019two} we  model a cloud of particles using
a method of averaging starting with the Reynolds decomposition
of any instantaneous particle variable $\eta$ into its average and fluctuating component according to $\eta=\overline{\eta}+\eta^\prime$, where the average is defined 
by its ensemble average
\begin{align}
    \overline { \eta  } =\frac { 1 }{ { N }_{ p } } \sum _{ i=1 }^{ { N }_{ p } }{ { \eta  }_{ i } } ,
\end{align}
for $N_{p}$ particles within a cloud.
We define the relative velocity $ \boldsymbol a$ and the relative temperature $b$ as 
\begin{subequations}\label{eq: a_b_definitions}
\begin{align}
    \boldsymbol{a} = \boldsymbol{u}-{\boldsymbol u_p}, 
    \label{eq: a_definition} \\ 
    b = T-T_p, 
    \label{eq: b_definition}
\end{align}
\end{subequations}
and we let the correction functions of the forcing depend on the relative velocity so that $f_1=f_1(\boldsymbol{a})$ and $f_2=f_2(\boldsymbol{a})$.
Consequently, we Taylor expand this function
around the mean velocity of the cloud, e.g. for $f_1$ this leads to
\begin{align}
     f_1 \left( \overline { \boldsymbol{a} } +\boldsymbol{a}^\prime \right) = f_1 \left( \overline { {\boldsymbol a} }  \right) +{ {\boldsymbol a}^\prime }^{ T }\nabla f_1\left( \overline { {\boldsymbol a} }  \right) +\frac { 1 }{ 2 } { {\boldsymbol a}^\prime }^{ T }\mathbf{H}_{f_1}\left( \overline { {\boldsymbol a} }  \right) { {\boldsymbol a}^\prime }+\mathcal O \left( {  {\boldsymbol a}^\prime  }^{ 3 } \right).
\end{align}
Here, $\boldsymbol{H}_{{f_1}}\left( \overline{ {\boldsymbol a}} \right)$ is the Hessian matrix of the function $f_1$ evaluated at $\overline{{\boldsymbol a} }$. 
Substituting  this and Reynolds decomposing~\eqref{eq: xp_up_Tp_nondimensional} one finds 
\begin{subequations}\label{eq: x_p_up_Tp_meanPlusPrime}
\begin{align}
\frac { d\overline { \boldsymbol { x } }_p  }{ dt }+\frac { d \boldsymbol { x }_{ p }^\prime   }{ dt } &= \overline{\boldsymbol u}_p+\boldsymbol u_p^\prime , 
\label{eq: dxpdt_meanPlusPrime} \\
\frac { d\overline { \boldsymbol{ u } }_p  }{ dt } +\frac { d \boldsymbol{ u }_{ p }^\prime }{ dt } &=\frac{1}{St}\left( f_1\left( \overline { \boldsymbol a }  \right) +{ \boldsymbol a^\prime }^{ T }\nabla f_1\left( \overline { \boldsymbol a }  \right) +\frac { 1 }{ 2 } { \boldsymbol a^\prime }^{ T }\mathbf{H}_{f_1}\left( \overline { \boldsymbol a }  \right) { \boldsymbol a^\prime } \right) \left( \overline { \boldsymbol a } + \boldsymbol a^\prime \right) , 
\label{eq: dupdt_meanPlusPrime} \\
\frac { d\overline T_p  }{ dt } +\frac { d T_p^\prime }{ dt } &= \frac{2c_r}{3Pr} \frac{1}{St}\left( f_2\left( \overline { \boldsymbol a }  \right) +{ \boldsymbol a^\prime }^{ T }\nabla f_2\left( \overline { \boldsymbol a }  \right) +\frac { 1 }{ 2 } { \boldsymbol a^\prime }^{ T }\mathbf{H}_{f_2}\left( \overline { \boldsymbol a }  \right) { \boldsymbol a^\prime } \right) \left( \overline { b } + b^\prime \right).
\label{eq: dTpdt_meanPlusPrime}
\end{align}
\end{subequations}
Averaging of this system leads to the non-dimensional SPARSE equations for the mean particle position, velocity and temperature
\begin{subequations}\label{eq: x_p_up_Tp_vectorial}
\begin{align}
\frac { d\overline { \boldsymbol{ x } }_p  }{ dt } &= \overline{ \boldsymbol u}_p ,
\label{eq: mean_xp_vectorial} \\
St \frac { d\overline {\boldsymbol { u } }_p  }{ dt } &=\overline { \boldsymbol a } f_1\left( \overline { \boldsymbol a }  \right) +
\overline { \boldsymbol a^{ \prime  }\cdot \left( { \boldsymbol a^{ \prime  } }^{ T }\nabla f_1\left( \overline { \boldsymbol a }  \right) \right) } 
+\frac { \overline { \boldsymbol a }  }{ 2 } \cdot \overline { { \boldsymbol a^{ \prime  } }^{ T }\mathbf{H}_{f_1}\left( \overline { \boldsymbol a }  \right) { \boldsymbol a^{ \prime  } } } +\frac { 1 }{ 2 } \overline { { \boldsymbol a^{ \prime  } } \cdot \left( {\boldsymbol  a^{ \prime  } }^{ T }\mathbf{H}_{f_1}\left( \overline { \boldsymbol a }  \right) {\boldsymbol a^{ \prime  } } \right)  } , 
\label{eq: mean_up_vectorial} \\
\frac{3Pr}{2c_r}St\frac { d\overline{T}_p   }{ dt } &=\overline {b } f_2\left( \overline { \boldsymbol a }  \right) +
\overline { b^\prime\cdot \left( { \boldsymbol a^{ \prime  } }^{ T }\nabla f_2 \left( \overline { \boldsymbol a }  \right) \right) } 
+\frac { \overline { b }  }{ 2 } \cdot \overline { { \boldsymbol a^{ \prime  } }^{ T }\mathbf{H}_{f_2}\left( \overline { \boldsymbol a }  \right) { \boldsymbol a^{ \prime  } } } +\frac { 1 }{ 2 } \overline { { b^{ \prime  } } \cdot \left( {\boldsymbol  a^{ \prime  } }^{ T }\mathbf{H}_{f_2}\left( \overline { \boldsymbol a }  \right) {\boldsymbol a^{ \prime  } } \right)  }. 
\label{eq: mean_Tp_vectorial}
\end{align}
\end{subequations}
In previous work, we omitted the third term on the right hand sides of \eqref{eq: mean_up_vectorial} and \eqref{eq: mean_Tp_vectorial} as they pertain to
derivatives of order higher than one of the forcing function which generally turns out to be smaller than the gradient.
However, because this term after averaging is $\mathcal O\left(\overline{{a^\prime}^2}\right)$ and thus potential similar in order of 
terms, we retain it here for completeness. 

\subsection{SPARSE with Second-Order Moments}

In order to close the SPARSE equations, we must first extend the model \eqref{eq: x_p_up_Tp_vectorial} to include equations that govern the second order moments. 
These can be obtained
following a standard procedure, i.e.,
first obtain equations for the fluctuating variables by subtracting 
the averaged equations from the instantaneous equations; then
multiplying or taking the inner product of the resulting system with the fluctuating variables and vectors, respectively; finally, averaging
and neglecting terms on the order of fluctuations to the third power or higher, we arrive at the following
\begin{subequations}\label{eq: SPARSE2}
\begin{align}
\frac{d {\overline{x}_p}_i}{dt} &= {\overline{u}_p}_i
\label{eq: SPARSE2_mean_xp}, \\
St \frac{d {\overline{u}_p}_i}{dt} &= \overline{a}_i \left( f_1 \left( \overline{\boldsymbol a} \right) + \frac{1}{2}\overline{{a_j^\prime}^2} \left.\frac{\partial^2 f_1}{\partial a_j^2} \right|_{\overline{\boldsymbol a}} \right) + \overline{ a^\prime_i a^\prime_j} \left. \frac{\partial f_1}{\partial a_j}\right|_{\overline{\boldsymbol a}}    
\label{eq: SPARSE2_mean_up}, \\
\frac{3Pr}{2c_r} St\frac{d \overline{T}_p}{dt} &= \overline{b} \left( f_2 \left( \overline{\boldsymbol a} \right) + \frac{1}{2}\overline{{a_i^\prime}^2} \left.\frac{\partial^2 f_2}{\partial a_i^2} \right|_{\overline{\boldsymbol a}} \right) + \overline{ b^\prime a^\prime_i} \left. \frac{\partial f_2}{\partial a_i}\right|_{\overline{\boldsymbol a}}    
\label{eq: SPARSE2_mean_Tp}, \\
\frac{d}{dt}\left( \overline{{x^\prime_p}_i{x_p^\prime}_j } \right) &= \overline{{x^\prime_p}_i{u_p^\prime}_j} + \overline{{x^\prime_p}_j{u_p^\prime}_i}
\label{eq: SPARSE2_xpxp}, \\
St \frac{d}{dt}\left( \overline{{u_p^\prime}_i {u_p^\prime}_j}\right)  &= \left(\overline{a_i^\prime {u_p^\prime}_j}+\overline{a_j^\prime {u_p^\prime}_i} \right)f_1\left(\overline{\boldsymbol a}\right) 
+\overline{a_i}\overline{{ u_p^\prime}_j a^\prime_k} \left. \frac{\partial f_1}{\partial a_k}\right|_{\overline{\boldsymbol a}} 
+\overline{a_j}\overline{{ u_p^\prime}_i a^\prime_k} \left. \frac{\partial f_1}{\partial a_k}\right|_{\overline{\boldsymbol a}}
\label{eq: SPARSE2_upup}, \\
\frac{3Pr}{2c_r}\frac{St}{2} \frac{d \overline{{T_p^\prime}^2}}{dt} &= \overline{T_p^\prime b^\prime}f_2\left(\overline{\boldsymbol a}\right) 
+\overline{b}\overline{T_p^\prime a^\prime_i} \left. \frac{\partial f_2}{\partial a_i}\right|_{\overline{\boldsymbol a}} 
\label{eq: SPARSE2_TpTp}, \\
\frac{d}{dt}\left( \overline{{x^\prime_p}_i{u_p^\prime}_j } \right) &= \overline{{u^\prime_p}_i{u_p^\prime}_j }+\frac{1}{St} \left( \overline{{x^\prime_p}_i a^\prime_j} f_1\left( \overline{\boldsymbol a} \right)+\overline{a_j}\overline{{ x_p^\prime}_i a^\prime_k} \left. \frac{\partial f_1}{\partial a_k}\right|_{\overline{ \boldsymbol a}} \right)
\label{eq: SPARSE2_xpup}, \\
\frac{d}{dt}\left( \overline{{x^\prime_p}_i{T_p^\prime} } \right) &= \overline{{u^\prime_p}_i T_p^\prime }+\frac{2c_r}{3Pr}\frac{1}{St} \left( \overline{{x^\prime_p}_i b^\prime} f_2\left( \overline{\boldsymbol a} \right)+\overline{b}\overline{{ x_p^\prime}_i a^\prime_j} \left. \frac{\partial f_2}{\partial a_j}\right|_{\overline{\boldsymbol a}} \right)
\label{eq: SPARSE2_xpTp}, \\
St\frac{d}{dt}\left( \overline{{u^\prime_p}_i T_p^\prime } \right) &= \overline{a^\prime_i T_p^\prime}f_1\left( \overline{\boldsymbol a}\right)+\overline{a}_i\overline{T_p^\prime a^\prime_j}\left.\frac{\partial f_1}{\partial a_j}\right|_{\overline{\boldsymbol a}}
+\frac{2c_r}{3Pr} \left( \overline{b^\prime {u_p^\prime}_i}f_2\left(\overline{\boldsymbol a} \right) +\overline{b} \overline{{u_p^\prime}_i a_j^\prime}\left. \frac{\partial f_2}{\partial a_j}\right|_{\overline{\boldsymbol a}} \right)
\label{eq: SPARSE2_upTp}.
\end{align}
\end{subequations}
Here, we have used index notation for briefness with $i=1,2,3$ and $j=1,2,3$.
Because we have  retained only terms on the order of fluctuations squared, this SPARSE formulation is a second order CIC model.
\subsection{Closed SPARSE} \label{sec: closure}

The second-order SPARSE formulation in~\eqref{eq: SPARSE2} is not yet closed as many of the terms have the form of a covariance of particle variables
with carrier phase variables or a covariance
between two carrier phase variables, and the carrier phase has an
unknown distribution within the cloud region.
To highlight those terms more explicitly, we 
make use of~\eqref{eq: a_definition} and~\eqref{eq: b_definition} to unroll the second moment terms as follows
\begin{subequations}\label{eq: ab_terms}
\begin{align}
    \overline{a}_i &= \boxed{\overline{u}_i}-{\overline{u}_p}_i , 
    \label{eq: mean_ai} \\
    \overline{b} &= \boxed{\overline{T}}-\overline{T}_p , 
    \label{eq: mean_b} \\
    \overline{{a_i^\prime}{a_j^\prime}} &= \boxed{\overline{u_i^\prime u_j^\prime}}+\overline{{u_p^\prime}_i{u_p^\prime}_j}-\boxed{\overline{u_i^\prime{u_p^\prime}_j}}-\boxed{\overline{u_j^\prime{u_p^\prime}_i}},  
    \label{eq: aa} \\
    \overline{{b^\prime}{a_i^\prime}} &= \boxed{\overline{T^\prime u_i^\prime}}+\overline{{T_p^\prime}{u_p^\prime}_i}-\boxed{\overline{T^\prime{u_p^\prime}_i}}-\boxed{\overline{T_p^\prime{u^\prime_i}}},  
    \label{eq: ba} \\
    \overline{{x_p^\prime}_i{a_j^\prime}} &= \boxed{\overline{{x_p^\prime}_i u_j^\prime}}-\overline{{x_p^\prime}_i {u_p^\prime}_j} , 
    \label{eq: xpa} \\
    \overline{{u_p^\prime}_i{a_j^\prime}} &= \boxed{\overline{{u_p^\prime}_i u_j^\prime}} - \overline{{u_p^\prime}_i {u_p^\prime}_j} ,
    \label{eq: upa} \\
    \overline{{T_p^\prime}{a_i^\prime}} &= \boxed{\overline{{T_p^\prime} u_i^\prime}} - \overline{{T_p^\prime} {u_p^\prime}_i} , 
    \label{eq: Tpa} \\
    \overline{b^\prime {x_p^\prime}_i} &= \boxed{\overline{{T^\prime} {x_p^\prime}_i} }- \overline{{T_p^\prime} {x_p^\prime}_i} , 
    \label{eq: bxp} \\
    \overline{b^\prime {u_p^\prime}_i} &= \boxed{\overline{{T^\prime} {u_p^\prime}_i}} - \overline{{T_p^\prime} {u_p^\prime}_i} , 
    \label{eq: bup} \\
    \overline{b^\prime {T_p^\prime}_i} &= \boxed{\overline{{T^\prime} {T_p^\prime}_i}} - \overline{{T_p^\prime}^2 },
    \label{eq: bTp}
\end{align}
\end{subequations}
where the boxed terms need closing.
To be consistent with the SPARSE framework, we need to account for the influence of the carrier phase at the mean location.  
We will rely on averaging and Taylor series expansions once more
by expanding the carrier velocity and temperature in~\eqref{eq: mean_ai} and~\eqref{eq: mean_b} around the mean cloud location, $\overline{\mathbf{x}}_p$, as follows
\begin{subequations}\label{eq: closure_means}
\begin{align}
    \overline{u}_i &\approx \overline{u_i \left( \overline{\boldsymbol x}_p \right) + {x_p^\prime}_j\left.\frac{\partial u_i}{\partial x_j} \right|_{\overline{\boldsymbol x}_p}  +\frac{1}{2}{{x_p^\prime}_j^2}\left.\frac{\partial^2 u_i}{\partial x^2_j} \right|_{\overline{\boldsymbol x}_p}} = u_i \left( \overline{\boldsymbol x}_p \right) +\frac{1}{2}\overline{ {x_p^\prime}_j^2 }  \left.\frac{\partial^2 u_i}{\partial x^2_j} \right|_{\overline{\boldsymbol x}_p},
    \label{eq: closure_mean_u} \\ 
    \overline{T} &\approx \overline{T \left( \overline{\boldsymbol x}_p \right) + {x_p^\prime}_i\left.\frac{\partial T}{\partial x_i} \right|_{\overline{\boldsymbol x}_p}  +\frac{1}{2}{{x_p^\prime}_i^2}\left.\frac{\partial^2 T}{\partial x^2_i} \right|_{\overline{\boldsymbol x}_p}} = T \left( \overline{\boldsymbol x}_p \right) +\frac{1}{2}\overline{ {x_p^\prime}_i^2 }  \left.\frac{\partial^2 T}{\partial x^2_i} \right|_{\overline{\boldsymbol x}_p}.
    \label{eq: closure_mean_T} 
\end{align}
\end{subequations}
Note that the second term on the right hand side are zero after averaging.

We close the second moments in a similar way. 
For example,
\begin{subequations}\label{eq: closure_means_uu}
\begin{align}
    \overline{u_i^\prime u_j^\prime} &=\overline{u_i^\prime \left(\overline{u}_j+u_j^\prime \right)} \approx \overline{u_i^\prime \left(   u_j \left( \overline{\boldsymbol x}_p \right) + {x_p^\prime}_k \left.\frac{\partial u_j}{\partial x_k} \right|_{\overline{\boldsymbol x}_p}  \right)} = \overline{u_i^\prime {x_p^\prime}_k } \left.\frac{\partial u_j}{\partial x_k} \right|_{\overline{\boldsymbol x}_p} ,
    \label{eq: closure_uu} \\ 
    \overline{T^\prime u_i^\prime} &=\overline{T^\prime \left(\overline{u}_i+u_i^\prime \right)} \approx \overline{T^\prime \left(   u_i \left( \overline{\boldsymbol x}_p \right) + {x_p^\prime}_j \left.\frac{\partial u_i}{\partial x_j} \right|_{\overline{\boldsymbol x}_p}  \right)} = \overline{T^\prime {x_p^\prime}_j } \left.\frac{\partial u_i}{\partial x_j} \right|_{\overline{\boldsymbol x}_p} ,
    \label{eq: closure_Tu} \\ 
    \overline{{T^\prime}^2} &=\overline{T^\prime \left(\overline{T}+T^\prime \right)} \approx \overline{T^\prime \left(   T \left( \overline{\boldsymbol x}_p \right) + {x_p^\prime}_i \left.\frac{\partial T}{\partial x_i} \right|_{\overline{\boldsymbol x}_p}  \right)} = \overline{T^\prime {x_p^\prime}_i } \left.\frac{\partial T}{\partial x_i} \right|_{\overline{\boldsymbol x}_p} ,
    \label{eq: closure_TT} 
\end{align}
\end{subequations}
close the Reynolds stress and the heat flux on the sub-cloud scale.
Further closures of covariances are as follows
\begin{subequations}\label{eq: closure_means_xpu}
\begin{align}
    \overline{{x_p^\prime}_i u_j^\prime} &\approx \overline{{x_p^\prime}_i {x_p^\prime}_k } \left.\frac{\partial u_j}{\partial x_k} \right|_{\overline{\boldsymbol x}_p} ,
     \ \ \ \ \
    \overline{{u_p^\prime}_i u_j^\prime} \approx \overline{{u_p^\prime}_i {x_p^\prime}_k } \left.\frac{\partial u_j}{\partial x_k} \right|_{\overline{\boldsymbol x}_p} ,
    \label{eq: closure_upu} \\ 
    \overline{{T_p^\prime} u_i^\prime} &\approx \overline{{T_p^\prime} {x_p^\prime}_j } \left.\frac{\partial u_i}{\partial x_j} \right|_{\overline{\boldsymbol x}_p} ,
     \ \ \ \ \
    \overline{{x_p^\prime}_i T^\prime} \approx \overline{{x_p^\prime}_i {x_p^\prime}_j } \left.\frac{\partial T}{\partial x_j} \right|_{\overline{\boldsymbol x}_p} ,
    \label{eq: closure_xpT} \\
    \overline{{u_p^\prime}_i T^\prime} &\approx \overline{{u_p^\prime}_i {x_p^\prime}_j } \left.\frac{\partial T}{\partial x_j} \right|_{\overline{\boldsymbol x}_p} ,
     \ \ \ \ \
    \overline{{T_p^\prime} T^\prime} \approx \overline{{T_p^\prime} {x_p^\prime}_i } \left.\frac{\partial T}{\partial x_i} \right|_{\overline{\boldsymbol x}_p}.
    \label{eq: closure_TpT} 
\end{align} 
\end{subequations}
Substituting \eqref{eq: ab_terms}--\eqref{eq: closure_means_xpu} into \eqref{eq: SPARSE2} closes the model.



The resulting complete, closed SPARSE formulation is as follows:

\begin{align}
\begin{split}
\frac{d {\overline{x}_p}_i}{dt} &= {\overline{u}_p}_i ,
\end{split}
\label{eq: SPARSE2_closed_mean_xp_app} \\
\begin{split}
St \frac{d {\overline{u}_p}_i}{dt} &= \left(\overline{u}_i-{\overline{u}_p}_i\right) \left[ f_1 \left( \overline{\boldsymbol a} \right) + \frac{1}{2} \left(  \overline{{x^\prime_p}_k{x^\prime_p}_q}\left.\frac{\partial u_j}{\partial x_k} \right|_{\overline{\boldsymbol{x}}_p}\left.\frac{\partial u_j}{\partial x_q} \right|_{\overline{\boldsymbol{x}}_p}-2\overline{{u^\prime_p}_j{x^\prime_p}_k}\left. \frac{\partial u_j}{\partial x_k}\right|_{\overline{\boldsymbol{x}}_p} +\overline{{u^\prime_p}_j^2} \right)   \left.\frac{\partial^2 f_1}{\partial a_j^2} \right|_{\overline{\boldsymbol a}} \right] \\
&+ \left(\overline{{x^\prime_p}_k{x^\prime_p}_q}\left.\frac{\partial u_i}{\partial x_k} \right|_{\overline{\boldsymbol{x}}_p}\left.\frac{\partial u_j}{\partial x_q} \right|_{\overline{\boldsymbol{x}}_p} -\overline{{u^\prime_p}_j{x^\prime_p}_k}\left. \frac{\partial u_i}{\partial x_k}\right|_{\overline{\boldsymbol{x}}_p}-\overline{{u^\prime_p}_i{x^\prime_p}_k}\left. \frac{\partial u_j}{\partial x_k}\right|_{\overline{\boldsymbol{x}}_p} +\overline{{u^\prime_p}_i{u^\prime_p}_j} \right) \left. \frac{\partial f_1}{\partial a_j}\right|_{\overline{\boldsymbol a}}    ,
\end{split}
\label{eq: SPARSE2_closed_mean_up_app} \\
\begin{split}
\frac{3Pr}{2c_r} St\frac{d \overline{T}_p}{dt} &= \left(\overline{T}-\overline{T}_p\right) \left( f_2 \left( \overline{\boldsymbol a} \right) + \frac{1}{2}\left(  \overline{{x^\prime_p}_j{x^\prime_p}_k}\left.\frac{\partial u_i}{\partial x_j} \right|_{\overline{\boldsymbol{x}}_p}\left.\frac{\partial u_i}{\partial x_k} \right|_{\overline{\boldsymbol{x}}_p}-2\overline{{u^\prime_p}_i{x^\prime_p}_j}\left. \frac{\partial u_i}{\partial x_j}\right|_{\overline{\boldsymbol{x}}_p} +\overline{{u^\prime_p}_i^2} \right) \left.\frac{\partial^2 f_2}{\partial a_i^2} \right|_{\overline{\boldsymbol a}} \right) \\
&+ \left( \overline{{x^\prime_p}_j{x^\prime_p}_k}\left.\frac{\partial T}{\partial x_k}\right|_{\overline{\boldsymbol{x}}_p}\left.\frac{\partial u_i}{\partial x_j}\right|_{\overline{\boldsymbol{x}}_p}+\overline{{u^\prime_p}_i{x^\prime_p}_j}\left.\frac{\partial T}{\partial x_j}\right|_{\overline{\boldsymbol{x}}_p} +\overline{T^\prime_p{x^\prime_p}_j}\left.\frac{\partial u_i}{\partial x_j}\right|_{\overline{\boldsymbol{x}}_p} +\overline{T^\prime_p{u^\prime_p}_i} \right) \left. \frac{\partial f_2}{\partial a_i}\right|_{\overline{\boldsymbol a}}   , 
\end{split}
\label{eq: SPARSE2_closed_mean_Tp_app} 
\end{align}

\begin{align}
\begin{split}
\frac{d}{dt}\left( \overline{{x^\prime_p}_i{x_p^\prime}_j } \right) &= \overline{{x^\prime_p}_i{u_p^\prime}_j} + \overline{{x^\prime_p}_j{u_p^\prime}_i} ,
\end{split}
\label{eq: SPARSE2_closed_xpxp_app} \\
\begin{split}
St \frac{d}{dt}\left( \overline{{u_p^\prime}_i {u_p^\prime}_j}\right)  &= \left( \overline{{u^\prime_p}_i{x^\prime_p}_k} \left.\frac{\partial u_j}{\partial x_k} \right|_{\overline{\boldsymbol x}_p}+\overline{{u^\prime_p}_j{x^\prime_p}_k} \left.\frac{\partial u_i}{\partial x_k} \right|_{\overline{\boldsymbol x}_p} -2\overline{{u^\prime_p}_i{u^\prime_p}_j}\right)f_1\left(\overline{\boldsymbol a}\right) \\ 
&+\left( \overline{u}_i-{\overline{u}_p}_i\right) \left( \overline{{u^\prime_p}_j{x^\prime_p}_q}\left.\frac{\partial u_k}{\partial x_q} \right|_{\overline{\boldsymbol x}_p} -\overline{{u^\prime_p}_j{u^\prime_p}_k}  \right)\left. \frac{\partial f_1}{\partial a_k}\right|_{\overline{\boldsymbol a}} +\left( \overline{u}_j-{\overline{u}_p}_j\right) \left( \overline{{u^\prime_p}_i{x^\prime_p}_q}\left.\frac{\partial u_k}{\partial x_q} \right|_{\overline{\boldsymbol x}_p} -\overline{{u^\prime_p}_i{u^\prime_p}_k}  \right)\left. \frac{\partial f_1}{\partial a_k}\right|_{\overline{\boldsymbol a}} ,
\end{split}
\label{eq: SPARSE2_closed_upup_app} \\
\begin{split}
\frac{3Pr}{2c_r}\frac{St}{2} \frac{d \overline{{T_p^\prime}^2}}{dt} &= \left( \overline{T^\prime_p {x^\prime_p}_i} \left. \frac{\partial T}{\partial x_i}\right|_{\overline{ \boldsymbol x}_p}-\overline{{T^\prime_p}^2} \right)f_2\left(\overline{\boldsymbol a}\right) 
+\left( \overline{T}-\overline{T}_p\right)\left( \overline{T^\prime_p {x^\prime_p}_j}\left. \frac{\partial u_i}{\partial x_j} \right|_{\overline{\boldsymbol x}_p}-\overline{{u^\prime_p}_i T^\prime_p}  \right) \left. \frac{\partial f_2}{\partial a_i}\right|_{\overline{\boldsymbol a}} ,
\end{split}
\label{eq: SPARSE2_closed_TpTp_app} \\
\begin{split}
\frac{d}{dt}\left( \overline{{x^\prime_p}_i{u_p^\prime}_j } \right) &= \overline{{u^\prime_p}_i{u_p^\prime}_j } \\
&+\frac{1}{St} \left[ \left( \overline{{x^\prime_p}_i{x^\prime_p}_k}\left.\frac{\partial u_j}{\partial x_k}  \right|_{\overline{\boldsymbol x}_p} -\overline{{x^\prime_p}_i{u^\prime_p}_j} \right) f_1\left( \overline{\boldsymbol a} \right)+\left(\overline{u}_j-{\overline{u}_p}_j\right)\left(\overline{{x^\prime_p}_i{x^\prime_p}_q}\left. \frac{\partial u_k}{\partial x_q} \right|_{\overline{ \boldsymbol x}_p}-\overline{{x^\prime_p}_i{u^\prime_p}_k}    \right) \left. \frac{\partial f_1}{\partial a_k}\right|_{\overline{\boldsymbol a}} \right] ,
\end{split}
\label{eq: SPARSE2_closed_xpup_app} \\
\begin{split}
\frac{d}{dt}\left( \overline{{x^\prime_p}_i{T_p^\prime} } \right) &=  \overline{{u^\prime_p}_i T_p^\prime } \\
&+\frac{2c_r}{3Pr}\frac{1}{St} \left[ \left( \overline{{x^\prime_p}_i{x^\prime_p}_j}\left.\frac{\partial T}{\partial x_j} \right|_{\overline{\boldsymbol  x}_p}-\overline{{x^\prime_p}_iT^\prime_p}  \right)  f_2\left( \overline{\boldsymbol a} \right)+\left( \overline{T}-\overline{T}_p  \right)\left( \overline{{x^\prime_p}_i{x^\prime_p}_k}\left. \frac{\partial u_j}{\partial x_k} \right|_{\overline{\boldsymbol x}_p} -\overline{{x^\prime_p}_i{u^\prime_p}_j}  \right) \left. \frac{\partial f_2}{\partial a_j}\right|_{\overline{\boldsymbol a}} \right],
\end{split}
\label{eq: SPARSE2_closed_xpTp_app} \\
\begin{split}
St\frac{d}{dt}\left( \overline{{u^\prime_p}_i T_p^\prime } \right) &= \left(\overline{T^\prime_p {x^\prime_p}_j}\left.\frac{\partial u_i}{\partial x_j} \right|_{\overline{\boldsymbol{x}}_p} -\overline{{u^\prime_p}_i T^\prime_p} \right)f_1\left( \overline{\boldsymbol a}\right)+\left(\overline{u}_i-{\overline{u}_p}_i  \right)\left( \overline{T^\prime_p{x^\prime_p}_k}\left. \frac{\partial u_j}{\partial x_k}\right|_{\overline{\boldsymbol{x}}_p}-\overline{{u^\prime_p}_j T^\prime_p} \right)\left.\frac{\partial f_1}{\partial a_j}\right|_{\overline{\boldsymbol a}} \\
&+\frac{2c_r}{3Pr} \left[ \left( \overline{{u^\prime_p}_i{x^\prime_p}_j}\left. \frac{\partial T}{\partial x_j} \right|_{\overline{\boldsymbol x}_p}-\overline{T^\prime_p {u^\prime_p}_i}  \right)f_2\left(\overline{\boldsymbol a} \right) +\left(\overline{T}-\overline{T}_p \right) \left( \overline{{u^\prime_p}_i{x^\prime_p}_k}\left. \frac{\partial u_j}{\partial x_k} \right|_{\overline{\boldsymbol x}_p}-\overline{{u^\prime_p}_i{u^\prime_p}_j}  \right)\left. \frac{\partial f_2}{\partial a_j}\right|_{\overline{\boldsymbol a}} \right],
\end{split}
\label{eq: SPARSE2_closed_upTp_app}
\end{align}
where the only remaining terms to include are the averages of the flow magnitudes computed with \eqref{eq: closure_means}.

We remark the following:
\begin{itemize}
\item The convergence and accuracy of closed SPARSE depend on a number of factors,
including the truncation of the statistical moments that are greater than the second moments,
the truncation error of the Taylor expansion of the carrier field at the averaged cloud location which depends
on the carrier field gradients and size of the cloud, and finally the truncation
error of the Taylor expansion of the forcing which depends on the range of the subgrid relative velocity  and the gradients of the forcing function with respect to the relative velocity.
\item The Taylor expansions are valid for a limited range which leads to limitation on the cloud size of an accurate simulation using the SPARSE method. The test we conduct in the verification and validation section below address these limitations heuristically. 
\item The accuracy of closed SPARSE as compared to PSIC is not directly dependent on the number of PSIC particles per SPARSE cloud provided that the initial conditions are set according to moments of the discrete particles within the cloud. For example, even a single point-particle is accurately traced by the closed SPARSE method provided that all second-order moments in SPARSE are set to zero initially. The number of samples can affect the accuracy of the evolution of the third moments, which determine in part the accuracy of closed SPARSE (see first bullet point) and which are captured (implicitly through sampling) with greater accuracy for an increased number of PSIC samples within a given cloud area.
\item Because of  Taylor expansion of the carrier flow field, smooth changes of the carrier flow velocity and temperature within the cloud region are assumed and required.
For flows in which high gradients are present within the cloud, subdividing or splitting into more macro-particles help to smoothen gradients inside each macro-particle.
 
\item Time accuracy is limited because of the stretching of clouds and the increase with it of higher-order moments that are not included. 
Tests show that the model is accurate over significant time intervals. 
At that point an adaptive cloud model is necessary that is beyond the scope of the current paper. 
We plan to report on this in future work.
\item The SPARSE method can be interpreted as more than just a model reduction through cloud tracing, it also captures the effect of subgrid scale physics and the size. We note that the cloud size does not necessarily have to be bounded by the Kolmogorov scale. 

\end{itemize}


\section{Verification Tests} \label{sec: tests}

For verification purposes, we repeat the one-dimensional tests with  constant and linear forcings as described in Davis \textit{et al.}~\cite{davis2017sparse}. 
We consider an additional one-dimensional test with a known empirical forcing that includes a more realistic dependence of the forcing correction with respect to the relative velocity.
The parameters of the resulting four test cases, including the carrier phase velocity fields, Stokes numbers and correction factor functions  are summarized in   Table~\ref{tab:test_cases}. 


All test cases are computed with a total number of particles of $N_p=10,000$ and with initial conditions for position and velocity given according to the uniform density distributions  $x_p(0)\sim \mathcal{U}(-1,1)$ and $u_p(0)\sim \mathcal{U}(-5,15)$. 
Here, $\mathcal{U}$ denotes a uniform distribution function and the arguments give the minimum and maximum value of the distribution.
PSIC simulations are conducted to obtain the reference solution.
For each test case, we compare the closed SPARSE method with the SPARSE method from Davis \textit{et al.}\cite{davis2017sparse} which was closed \textit{a priori} and only traces the averages of the particle cloud. 
We refer to this model as "SPARSE \textit{a priori}". 
We  did not use subdivisions of a global group of particles into sub-clouds for this \textit{a priori} closed model. 
For predictions with the closed SPARSE formulation, however, we have subdivided into $M_p$ number of sub-clouds to ensure accuracy and convergence. 
The global averages and variances of the combined set of sub-clouds are given by
\begin{subequations}\label{eq: join_moments}
\begin{align}
\overline{\phi} &= \sum_{k=1}^{M_p}w_k \overline{\phi}_k, 
\label{eq: join_moments_mean} \\
\overline{\eta^\prime \theta^\prime} &= \sum_{k=1}^{M_p}w_k \overline{\eta^\prime \theta^\prime}_k+\sum_{k=1}^{M_p}w_k(\overline{\eta}_k-\overline{\eta})(\overline{\theta}_k-\overline{\theta}), 
\label{eq: join_moments_cm2}
\end{align}
\end{subequations}
for arbitrary solution variables $\phi$, $\eta$ and $\theta$. 
The weight $w_k$, of the $k$-th cloud represents the number of particles per cloud. Here, we take it as the ratio between the number of point-particles in the cloud $k$ denoted by ${N_p}_k$ and the total number of particles so that $w_k={N_p}_k/N_p$. 

SPARSE reduces the computational expense as compared to PSIC simulations when tracing a cloud of point-particles as a point. 
The PSIC description in three-dimensions of a non-isothermal cloud with $N_p$ particles requires the solution of $7N_p$ equations according to the system \eqref{eq: xp_up_Tp_nondimensional}. 
For the same case, the closed SPARSE method with $M_p$ macro-particles solves $35M_p$ equations as described by the system~\eqref{eq: SPARSE2}. 
Generally, $M_p \ll N_p$ to reproduce accurate mean and variances of the cloud and the computational savings is on the order of $35M_p/(7N_p)$.

To determine the difference between PSIC and SPARSE methods, we normalize the $L_2$ norm of the moment difference with the $L_\infty$ norm of the PSIC result as follows
\begin{align}
    \varepsilon(\cdot) = \frac{\| (\cdot)^{\mbox{SPARSE}}-(\cdot)^{\mbox{PSIC}} \|_2}{\| (\cdot)^{\mbox{PSIC}}\|_{\infty} }.
    \label{eq: error}
\end{align}
The convergence rate of SPARSE is affected by the moment accuracy, and thus the inverse square root of the sub-cloud sample size,
as well as the second order Taylor truncation of the forcing and the carrier-phase field.

\begin{table}[htbp]
\centering
\begin{tabular}{c|c|c|c}
\hline
\hline
{Case Number} & \textbf{$u(x)$} & \textbf{$St$} & \textbf{$f_1(u-u_p)$} \\ \hline \hline
1             & $10$                  & $10$                     & $u-u_p$                      \\ \hline
2             & $10$                  & $10$                     & $|u-u_p|$                    \\ \hline
3             & $x+5+\cos(\pi(x+5))$  & $1/24$                   & $1$                          \\ \hline
4             & $9+\cos(\pi x/10)$     & $1/2$                    & $1+0.15 \left(0.9487(|u-u_p|)\right)^{0.687}$           \\ \hline
\end{tabular}
\caption{Summary of the one-dimensional carrier-flow velocity fields, $u(x)$, Stokes number, $St$, and forcing correction factor function, $f_1(u-u_p)$ for four one-dimensional test cases.}
\label{tab:test_cases}
\end{table}

\subsection{Linear Forcing in Constant Carrier Velocity Field, Case 1} 

In Case 1  the carrier-phase velocity is taken constant $u(x)=10$. The Stokes number is set to $St=10$, and the  correction factor of the forcing is linearly dependent on the relative velocity.
In the constant carrier-phase velocity field,  the fluctuations of the velocity field are zero. Thus  the sub-cloud carrier-phase fluctuations   are zero, $u^\prime=0$, and the covariance terms involving the carrier-phase variables in the SPARSE formulation are zero also and cannot affect the (SPARSE) solution.
The first derivative of the linear correction factor function, $f_1(u-u_p)$,  with respect to the relative particle  velocity is unity and the second derivative is zero. Therefore, terms with a second derivative of the forcing in~\eqref{eq: SPARSE2_mean_up} and thus the the Taylor series expansion has no 
effect on the model accuracy for this case.
In the constant carrier-flow velocity, the cloud accelerates towards this velocity 
(Fig.~\ref{fig: 1Dc1_mean}) and translates  and widens correspondingly in the positive  $x$-coordinate (Figs.~\ref{fig: 1Dc1_mean}  and
\ref{fig: 1Dc1_sigma}, respectively).
The averages and variances of the particle position and velocity determined with $N_p=10^4$ PSIC particles are in excellent comparison with a SPARSE  results that uses only $M_p=16$ clouds. 
Moreover, a comparison of the moment differences (Fig.~\ref{fig: 1Dc1_epsilon}) between the PSIC and SPARSE 
show a monotonic error reduction with an increasing number of clouds, providing evidence of a consistent convergence.

\subsection{Positive Linear Forcing in Constant Carrier Velocity Field, Case 2}

\begin{figure}[htbp]
	\centering
	\subfloat[]{
		\label{fig: 1Dc1_mean}
		\includegraphics[width=0.32\textwidth]{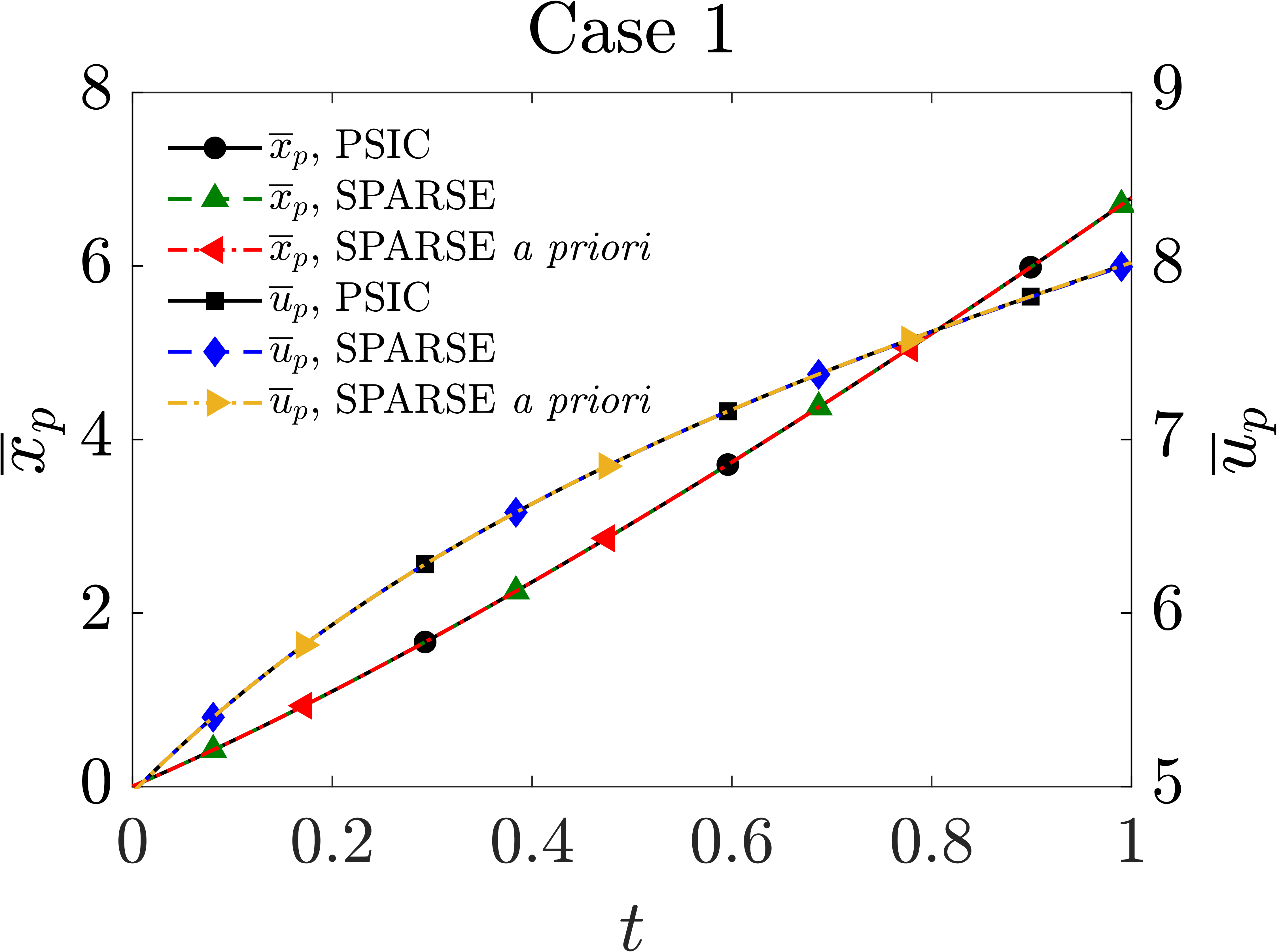}}
        \hfill	
    \subfloat[]{
		\label{fig: 1Dc1_sigma}
		\includegraphics[width=0.33\textwidth]{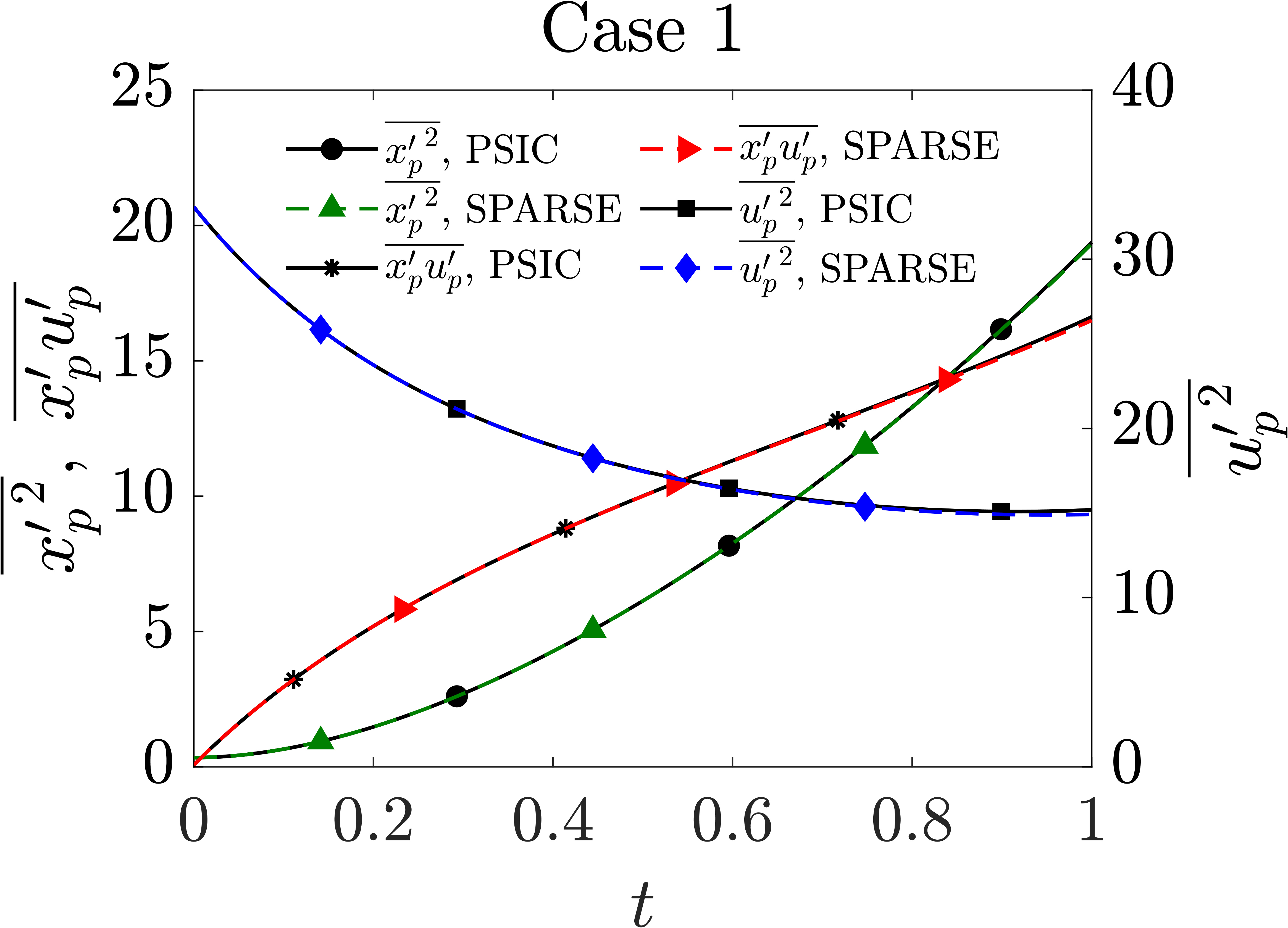}} 
		\hfill
	\subfloat[]{
		\label{fig: 1Dc1_epsilon}
		\includegraphics[width=0.3\textwidth]{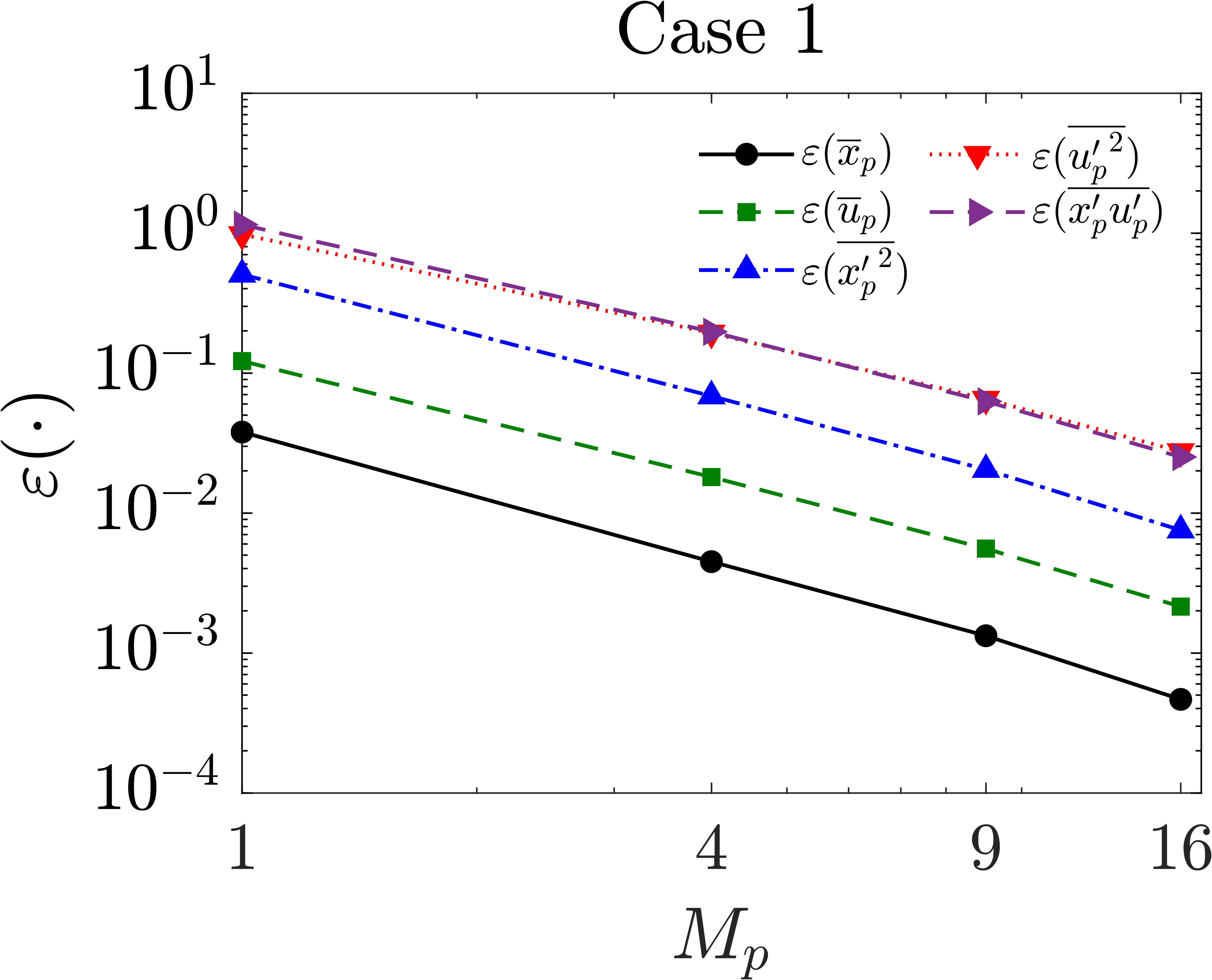}} \\
	\subfloat[]{
		\label{fig: 1Dc2_mean}
		\includegraphics[width=0.32\textwidth]{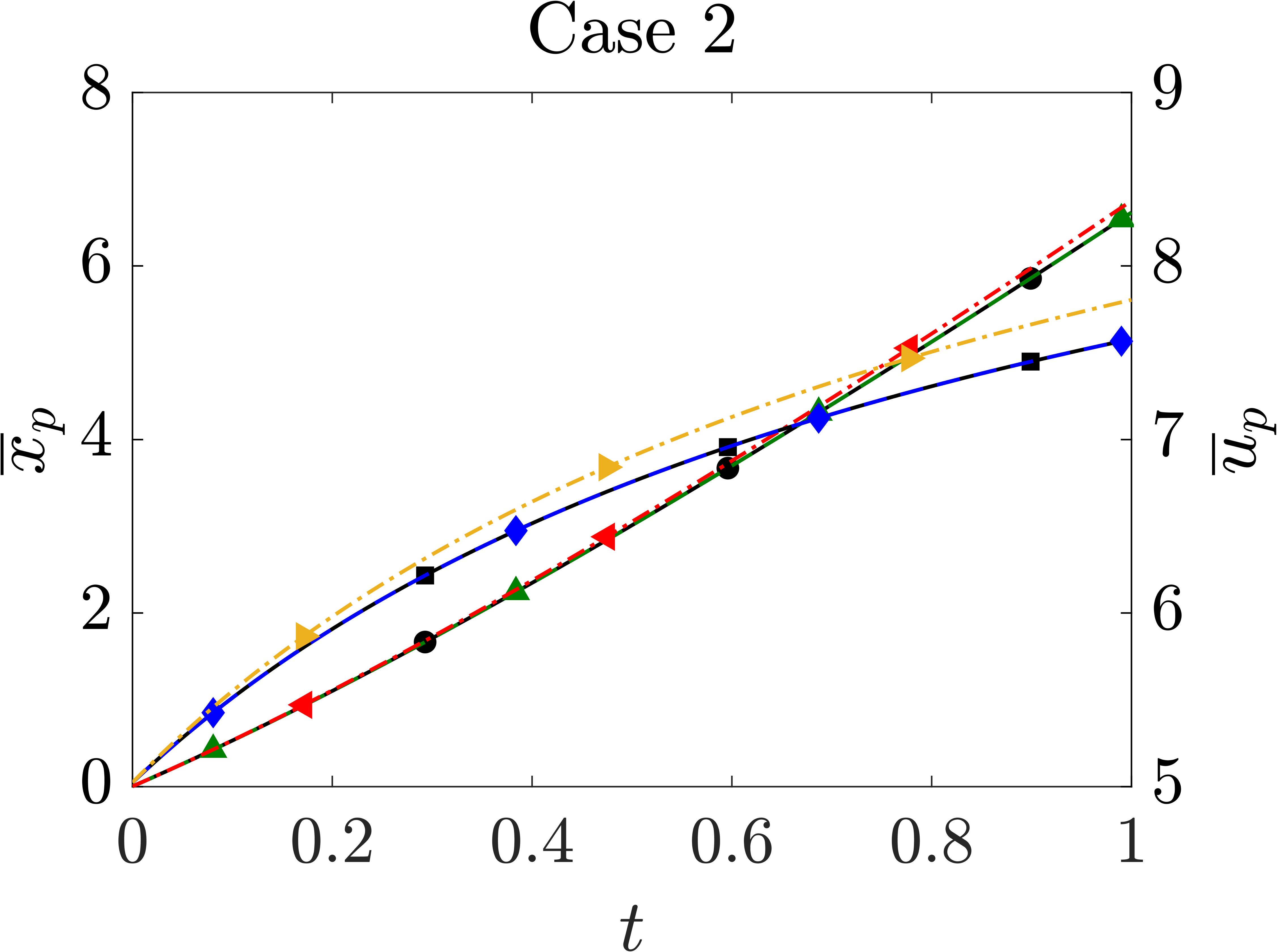}}
        \hfill	
    \subfloat[]{
		\label{fig: 1Dc2_sigma}
		\includegraphics[width=0.33\textwidth]{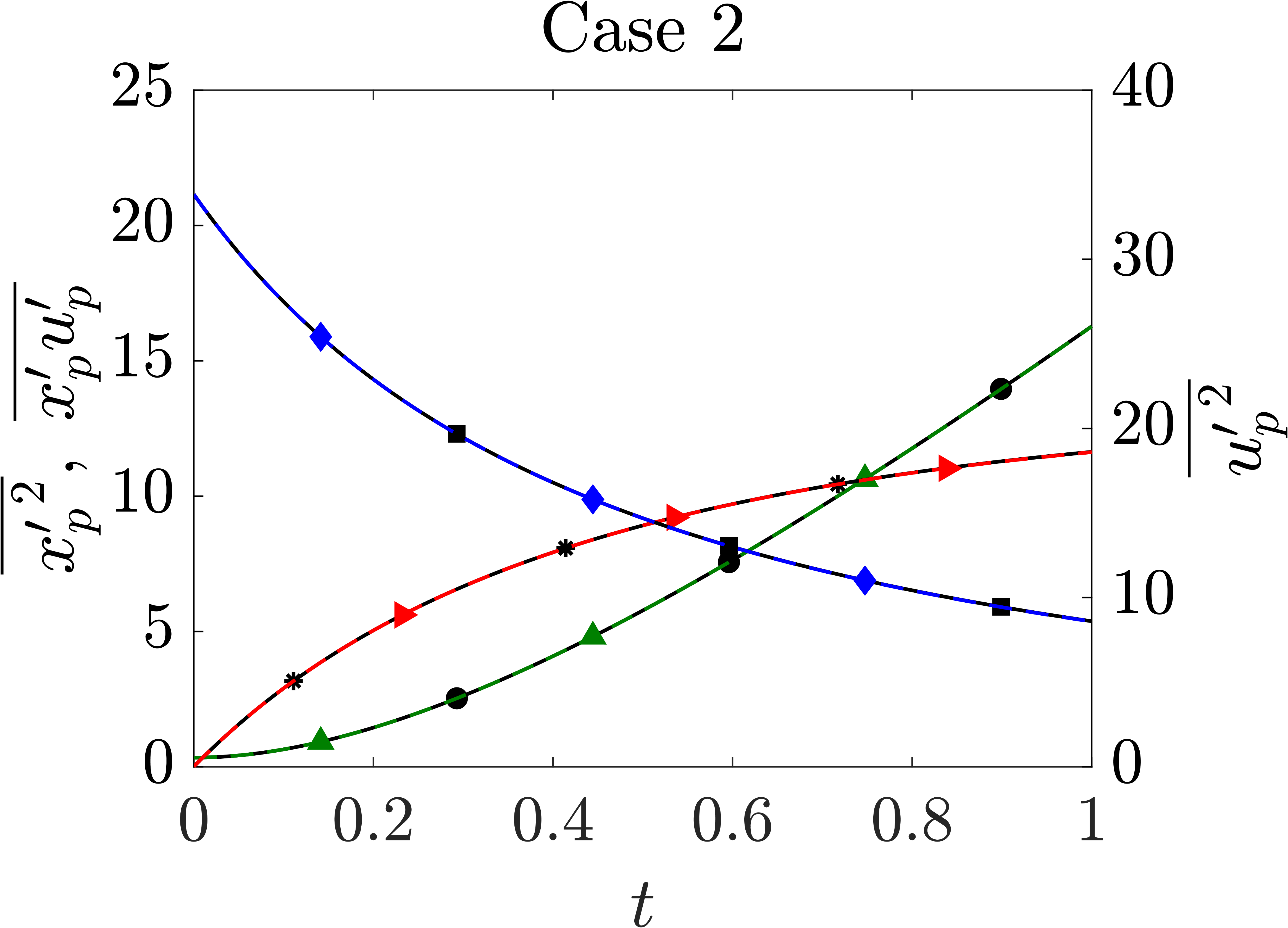}} 
		\hfill
	\subfloat[]{
		\label{fig: 1Dc2_epsilon}
		\includegraphics[width=0.3\textwidth]{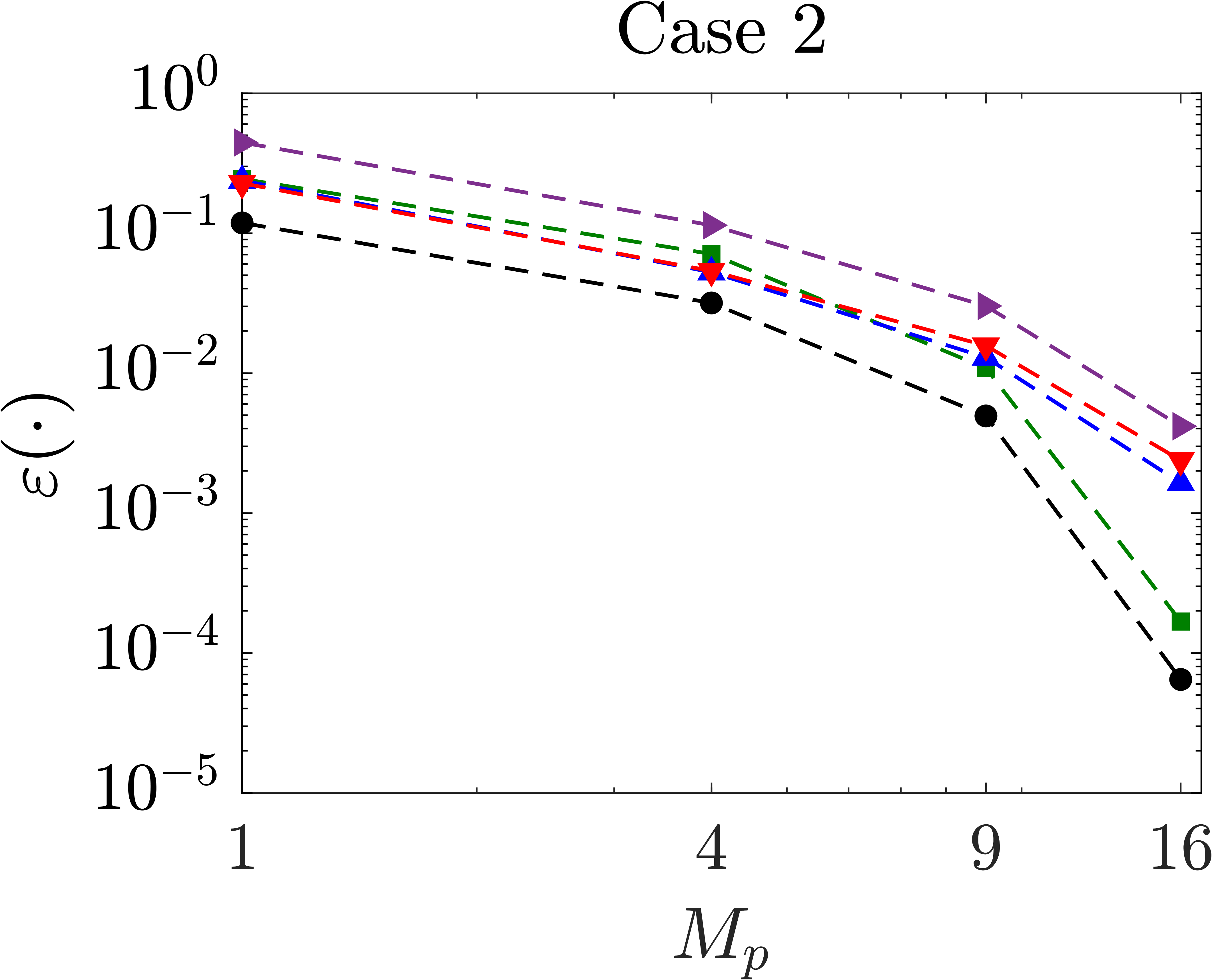}} \\
			\subfloat[]{
		\label{fig: 1Dc3_mean}
		\includegraphics[width=0.32\textwidth]{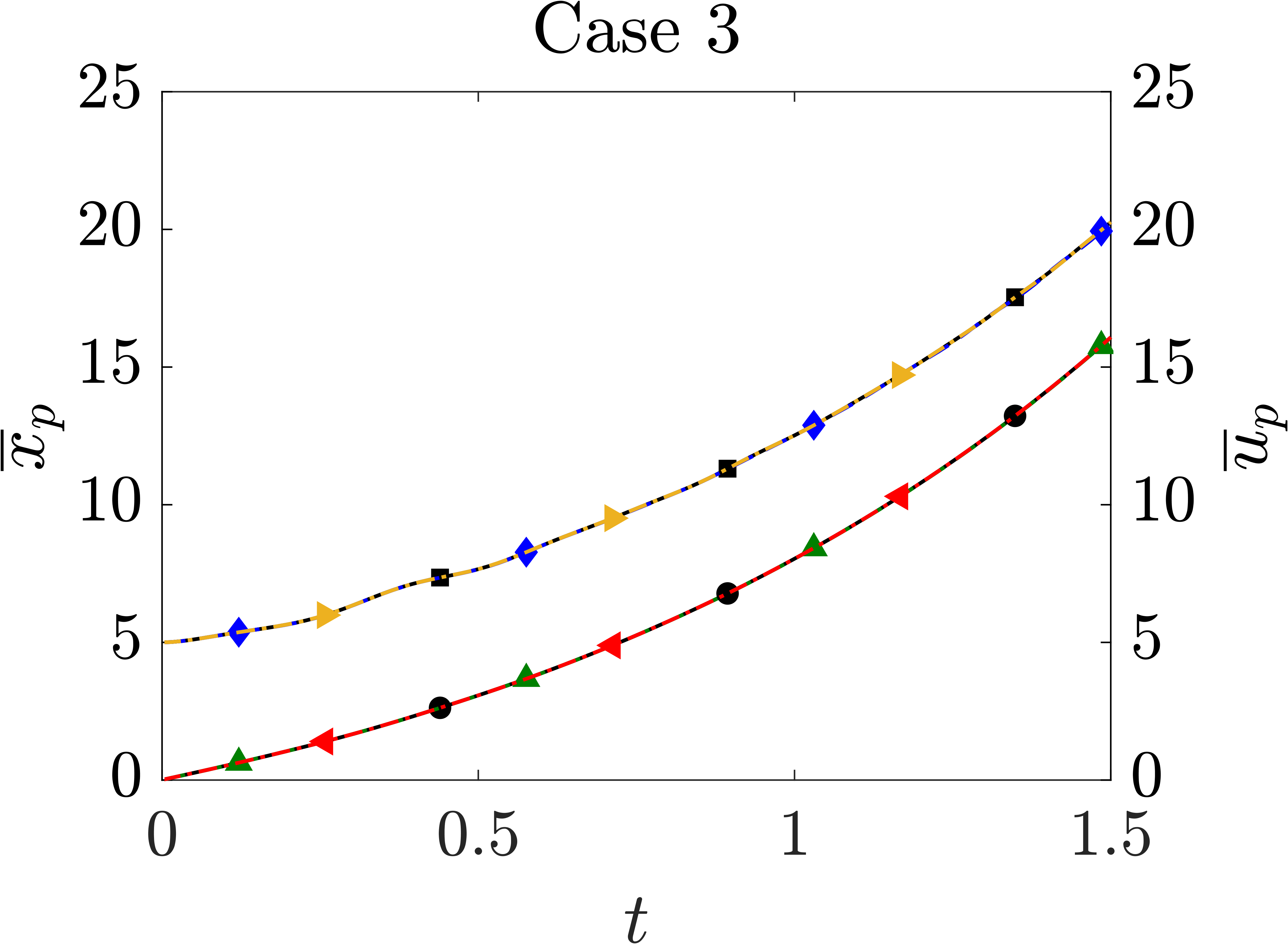}}
        \hfill	
    \subfloat[]{
		\label{fig: 1Dc3_sigma}
		\includegraphics[width=0.33\textwidth]{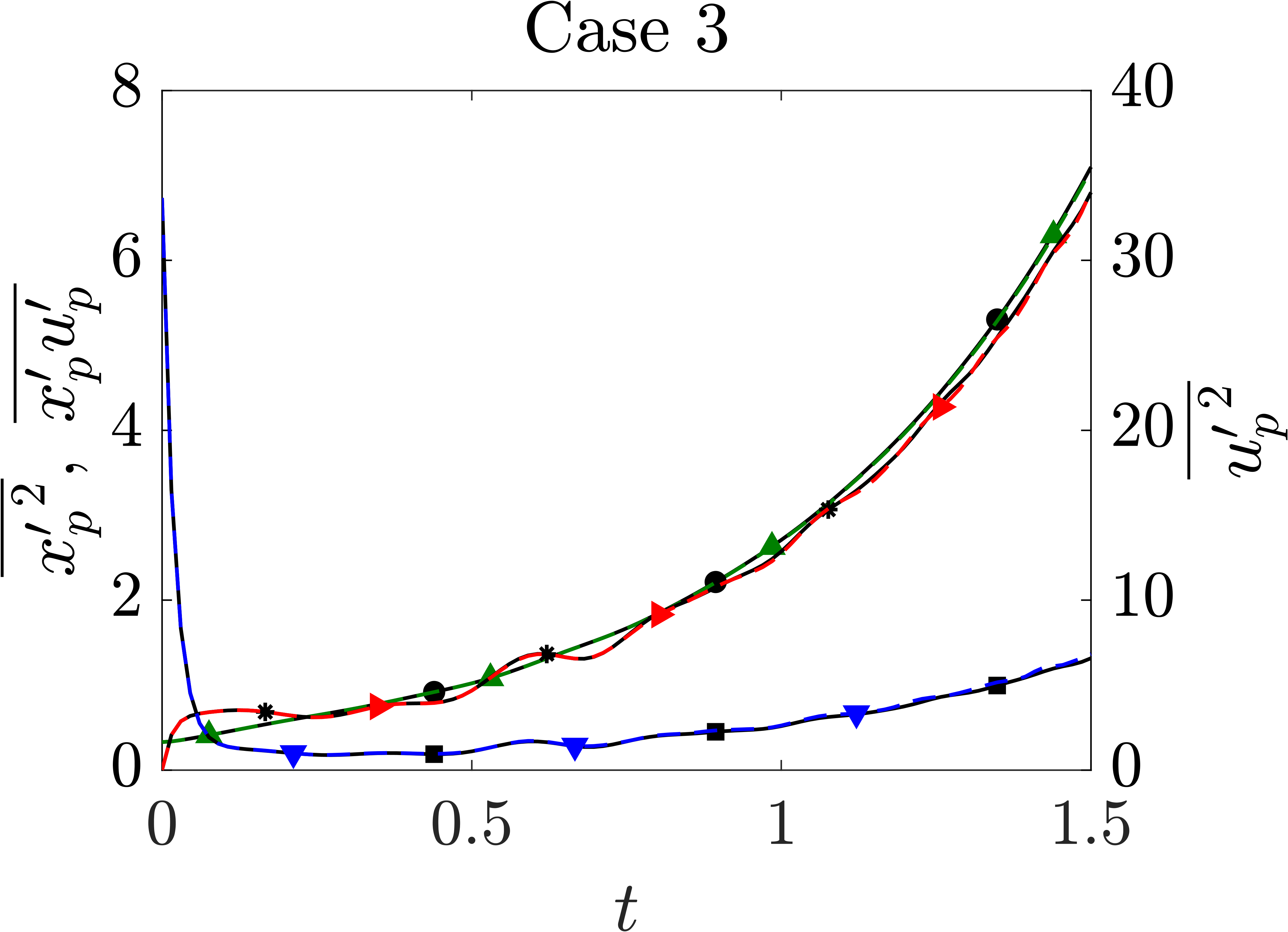}} 
		\hfill
	\subfloat[]{
		\label{fig: 1Dc3_epsilon}
		\includegraphics[width=0.3\textwidth]{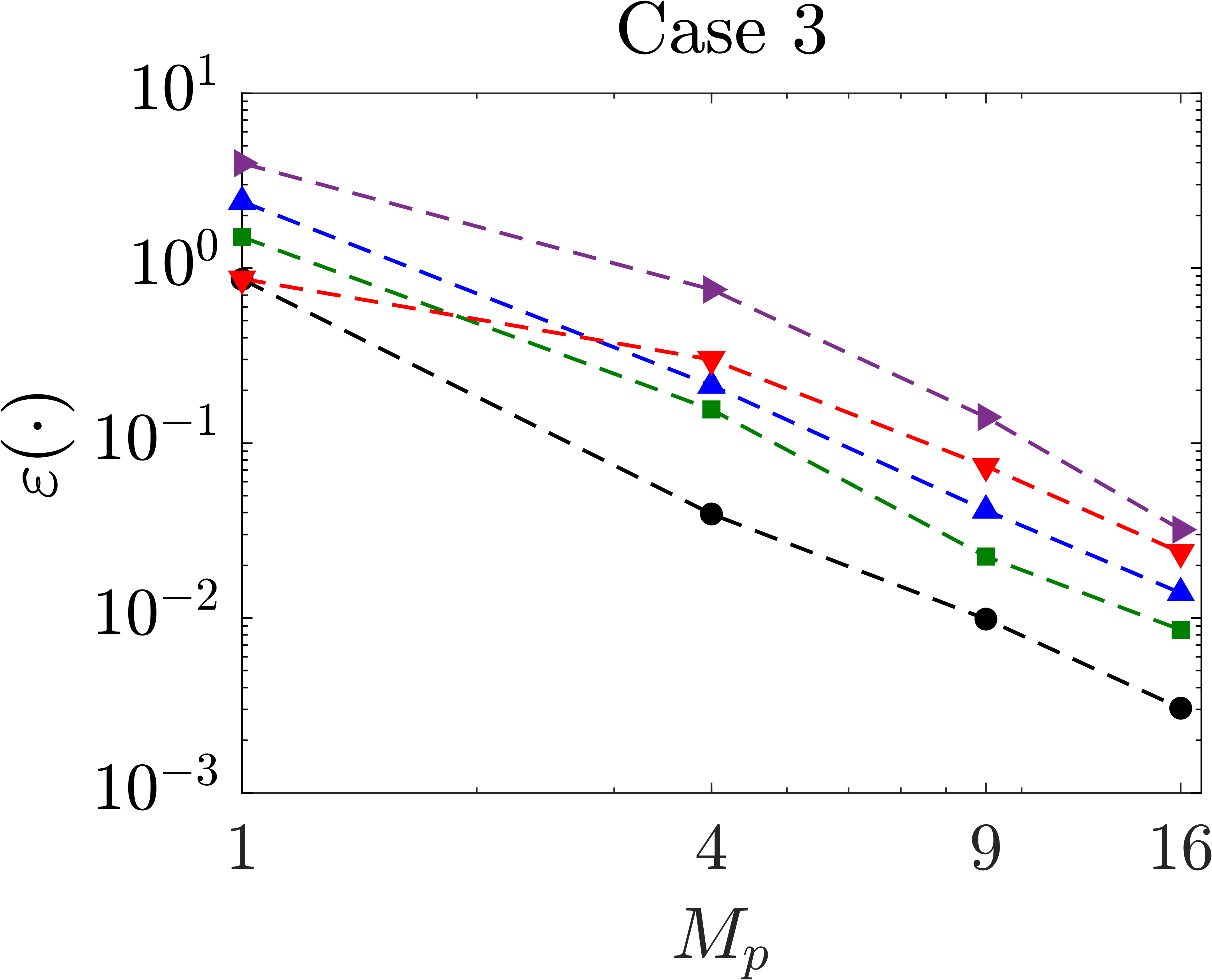}} \\
			\subfloat[]{
		\label{fig: 1Dc4_mean}
		\includegraphics[width=0.32\textwidth]{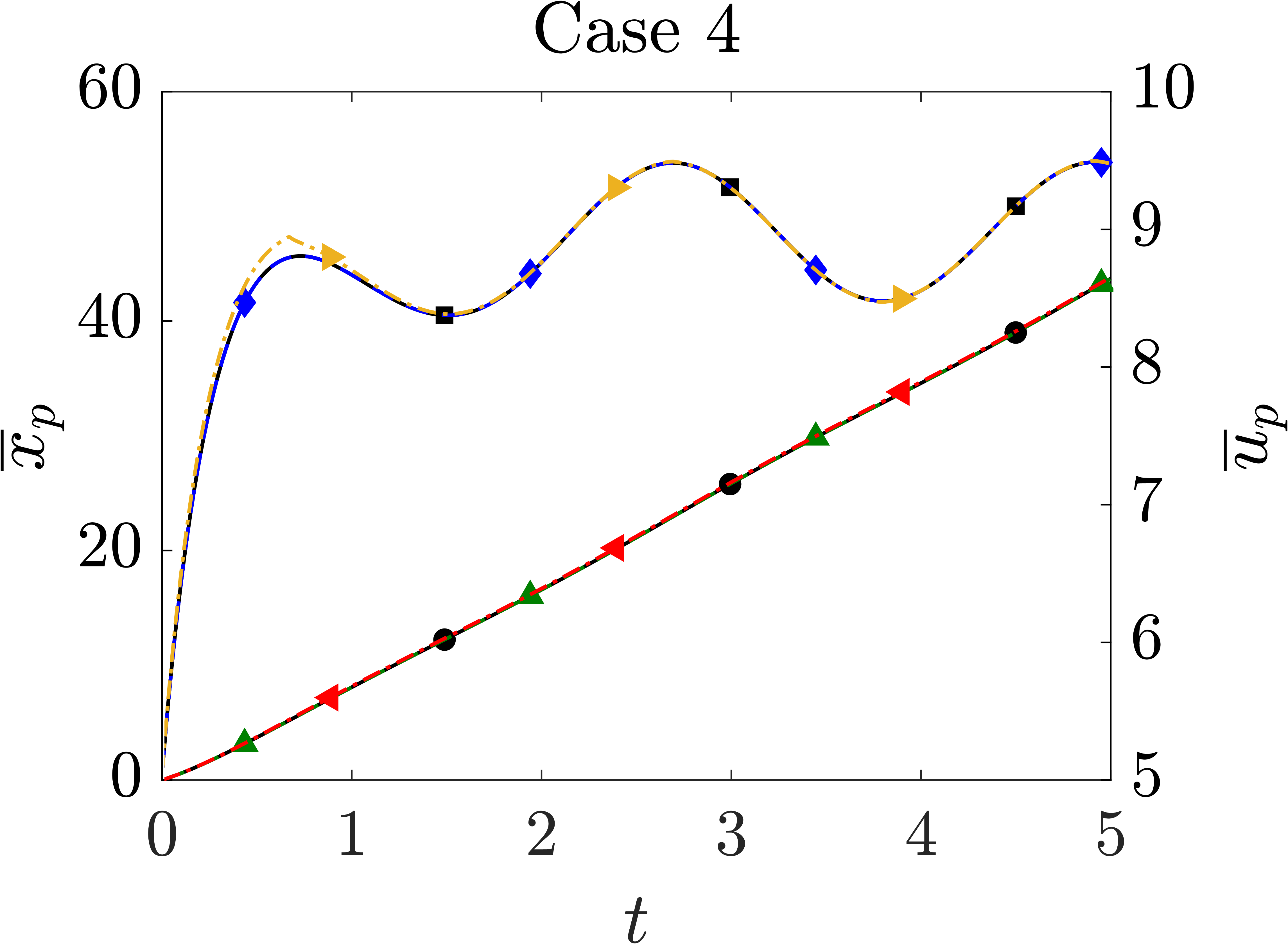}}
        \hfill	
    \subfloat[]{
		\label{fig: 1Dc4_sigma}
		\includegraphics[width=0.33\textwidth]{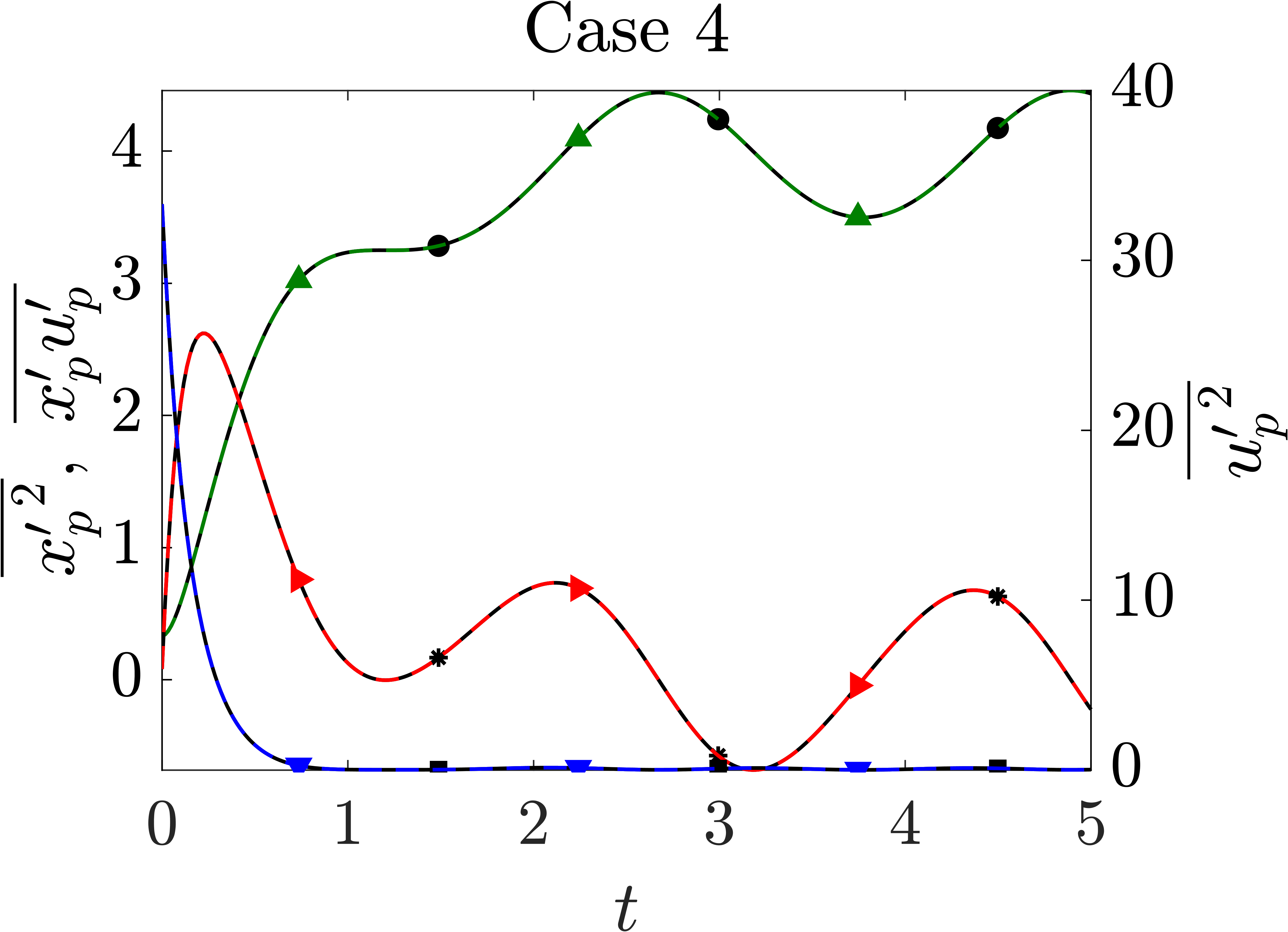}} 
		\hfill
	\subfloat[]{
		\label{fig: 1Dc4_epsilon}
		\includegraphics[width=0.3\textwidth]{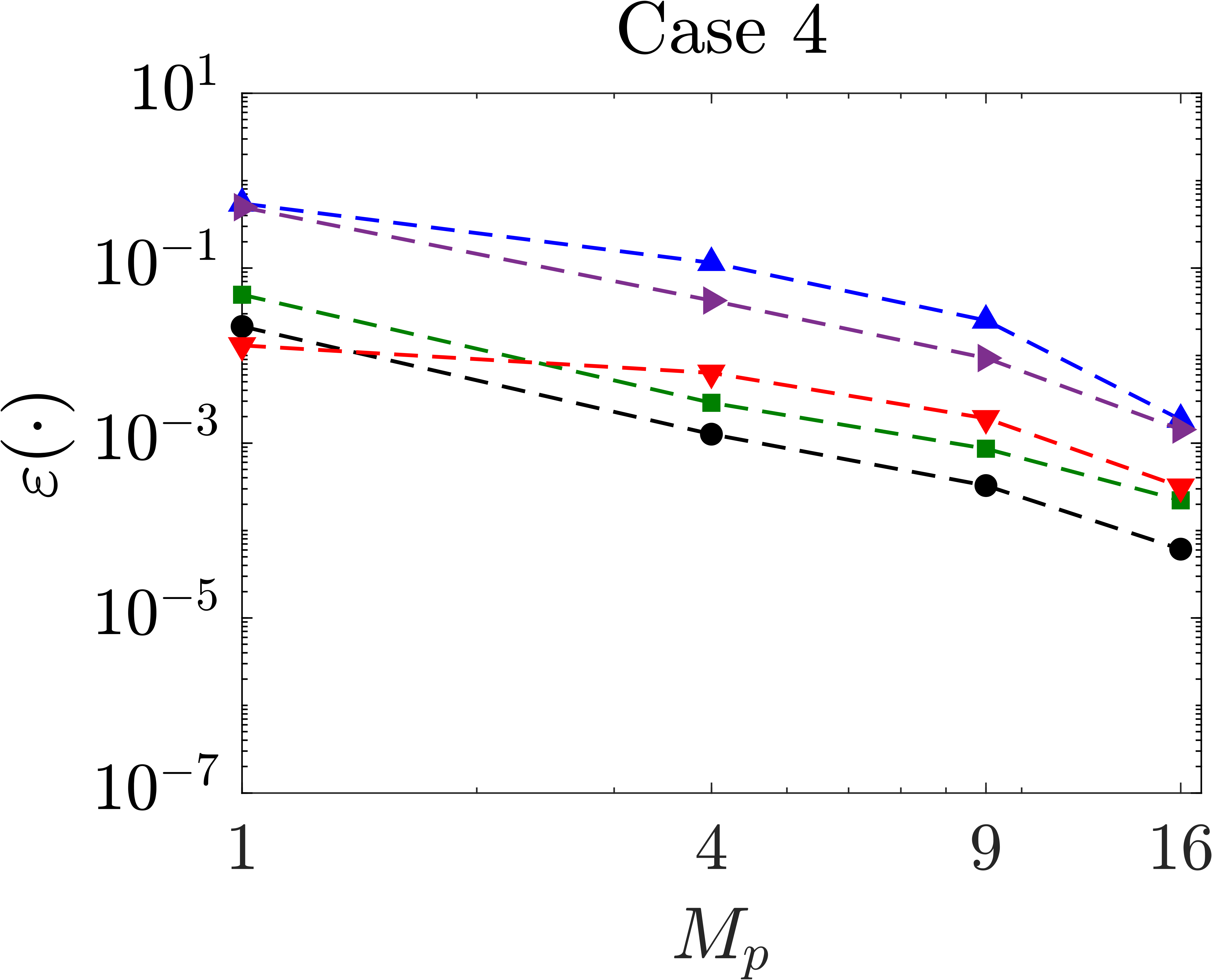}} 
	\caption[]{Comparison of the first (first column) and second (second column) moment results versus time between the closed SPARSE method with $M_p$=16, SPARSE \textit{a priori}, and PSIC for the  four, one-dimensional test cases, Case 1 (a,b), Case 2 (d,e), Case 3 (g,h) and Case 4 (j, k) as summarized in Table \ref{tab:test_cases}. The third column of figures show the error, $\epsilon$, as defined in ~\eqref{eq: error} versus the number of sub-divisions, $M_p$, for SPARSE.
	
	}
	\label{fig: UF_cases_1_and_2}
\end{figure}

Case 2 differs from Case 1 only in the  forcing function, 
which is is selected to be proportional to the \textit{absolute} value of the relative velocity. This results in a more realistic  positive (drag-like) forcing for negative relative velocities.
The first and second moment trends for Case 2 are generally
similar  to Case 1 as shown in Figures~\ref{fig: 1Dc2_mean} and~\ref{fig: 1Dc2_sigma}. Differences, such as the smaller position variance, $\overline{{x_p^\prime}^2}$, for Case 2 as compared to Case 1, can be easily attributed
to the  changes in the positive forcing function, which moves all particles in the positive $x$-coordinate.
The "SPARSE \textit{a priori}" method is showing visible inaccuracies in the mean trace (Fig.~\ref{fig: 1Dc2_mean}), whereas the SPARSE result compares well with PSIC. 
This is related to the subdivisions into sub-clouds for the closed SPARSE method. 
The
"SPARSE \textit{a priori}" method results are generated for a single global cloud without subdivisions.
The closed SPARSE method's subdivision reduces the magnitude of the truncated third order correlation terms per sub-cloud in~\eqref{eq: SPARSE2} 
and thus improves the accuracy of the global mean.  
Without the subdivision the third-order correlation leads to the difference observed in  the "SPARSE \textit{a priori}" model.

\subsection{Constant Forcing in a Harmonically Varying Carrier Velocity Field, Case 3}

In a third one-dimensional test, Case 3,
we take the correction factor constant ($f_1=1$) and specify the carrier-phase  velocity field according to a growing, oscillating function, which lets us investigate 
the effect of the velocity variance in the carrier-phase velocity field.
Specifically, in zeroth order models, the average flow field is poorly approximated at the average particle location such that $\overline{u(x_p)} \simeq {u}(\overline{x}_p)$, as discussed in Davis \textit{et al.}~\cite{davis2013coupling}.
In the closed SPARSE formulation, these fields are determined according to the closure model~\eqref{eq: ab_terms}--\eqref{eq: closure_means_xpu}.

The particle velocity variance trend in Figure~\ref{fig: 1Dc3_sigma} shows a sharp initial drop
in a time interval that is on the order of the Stokes number, after which it gradually grows while the particle phase is accelerated in the increasing carrier-phase  velocity field. At later times ($t>St$), the variance trends also  show  a dominant harmonic mode   of similar frequency as the the oscillating carrier-flow velocity.
This oscillatory effect can also be observed in the  particle location variance.
The average particle velocity trend in  Figure \ref{fig: 1Dc3_mean} closely follows the carrier-flow velocity field at the average particle location because of the small Stokes number and the inherent particles'  fast response to the carrier-flow.

The results are in excellent agreement with the PSIC  and the SPARSE \textit{a priori} results.
The error trends in Figure \ref{fig: 1Dc3_epsilon}  show a monotonic convergence, an indication that
the truncated  terms in the Taylor expansion in ~\eqref{eq: closure_means}--\eqref{eq: closure_means_xpu} are smaller with an increased number of subclouds (per expectation).


\subsection{Empirically Forced Particle Tracers in a Harmonically Varying Carrier Velocity Field, Case 4} \label{sec: tests_1D}

In a final, most demanding, one-dimensional test, Case 4, we assume both the forcing function and the velocity field to have non-trivial, non-linear dependencies per Table \ref{tab:test_cases}. The forcing is set by  the well-known function of Schiller and Naumann~\cite{bird2006transport}
\begin{align}
    f_1=1+0.15 Re_p^{0.687}, 
    \label{eq: f1_SN}
\end{align}
which is accurate for  particle Reynolds numbers of $Re_p=Re_{\infty}|u-u_p|d_p<1000$.
The reference Reynolds number is set to $Re_\infty=10^2$, the relative particle density to $\rho_p=10^3$ and the non-dimensional particle diameter as $d_p=9.478\times10^{-3}$. 
For this case both the forcing and statistical truncation affect the accuracy of the SPARSE solution.

\begin{figure}[htbp] 
	\centering
		\includegraphics[width=0.45\textwidth]{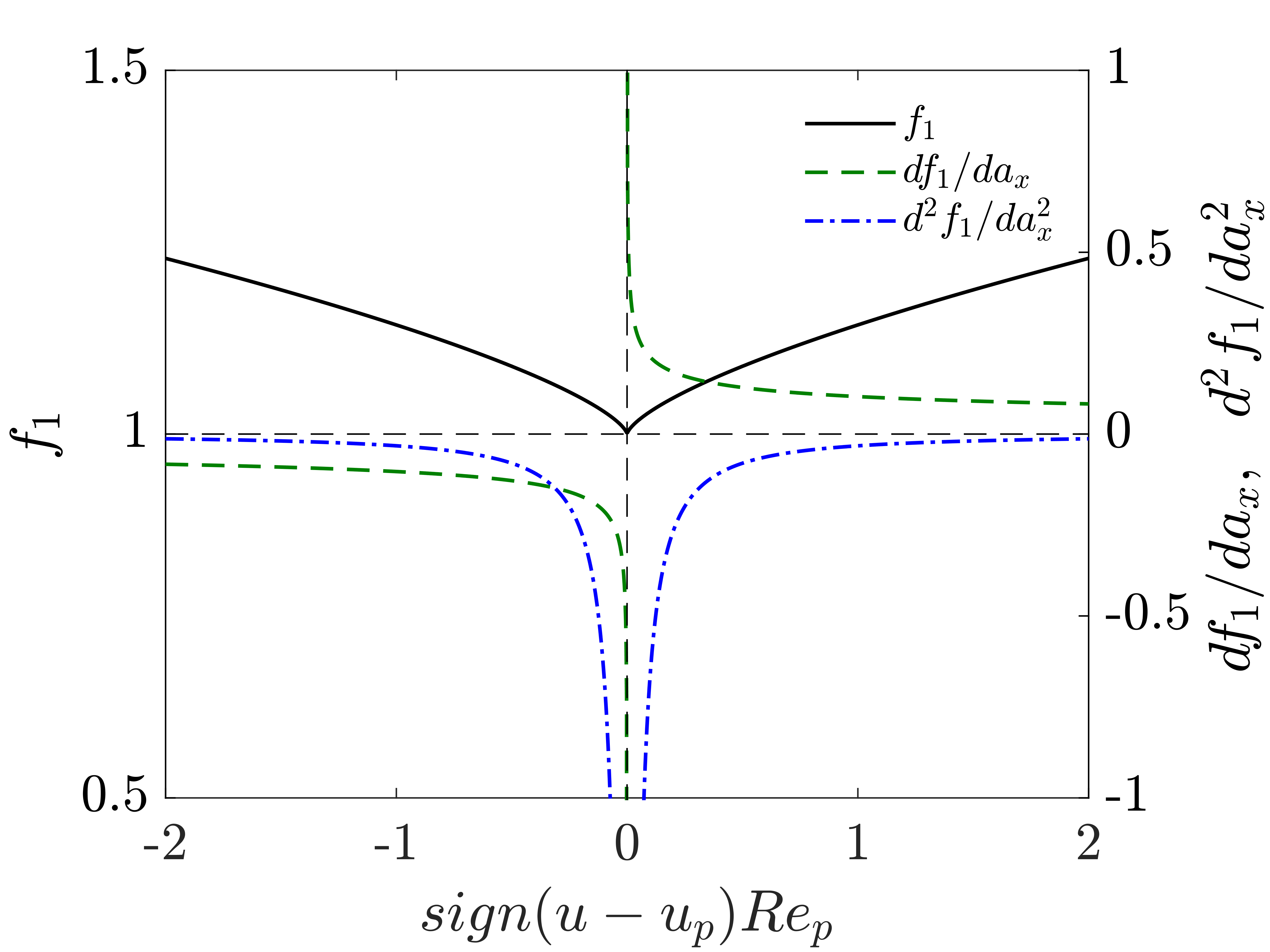}
	\caption[]{Drag coefficient correction factor $f_1$ in terms of the particle Reynolds number $Re_p$ with the sign of the relative velocity.}
	\label{fig: f1_corr_2}
\end{figure}

Before we discuss the SPARSE results, we make a few remarks on the  Schiller and Naumann correction factor which is plotted versus the particle Reynolds number multiplied
by the sign of the relative velocity in
Figure~\ref{fig: f1_corr_2}. Also plotted are its first two derivatives with respect to the relative velocity $a_x=u-u_p$. The second derivative  shows a singularity in the zero limit
of the particle Reynolds number. This singularity can negatively affect 
 accuracy through the terms that involves a second derivative in~\eqref{eq: SPARSE2}.
This can occur if a SPARSE cloud experiences a change from acceleration to deceleration along its  trajectory.
To avoid the singularity we neglect the drag force correction effect and its derivatives by setting $f_1$ to unity for $Re_p<0.1$ leading to the Stokes drag.

The particle phase's mean and variance trends are plotted in Figures \ref{fig: 1Dc4_mean} and \ref{fig: 1Dc4_sigma}, respectively, and show that
the particle cloud accelerates initially over a time proportional to the particle response time, until it reaches an oscillating plateau.
Coinciding with this acceleration,  the particle velocity variance reduces from its initial value to an oscillating trend with minima of approximately zero.
The cloud size, proportional to the particle location variances,  changes with the changes in the average relative velocity $\overline{u}-\overline{u}_p$: the cloud grows when $\overline{u}-\overline{u}_p>0$ and shrinks when $\overline{u}-\overline{u}_p<0$.
In transitioning from acceleration to deceleration, the front of the cloud decelerates faster than the tail, causing a switch in the relative velocity of the particles in the front with respect to the ones in the tail. 
Eventually, the cloud reaches a state in which the average relative velocity of the cloud is zero and the deviation of the particle velocity experiences a minimum, associated with a zero rate of change of the cloud size for that instant of time.

The SPARSE results are in excellent agreement with PSIC and SPARSE \textit{a priori}, 
verifying the closed SPARSE method.
The error reduces once again monotonically with an increase number of subdivisions of the cloud as shown in Figure~\ref{fig: 1Dc4_epsilon}. 
The convergence rate is slightly smaller as compared to previous cases, as the Case 4 requires convergence of  \textit{both}  the truncated Taylor series terms and the truncated, higher-order moment terms, where the accuracy for Case 1-3 is impacted by only one of the two truncations.

\section{Two- and Three-Dimensional,  One-Way Coupled, Particle-Laden Flow Tests} \label{sec: 2and3Dtests}

\subsection{Stagnation Flow}\label{sec: tests_SF}

To test the two-dimensional closed SPARSE formulation, we first consider a cloud traced in a carrier-phase velocity field according to the analytical stagnation flow solution of Hiemenz~\cite{hiemenz1911grenzschicht} for an inviscid irrotational fluid, in the domain $x\in [-\infty,0]$ as follows
\begin{subequations}\label{eq: SF_flow}
\begin{align}
    u &= -k x, 
    \label{eq: SF_flow_u} \\ 
    v &= k y,
    \label{eq: SF_flow_v}
\end{align}
\end{subequations}
where $y$ is the coordinate perpendicular to the flow direction, and $k$ is
constant set to unity $k=1$.
To initialize a cloud of particles at rest, we sample from a uniform probability density distribution function with average location $\overline{x}_p=-1$ and $\overline{y}_p=0$ and with deviations in space given by $\sigma_{x_p}=\sigma_{y_p}=0.05$.
Because the particles are initialized at rest, the average and variance of the vertical and horizontal velocity components are zero, as well as any other moment involving a velocity component.
Particles in the cloud are forced according to the Stokes drag corrected with the Schiller and Naumann correlation  in~\eqref{eq: f1_SN}.
The reference Reynolds number in~\eqref{eq: Rep_St_Pr} is set to $Re_\infty=10^4$.
The Stokes number is selected to be unity $St=1$  and the particle to fluid density ratio is set to $\rho_p=10^3$.

A PSIC computation is performed for reference to determine the error of the closed SPARSE formulation. 
Because of the sampling error, well known to be proportional to $1/\sqrt{N_p}$, the moments of the sampled cloud differ from the uniform distribution used for the seeding. 
The average location of the sampled initial condition is $\overline{x}_p=-0.998$, $\overline{y}_p=1.23\times10^{-3}$ and the deviations $\sigma_{x_p}=4.98\times10^{-2}$ and $\sigma_{y_p}=4.95\times10^{-2}$.
The correlation is  $\overline{x_p^\prime y_p^\prime}=2.927\times10^{-5}$ at time zero.
The remainder of the moments are zero  because the cloud is at rest initially.
To initialize a single SPARSE cloud ($M_p=1$) we specify the initial condition according 
and consistently with the PSIC moments. With this initial condition,  inaccuracies in the evolution of the third moment mostly affect the comparison between PSIC and closed SPARSE (see third bullet point in the Remarks on pp. 7), not in the least because the Taylor expansion of the linear velocity field in the stagnation flow case is exact and errors in the truncation of the Schiller and Naumann function are relatively small.

\begin{figure}[htbp]
    \centering
	\includegraphics[width=0.5\textwidth]{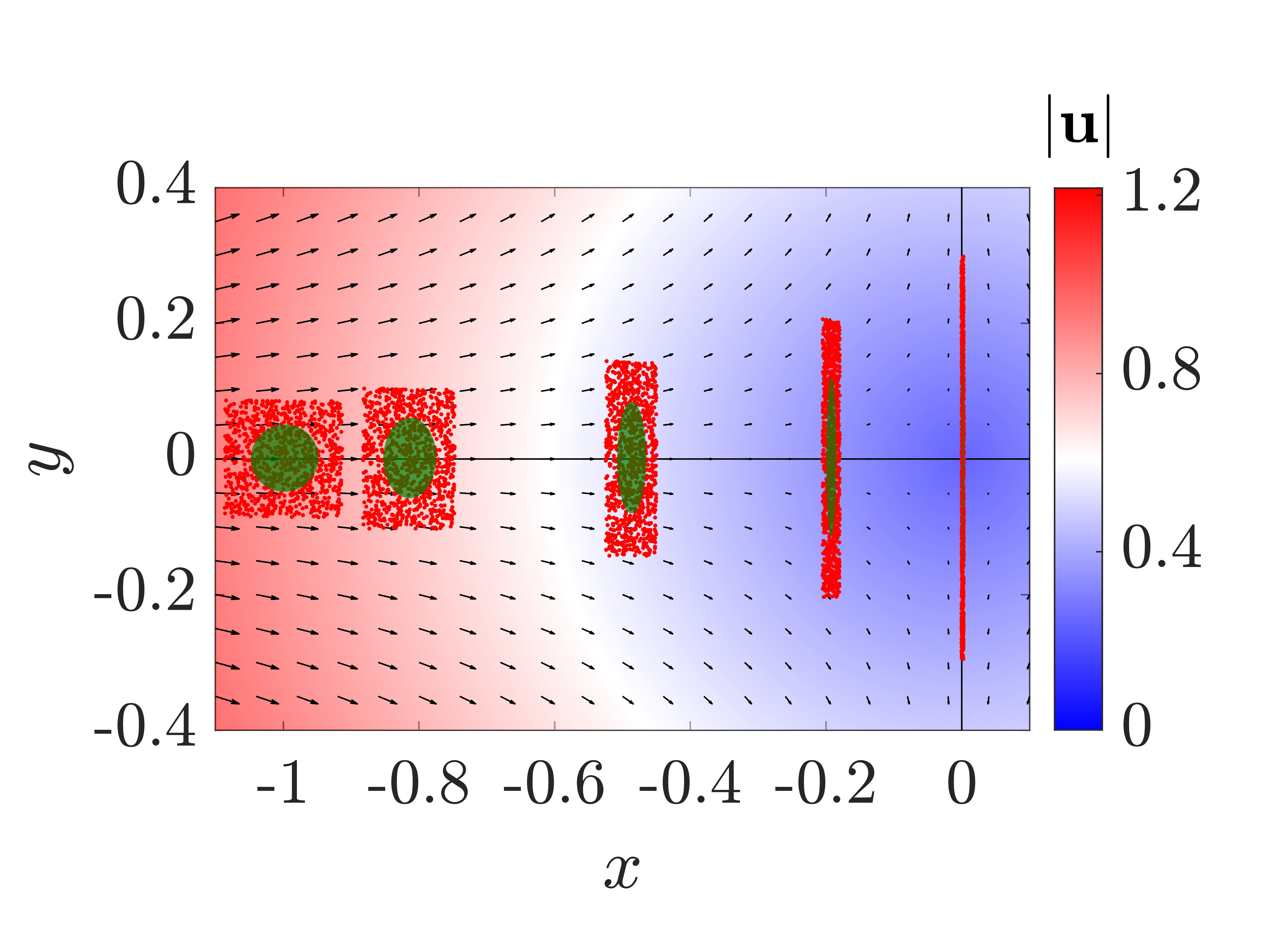}
    \caption[]{Evolution of the particle cloud immersed in the stagnation flow for different instants of time $t=[0, \ 0.5, \ 1.65, \ 2.2]$. The red dots represent the particles inside the cloud traced by a PSIC simulation and the green macro-particle is given by the SPARSE method. The background is colored according to the modulus of the stagnation flow field $\boldsymbol{u}=(u,v)^\top$.}
	\label{fig: SF_contour}
\end{figure}

\begin{figure}[htbp]
	\centering
	\subfloat[]{
		\label{fig: SF_mean_xpup}
		\includegraphics[width=0.31\textwidth]{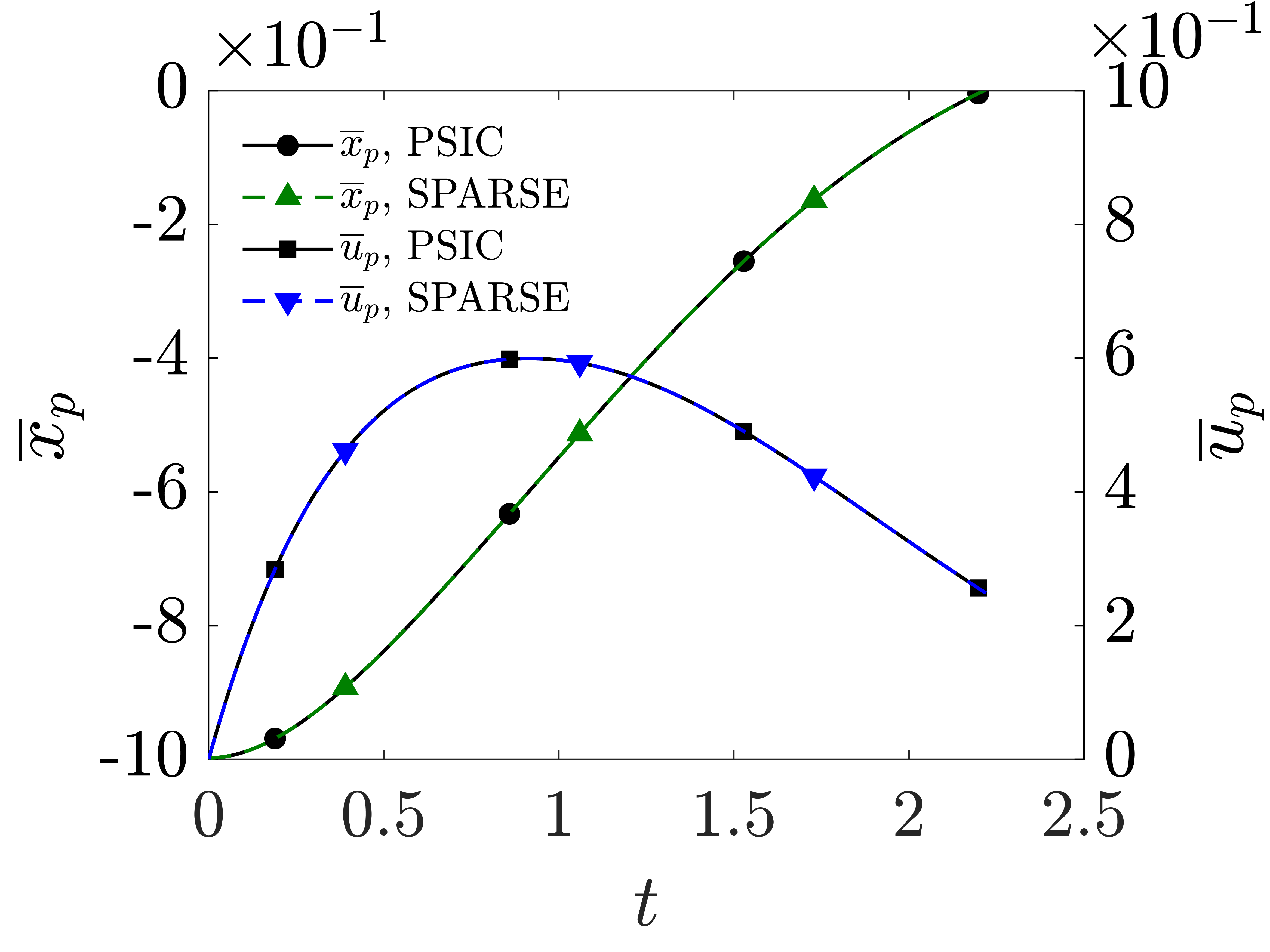}}
        \hfill	
	\subfloat[]{
		\label{fig: SF_mean_ypvp}
		\includegraphics[width=0.31\textwidth]{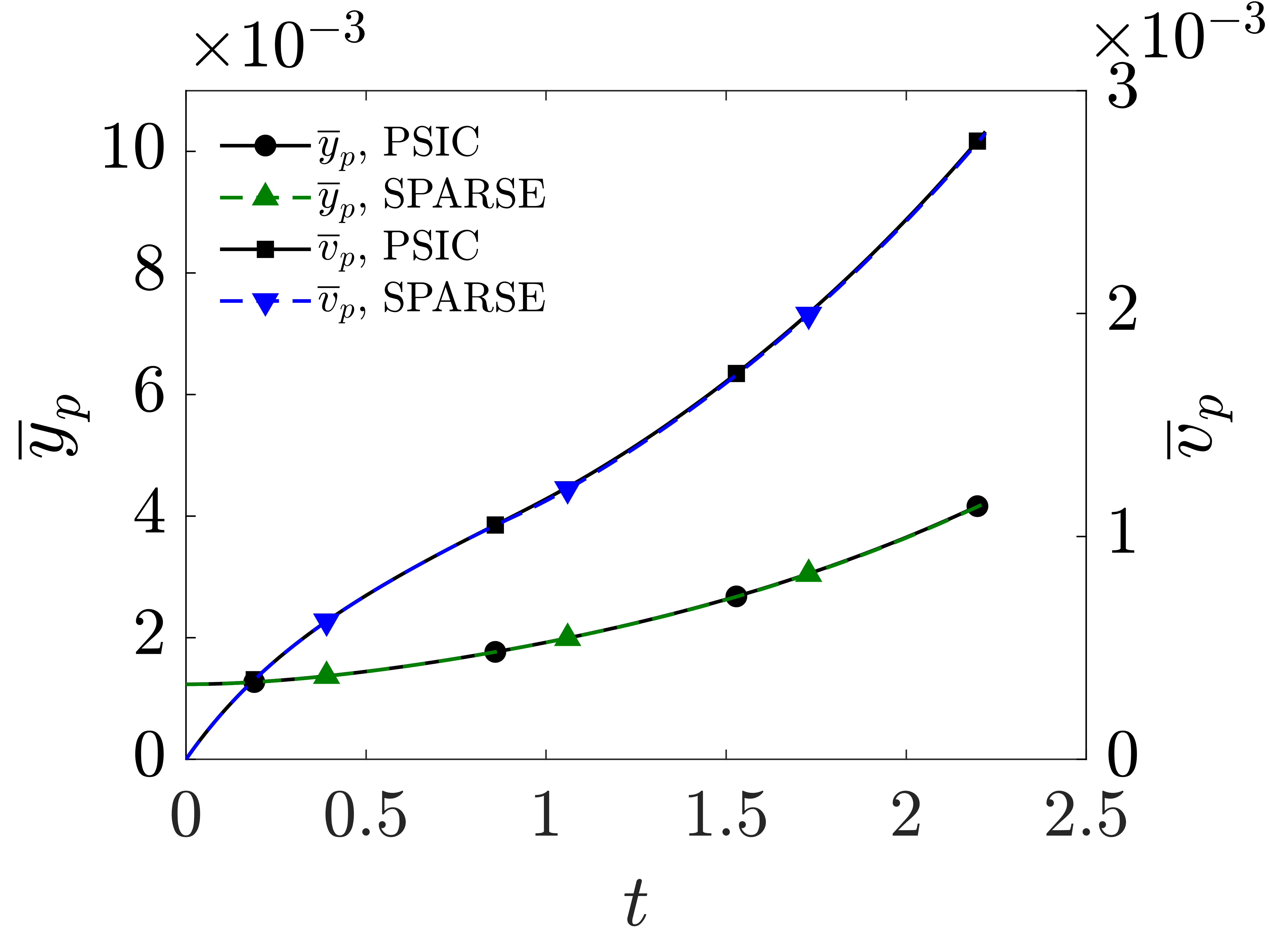}}  
		\hfill
	\subfloat[]{
		\label{fig: SF_cm2_xpypupvp}
		\includegraphics[width=0.31\textwidth]{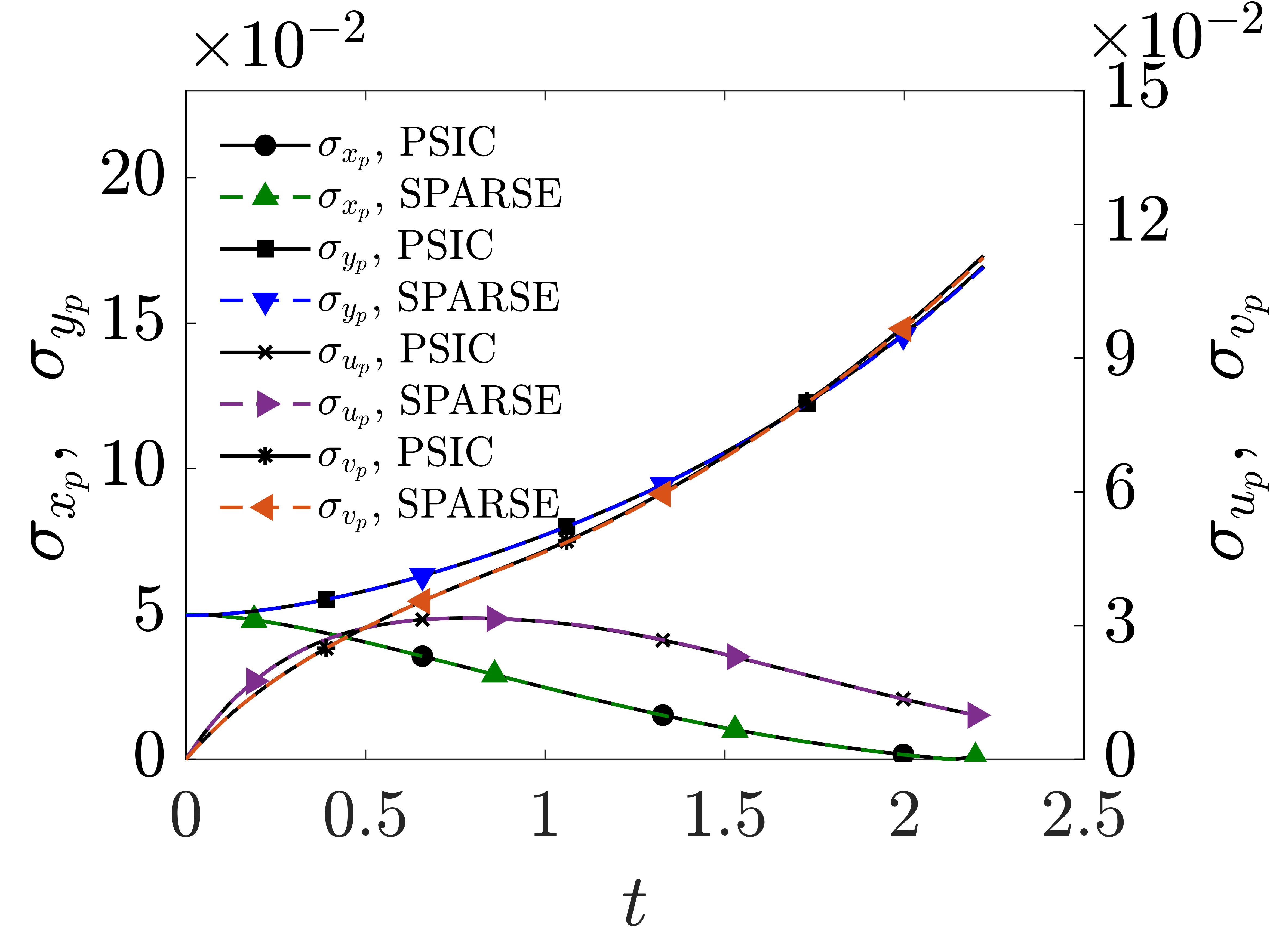}} \\  
	\subfloat[]{
		\label{fig: SF_xpup_ypvp}
		\includegraphics[width=0.31\textwidth]{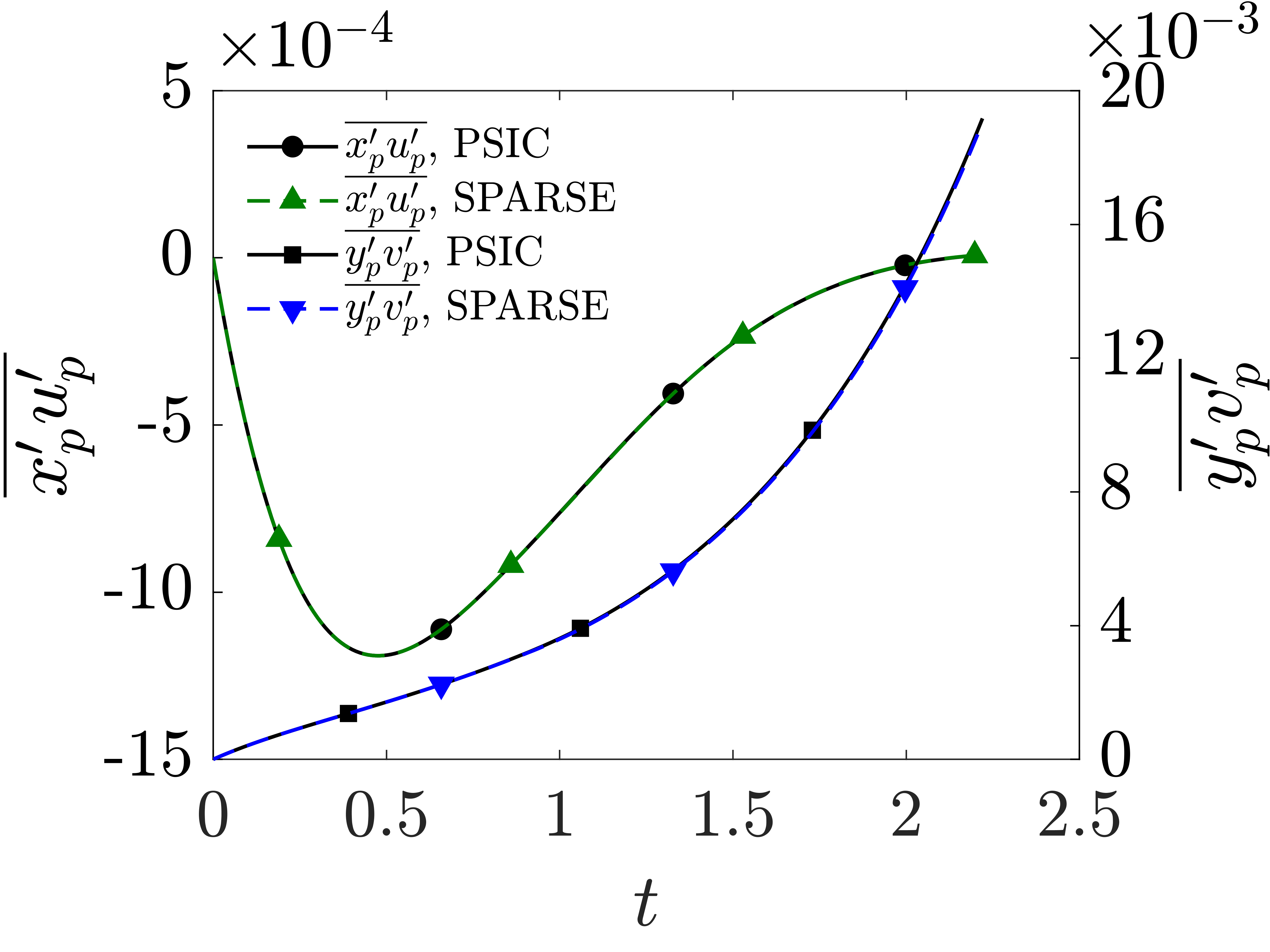}} 
		\hfill
	\subfloat[]{
		\label{fig: SF_xpyp_upvp}
		\includegraphics[width=0.31\textwidth]{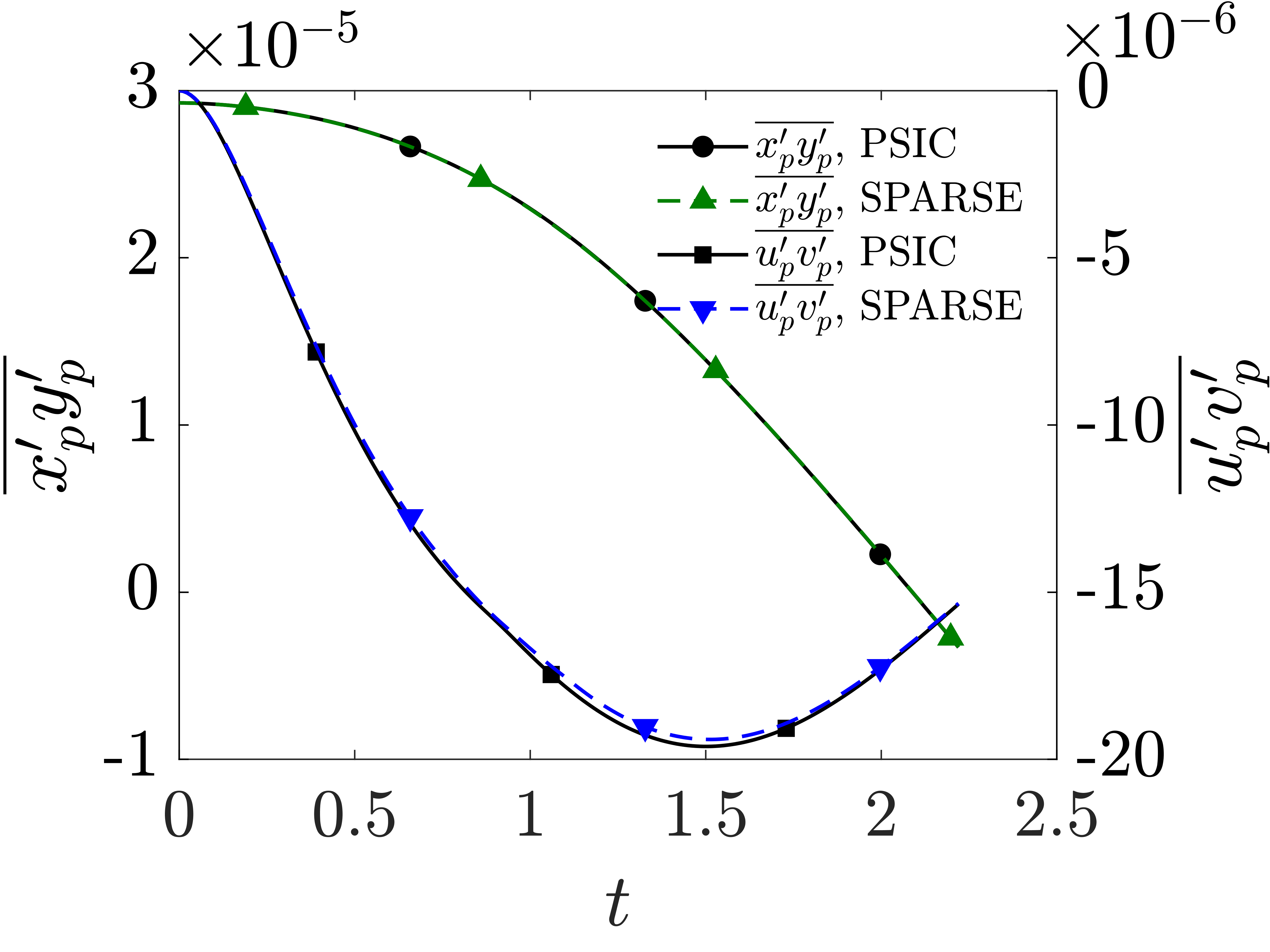}} 
	    \hfill
	\subfloat[]{
		\label{fig: SF_xpvp_ypup}
		\includegraphics[width=0.31\textwidth]{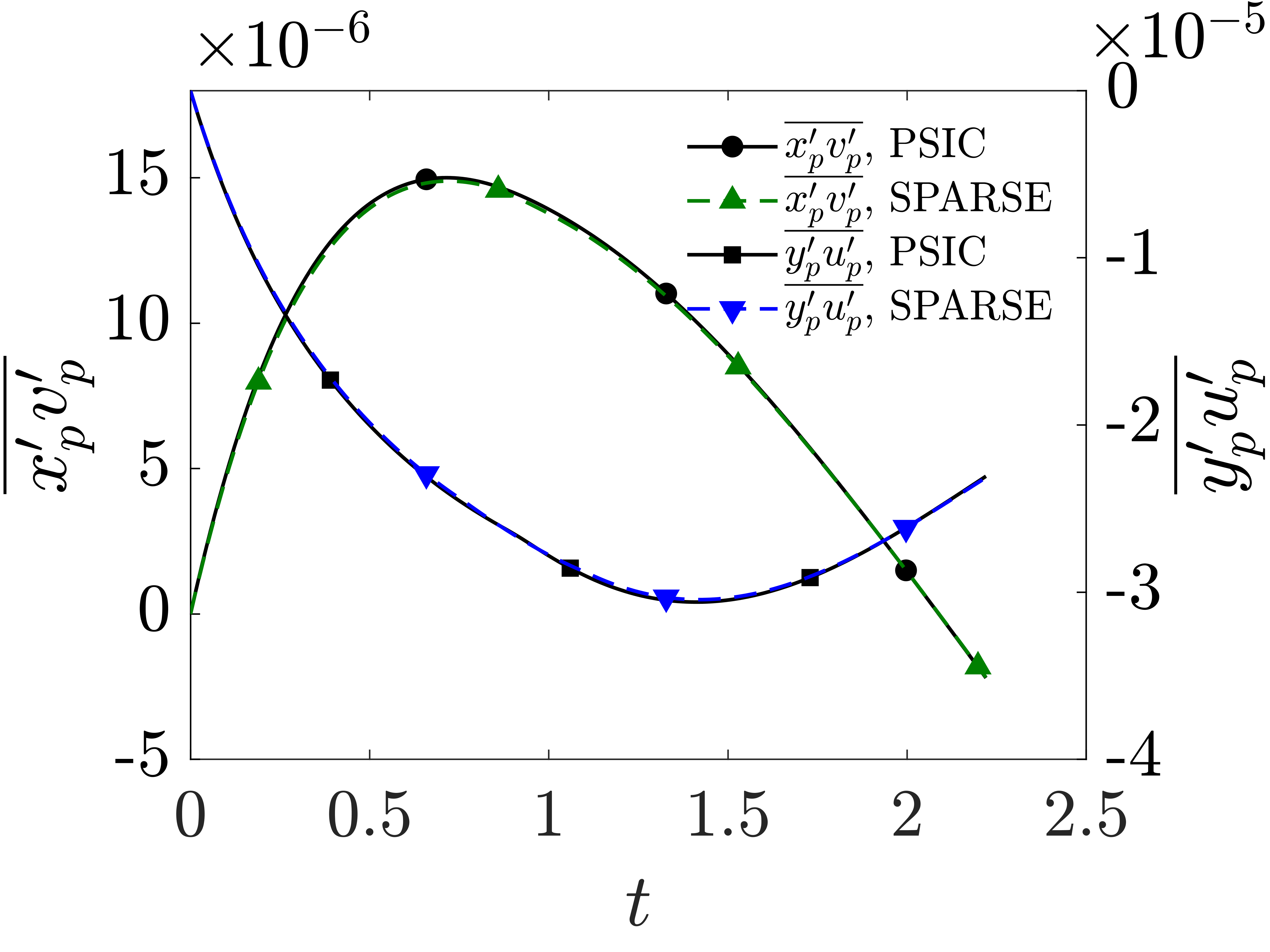}} 
	\caption[]{Averages (a) and deviations (b) of the particle phase for the closed SPARSE method and the PSIC method.}
	\label{fig: SF}
\end{figure}

The traces of PSIC particles (red dots) and SPARSE clouds (green ellipses) are compared in Figure~\ref{fig: SF_contour} for several instances of time.  
The radii of the ellipse and its orientation are set according to the eigenvalues and eigenvectors of the covariance matrix of the cloud's location in $x$- and $y$-direction computed from the SPARSE variables. 
The cloud compresses and expands in $x$- and $y$-direction, respectively, as it traverses the stagnating velocity field.
The evolution of the first two moments computed with both approaches are depicted in Figure~\ref{fig: SF}. 

The trends of the $x$-location of the cloud can be divided into three stages (see \cite{dominguez2021lagrangian} for a detailed discussion).
In a first stage, all the particles in the cloud accelerate with a positive relative velocity $u_i-{u_p}_i>0$ for $1\leq i \leq N_p$ towards a linearly decreasing carrier flow velocity.
At some point the carrier-flow velocity becomes less as compared to the velocity of some of the particles in the cloud. 
In this second stage, the cloud changes from having all the particles accelerating to all decelerating, producing a maximum in the average particle velocity $\overline{u}_p$ at approximately $t=1$ (see Figure~\ref{fig: SF_mean_xpup}). 
Correspondingly, the particle location trend changes from a parabolic increase to a linear increase.
After all the particles have crossed the zero relative velocity, all particles in the cloud decelerate towards the stagnation point, defining the third stage where $\overline{x}_p$ describes a parabolic downward trend.
The variances of the particle phase velocity in the $x-$direction follow a similar trend of increase and decrease as shown in Figure~\ref{fig: SF_cm2_xpypupvp} as its average counterpart.
The range of horizontal velocities grows as the cloud accelerates and decreases in the third stage when decelerating, showing a maximum in the second stage.
The horizontal size of the particle $\sigma_{x_p}$ decreases from its initial value as the cloud reaches the stagnation point.
Second  order correlations are depicted in Figures~\ref{fig: SF_xpup_ypvp}--\ref{fig: SF_xpvp_ypup}.
The three stages are once again observed.


Figures~\ref{fig: SF_contour} and~\ref{fig: SF} show that closed SPARSE 
is  accurate within  1.5\%  compared to the PSIC results for a time  period on the order of at least three characteristic time scales.
The relative error between the PSIC and SPARSE computations is related to the truncation of the third moments in the SPARSE equations and the Taylor expansion of the correction factor of the drag force $f_1$ given by~\eqref{eq: f1_SN}.
The matching between both approaches leads to a relative error less than $1\%$ for all moments except for the maximum relative error of $\overline{u_p^\prime v_p^\prime}$ which is $1.5\%$, which indicates that the most sensitive variable to third order moments in the cloud is the correlation between the velocity components.
The maximum relative error of the averages is $0.3\%$ for $\overline{v}_p$ and the one of the deviations is $0.8\%$ for $\sigma_{v_p}$.
Because of the very good match for a single SPARSE cloud with a small number of PSIC particles, we do investigate the effect of splitting clouds for  this case.

\subsection{ABC flow}\label{sec: tests_ABCflow}


Closed SPARSE is tested in three-dimensions by tracing clouds in the three-dimensional analytical velocity field of the so-called ABC flow. The ABC flow was introduced by Arnol'd~\cite{arnold1965topologie} as part of the family of Beltrami flows satisfying that $\nabla \wedge \boldsymbol{u}=\boldsymbol{u}$.
Any ABC flow is an exact steady solution of the Navier-Stokes equations
\begin{subequations}\label{eq: ABCflow}
\begin{align}
    \frac{\partial \boldsymbol{u}}{\partial t}+\boldsymbol{u}\cdot \nabla \boldsymbol{u}&=-\nabla p +\nu \nabla^2\boldsymbol{u}+\boldsymbol{f}, \\ 
    \nabla \cdot \boldsymbol{u}&=0,
\end{align}
\end{subequations}
where without loss of generality the density is assumed to be the unity, $p$ is the pressure, $\nu$ the dynamic viscosity and the forcing $\boldsymbol{f}$ is giving by
\begin{align}
    \boldsymbol{f}=\nu \left( A\sin{z}+C\cos{y}, \ B\sin{x}+A\cos{z}, \ C\sin{y}+B\cos{x} \right),
\end{align}
where for small Reynolds number, i.e., $\nu \gg 1$, the only stable solution is given by the ABC flow field $\boldsymbol{u}=(u,v,w)$ with
\begin{align}
    u &= A\sin{z}+C\cos{y}, \\
    v &= B\sin{x}+A\cos{z}, \\
    w &= C\sin{y}+B\cos{x}.
\end{align}
The ABC flow has been extensively used to study chaotic effects in turbulence~\cite{olivieri2020turbulence,rorai2013helicity} and non-linear dynamics~\cite{haller2001distinguished,froyland2012finite}.
Here, we use it to test SPARSE clouds immersed in the ABC carrier-phase flow.

We set the constants of the carrier flow field to $A=\sqrt{3}$, $B=\sqrt{2}$ and $C=1$.
A hundred thousand particles $N_p=10^5$ are released at rest with initial location $\overline{x}_p=\overline{y}_p=\overline{z}_p=\pi$ and variances $\sigma_{x_p}=\sigma_{y_p}=\sigma_{z_p}=0.02$ according to a uniform distribution where all variables are statistically independent at the initial time.
The initial velocity averages and variances are zero since all point-particles are at rest initially.
The correction factor $f_1$ used is taken according to~\eqref{eq: f1_SN} and the SPARSE simulation is computed with a single SPARSE cloud $M_p=1$.
Similar to the stagnation flow test case, the sampled initial condition for PSIC is used as initial condition for the closed SPARSE simulation.
The reference Reynolds number is set to unity $Re_\infty =1$ and the density
ratio is $\rho_p= 0.424\times10^4$. Four  clouds with different Stokes numbers $St=[1, \ 2, \ 5, \ 10]$ are traced. 

\begin{figure}[htbp]
	\centering
	\includegraphics[width=0.6\textwidth]{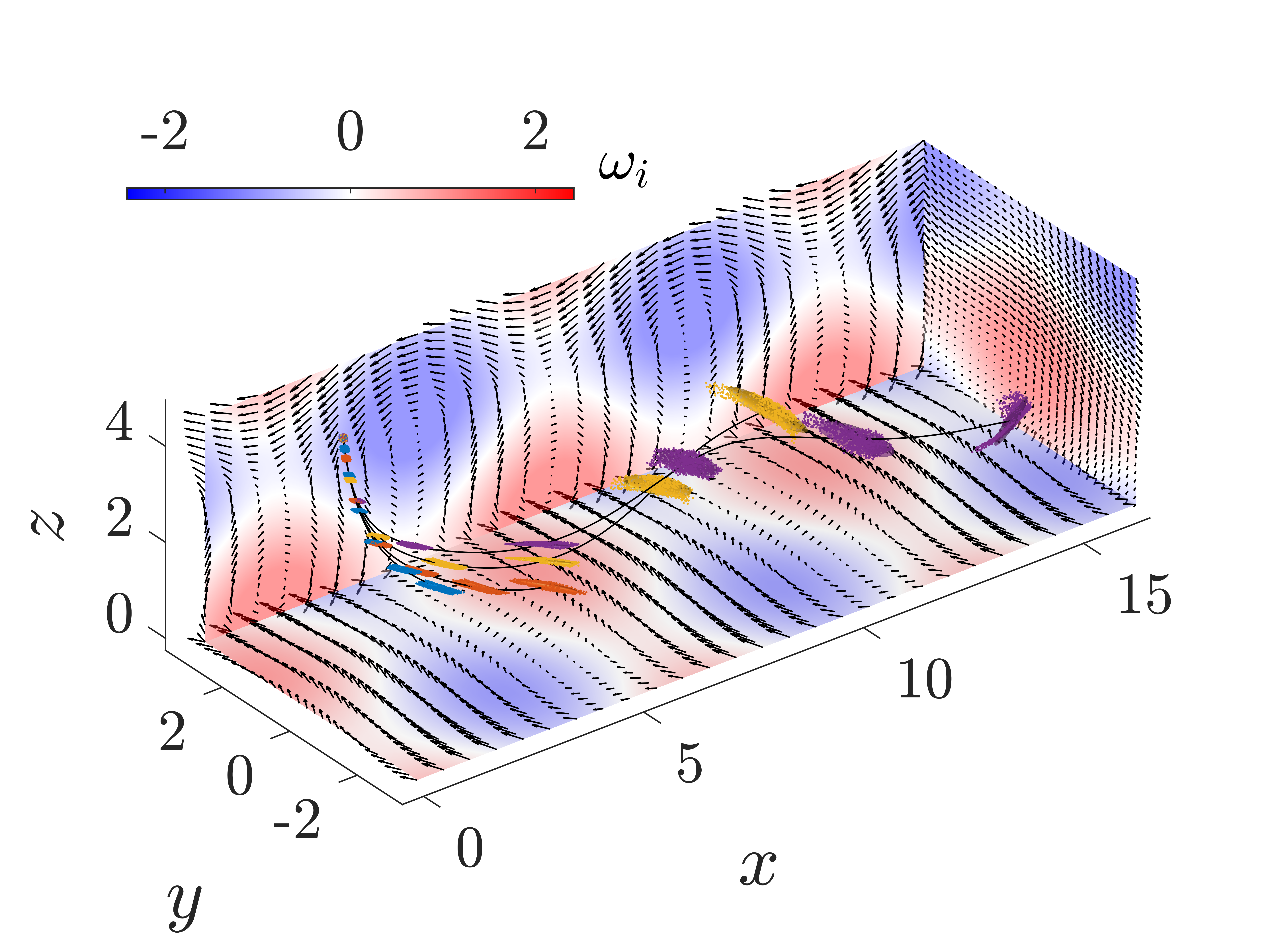}
	\caption[]{Four SPARSE cloud trajectories carried by an ABC flow for Stokes numbers $St=[1, \ 2, \ 5, \ 10]$ in colors purple, yellow, orange and blue, respectively. Along the path, the the SPARSE clouds are depicted with prolates scaled by the principle strains given by the second moments and the point-particles traced with PSIC are depicted as points for different instants of time $t=[0,\ 1.37,\ 2.67,\ 4.04,\ 5.03,\ 6.7,\ 8]$. Only $10^3$ out of the $10^5$ point-particles per cloud are plotted for a better visualization. The carrier-flow is visualized with vorticity contour plots and velocity vector fields.}
	\label{fig: ABC_3Dplot}
\end{figure}

\begin{figure}[htbp]
	\centering
	\subfloat[]{
	    \label{fig: ABC_St2_mean_xp}
	    \includegraphics[width=0.31\textwidth]{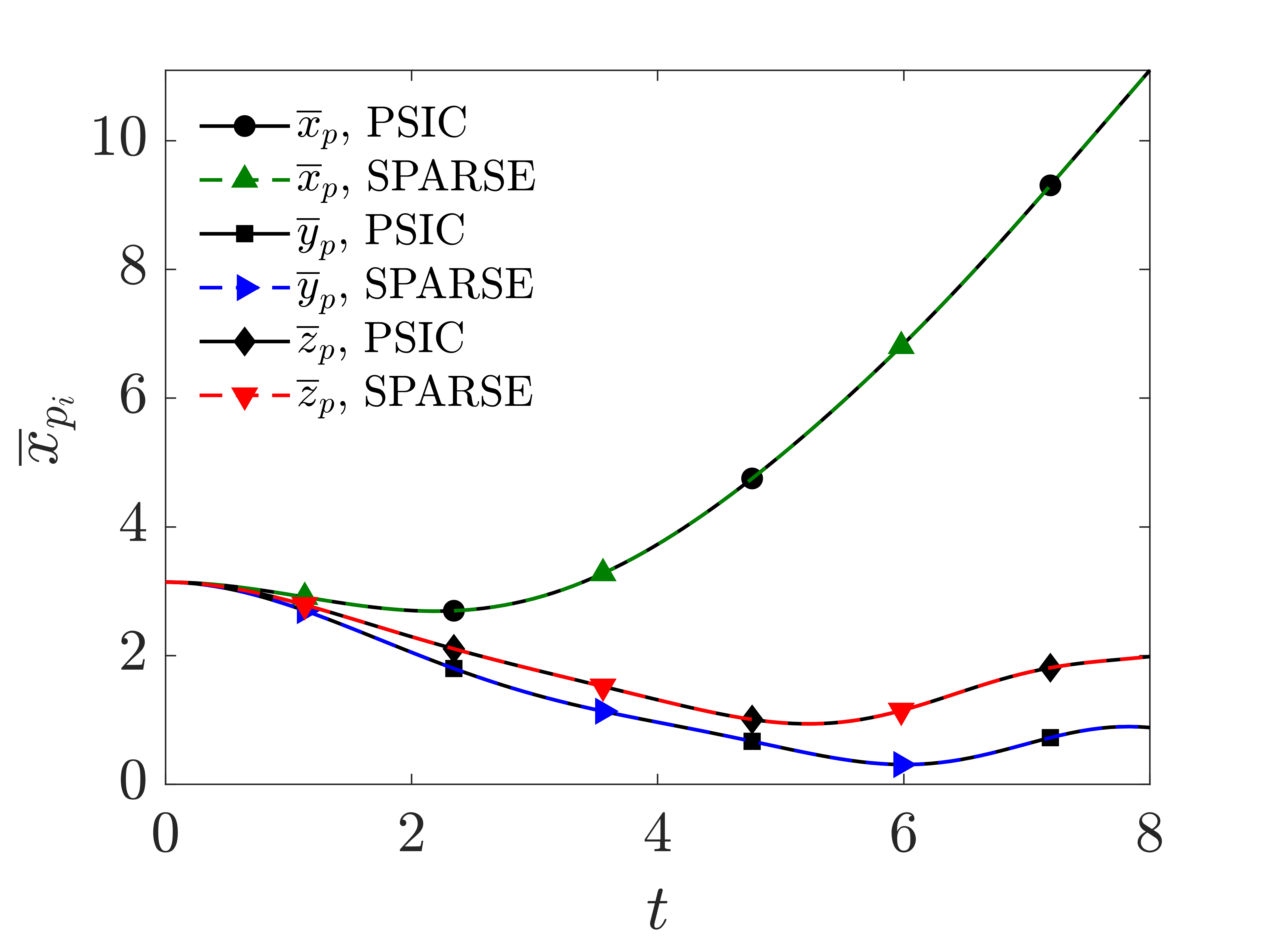}}
        \hfill	
	\subfloat[]{
		\label{fig: ABC_St2_mean_up}
		\includegraphics[width=0.31\textwidth]{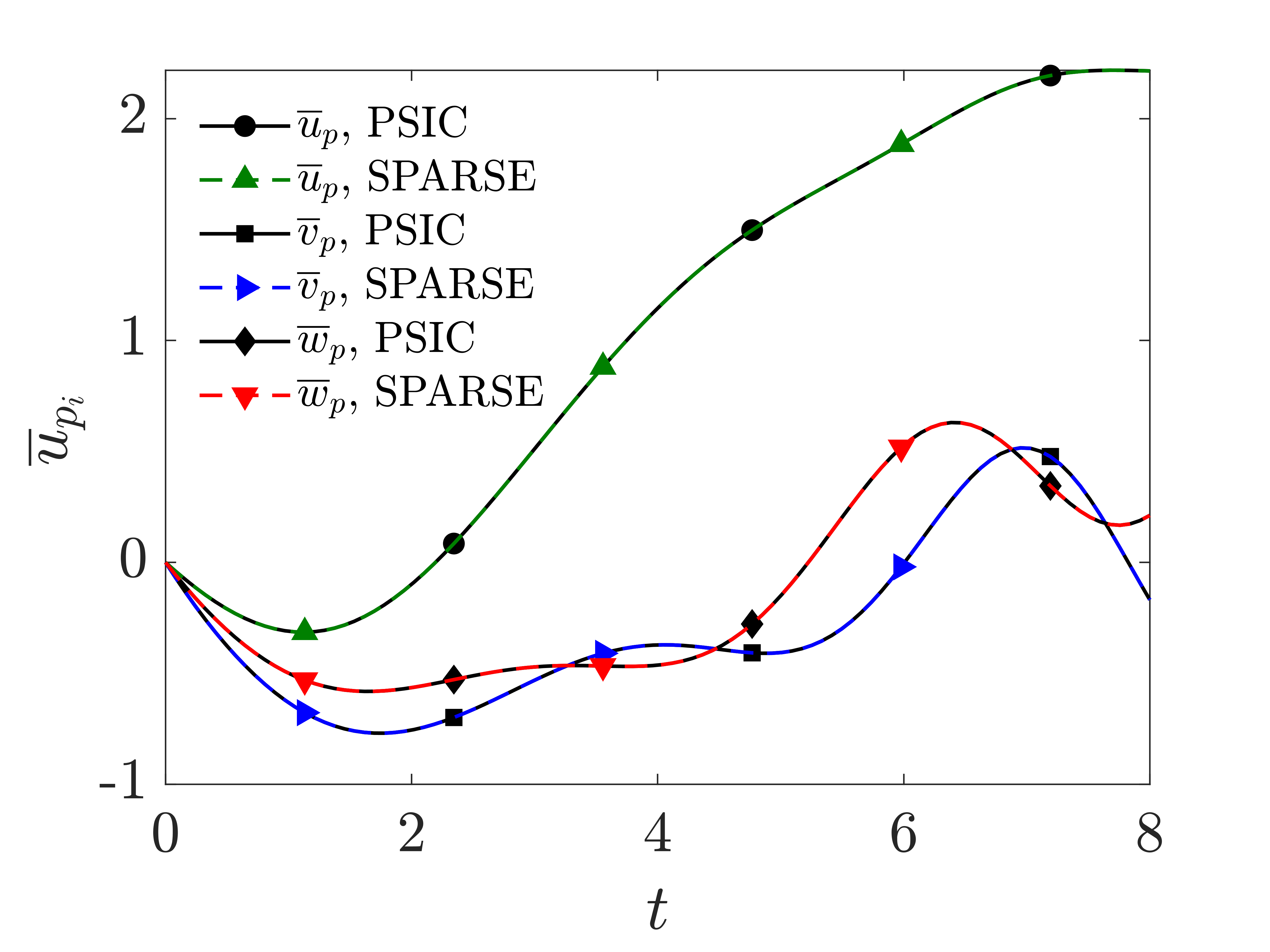}}
		\hfill	
	\subfloat[]{
		\label{fig: ABC_St2_cm2_xp}
		\includegraphics[width=0.31\textwidth]{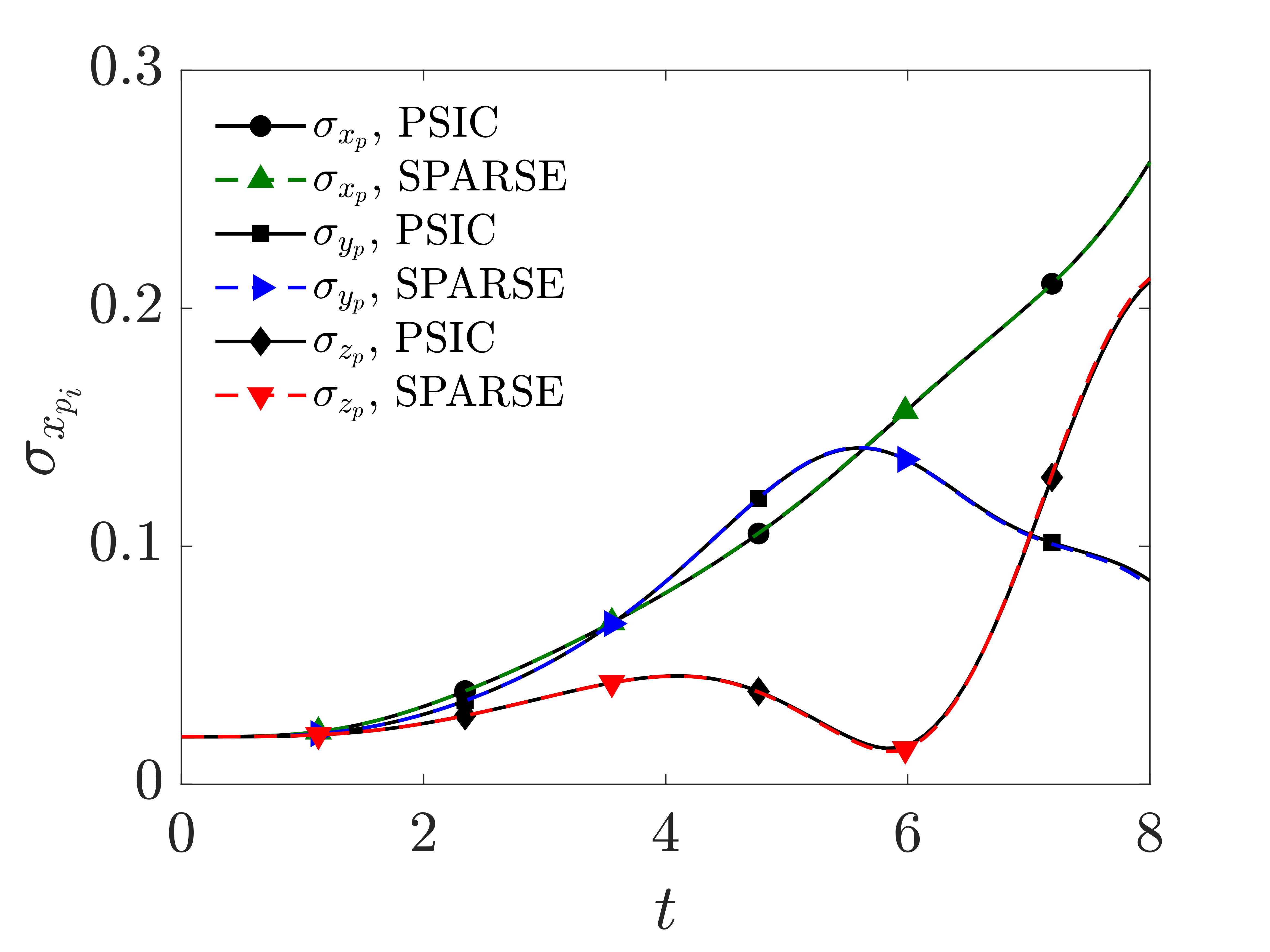}} \\
	\subfloat[]{
		\label{fig: ABC_St2_cm2_up}
		\includegraphics[width=0.31\textwidth]{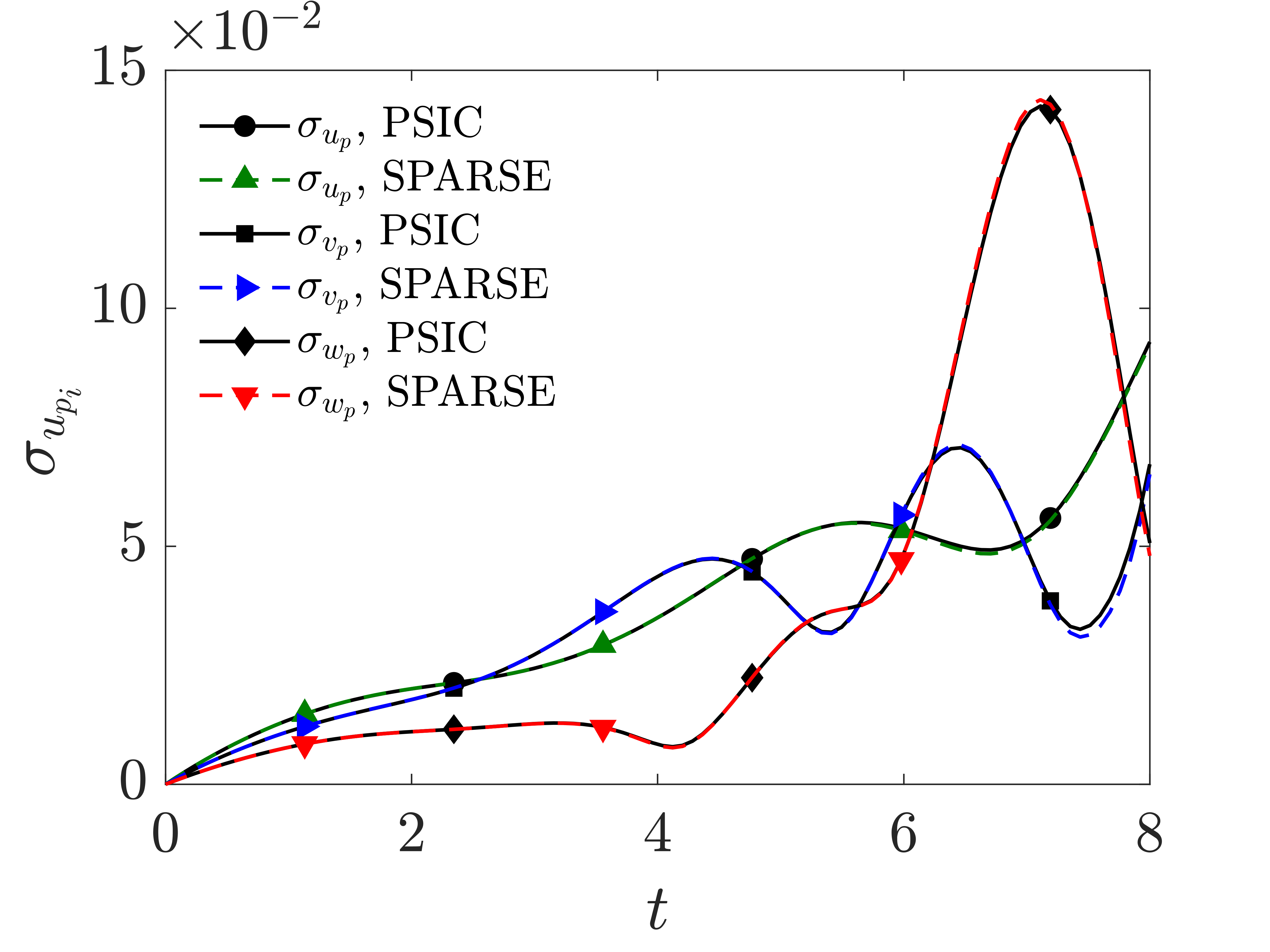}}
		\hfill
	\subfloat[]{
		\label{fig: ABC_St2_xpyp}
		\includegraphics[width=0.31\textwidth]{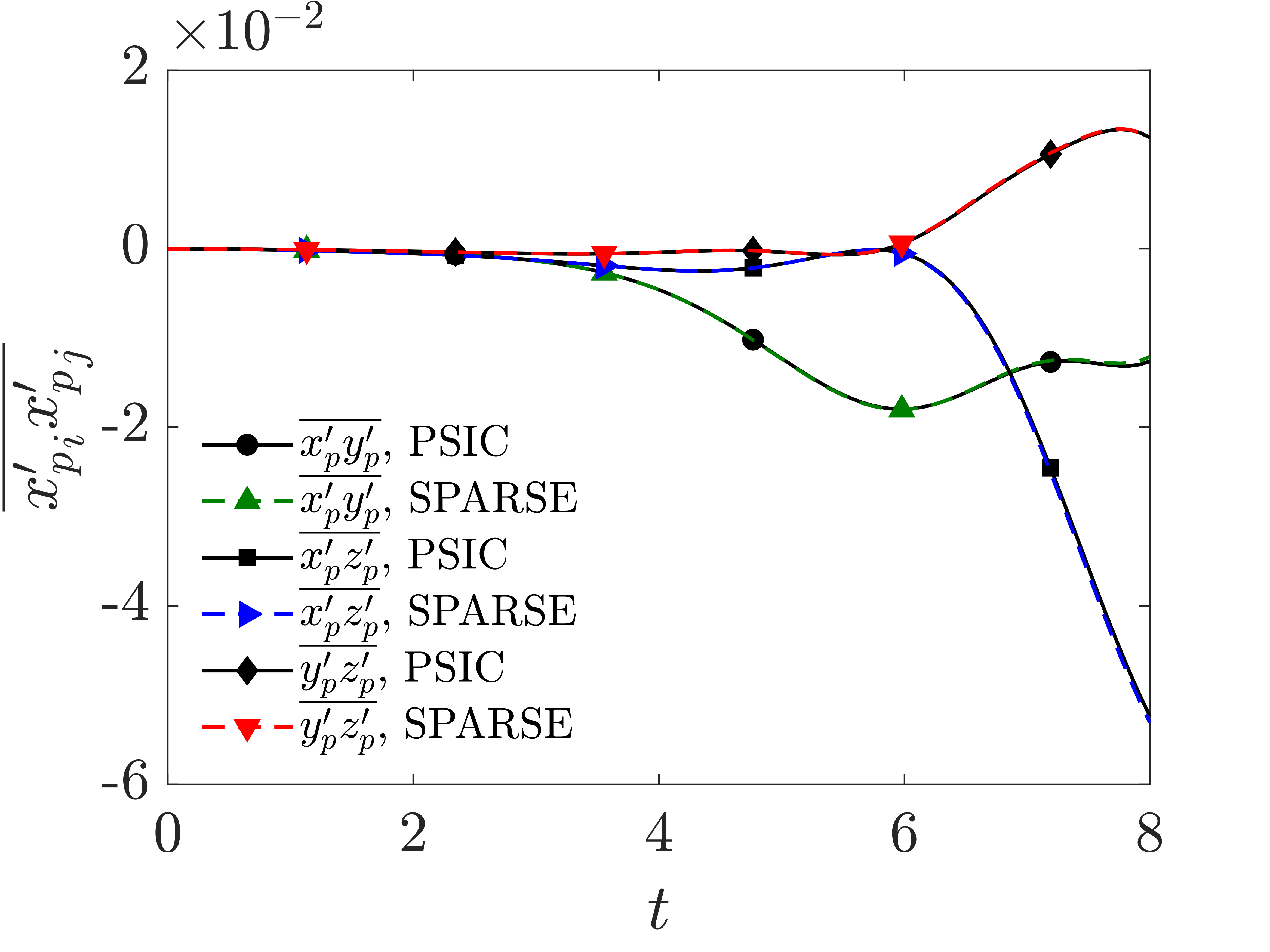}}
        \hfill	
	\subfloat[]{
		\label{fig: ABC_St2_upvp}
		\includegraphics[width=0.31\textwidth]{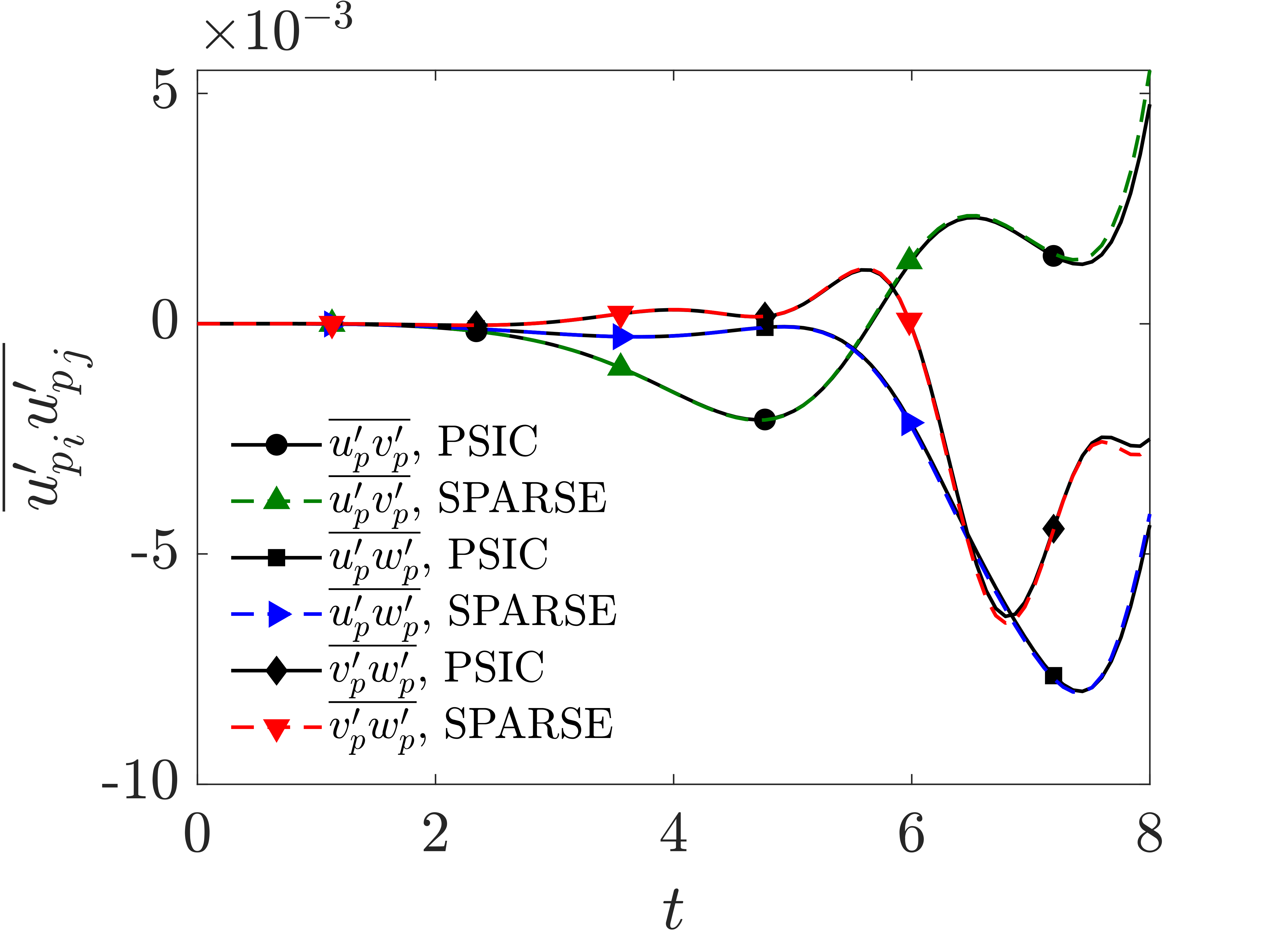}} \\
	\subfloat[]{
		\label{fig: ABC_St2_xpupi}
		\includegraphics[width=0.31\textwidth]{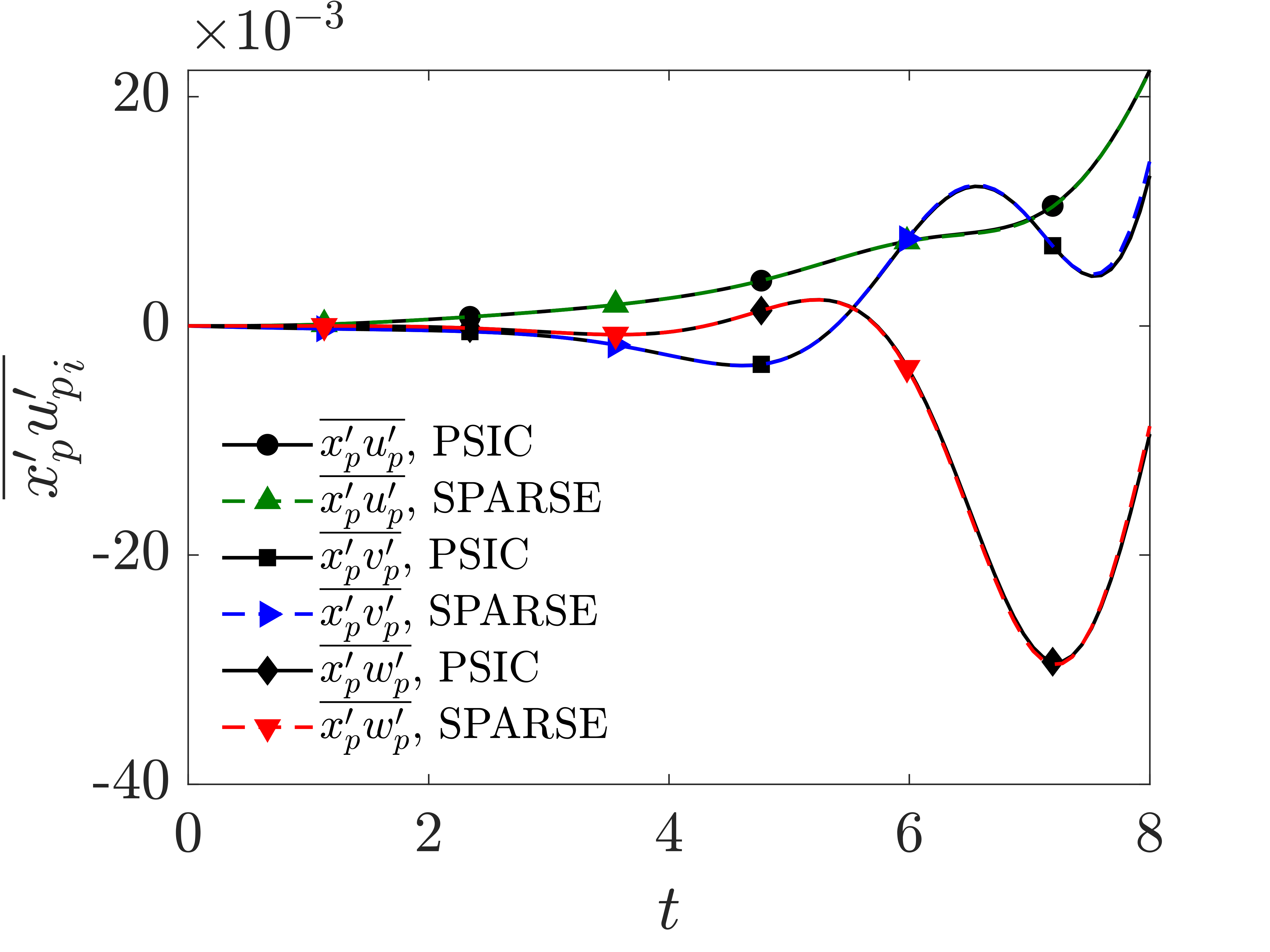}}
		\hfill
	\subfloat[]{
		\label{fig: ABC_St2_ypupi}
		\includegraphics[width=0.31\textwidth]{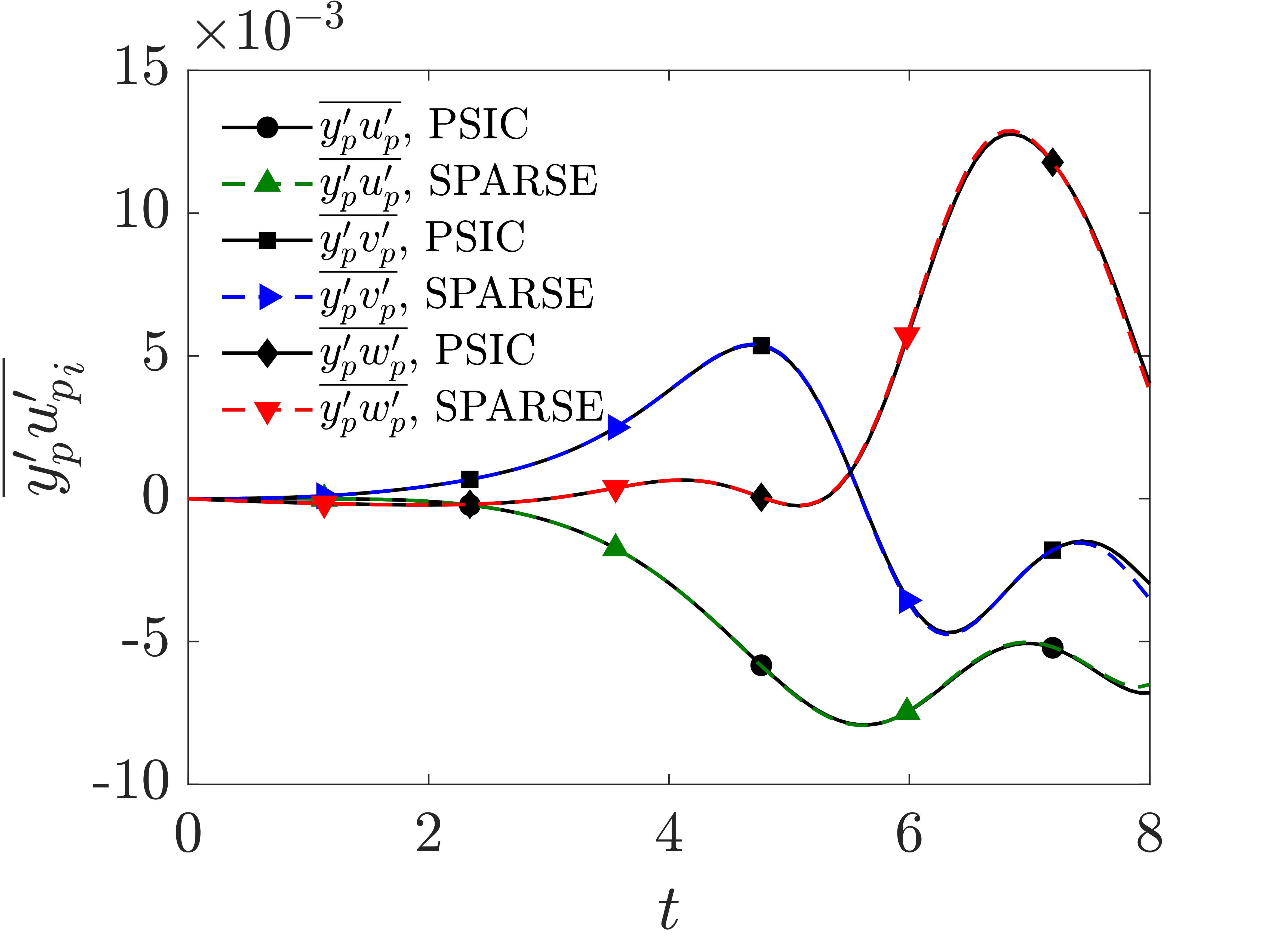}}
        \hfill	
	\subfloat[]{
		\label{fig: ABC_St2_zpupi}
		\includegraphics[width=0.31\textwidth]{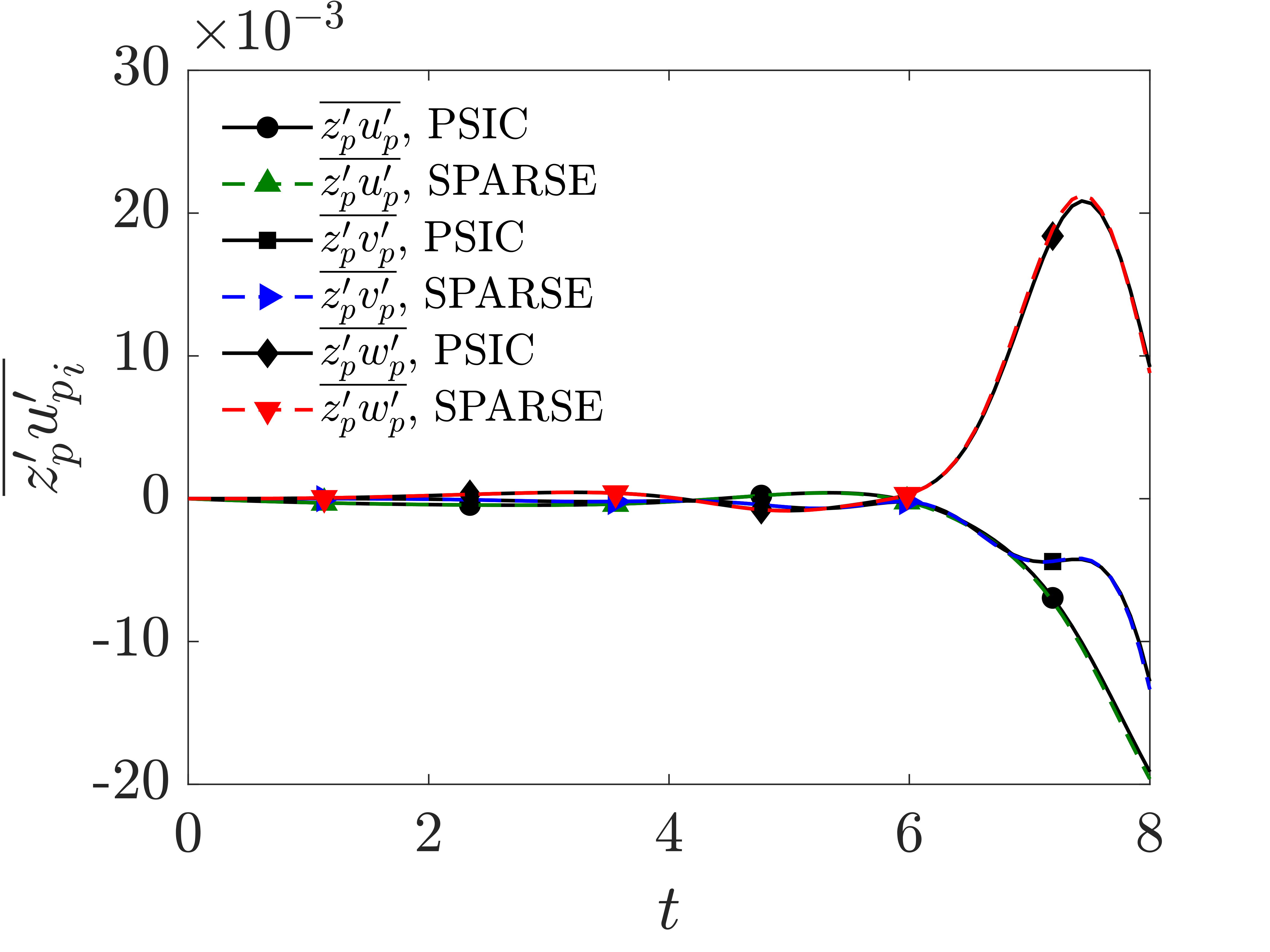}}
	\caption[]{Results of the SPARSE method compared with PSIC computations of the ABC flow for the average particle location (a), particle velocity (b), deviations in particle position (c) and particle velocity (d) as well as cross-terms in particle location (e) and in particle velocity (f) for $St=2$.}
	\label{fig: ABC_results}
\end{figure}

The average trajectories (solid black lines) of the four clouds  are visualized in  Figure~\ref{fig: ABC_3Dplot}. 
Three-dimensional prolates, whose axes are scaled with the principle strains, depict the cloud size. Single point-particles traced with the PSIC method are depicted as points for different instances of time.
The cloud trajectories are in large determined by the the coherent structures 
of the ABC flow as visualized by the vorticity contours in each plane of the boundaries of Figure~\ref{fig: ABC_3Dplot}.
The $x$ component of the particle velocity initially increases as the particle cloud is accelerated by one of these large flow structures.
After the initial acceleration, the clouds are transported primarily in $x-$direction and its lateral motion is affected only in a secondary manner by smaller vortices in the $x-y$ and $x-z$ planes.
Therefore, the velocity and locations along $x$ are greater (see Figures~\ref{fig: ABC_St2_mean_xp} and~\ref{fig: ABC_St2_mean_up}), whereas the average particle phase solutions in the perpendicular directions remain on the same order of magnitude.
As evidence of the excellent match between PSIC and the closed SPARSE method we select the case for $St=2$ to show the results of all moments of the cloud in Figure~\ref{fig: ABC_results}. 
The averages of the cloud location and velocity in Figures~\ref{fig: ABC_St2_mean_xp} and~\ref{fig: ABC_St2_mean_up} show no visible difference between both approaches.
The particle-phase variances are captured with the SPARSE method as shown in Figures~\ref{fig: ABC_St2_cm2_xp} and \ref{fig: ABC_St2_cm2_up}.
We note that at later time $t>7$, the variances show visible errors that are on the order of percentages, an indication  that at that instance the truncation errors are no longer
neglible.
The terms that correlate sub-cloud scale position and velocity fluctuations in multiple directions can be either positive or negative, indicating the combined grow of the cloud in the mixed direction.
These correlated second moments combining position and velocity are shown in Figures~\ref{fig: ABC_St2_xpyp}--\ref{fig: ABC_St2_zpupi} and show also relative error on the order of percentages for later times.

\begin{figure}[htbp]
	\centering
	\includegraphics[width=0.45\textwidth]{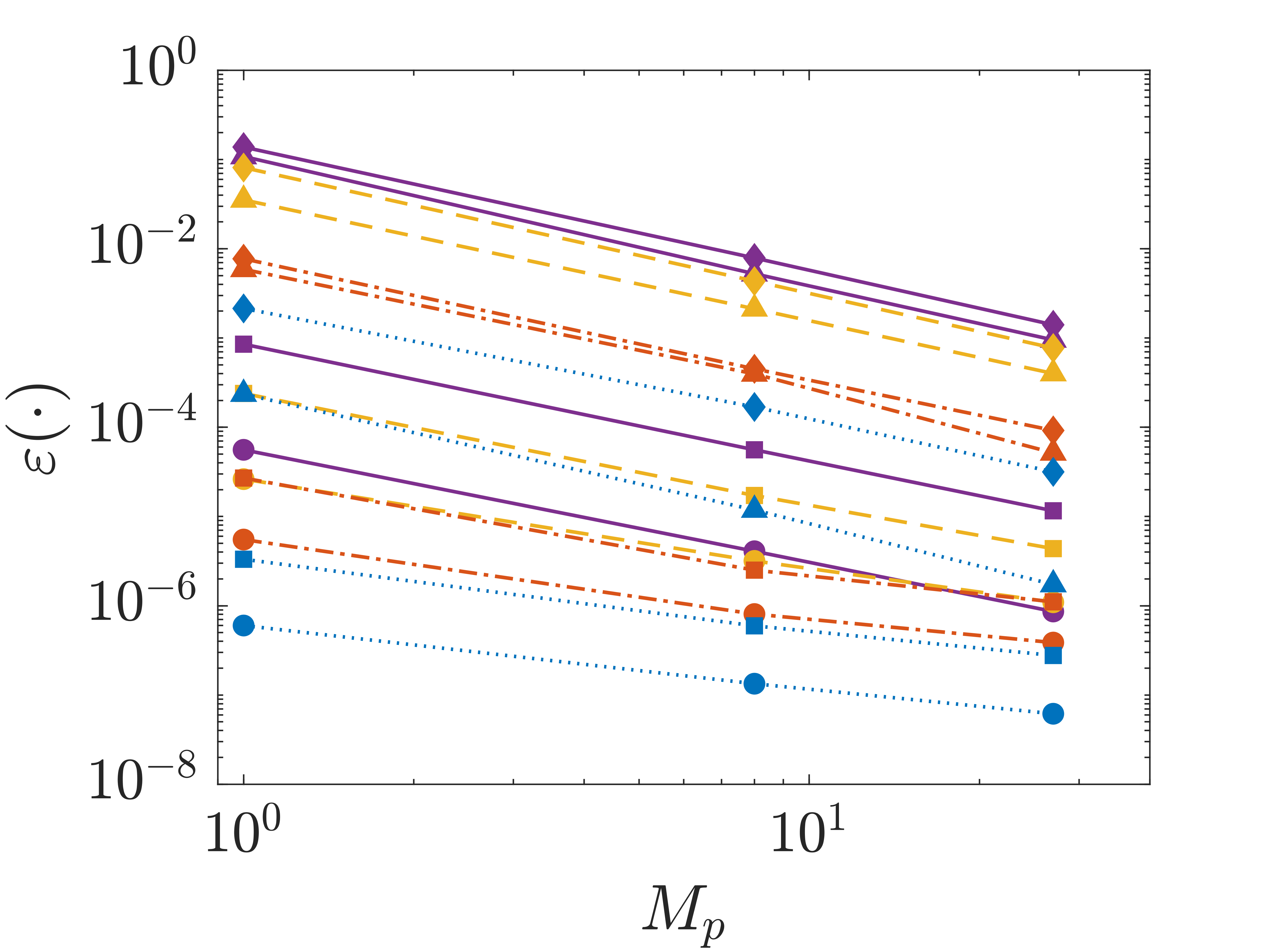}
	\caption[]{Convergence of the SPARSE method as compared with the PSIC approach for the average modulus of the particle location and velocity and representative particle deviation in position and velocity computed as in~\eqref{eq: 3D_measures}. The label is as follows: circles for $|{\bf \overline{x}}_p|$, squares for $|{\bf \overline{u}}_p|$, triangles for $\delta_{x_p}$ and diamonds for $\delta_{u_p}$ where the colors match the description in Figure~\ref{fig: ABC_3Dplot}  for the four clouds with $St=[1,\ 2,\ 5,\ 10]$. }
	\label{fig: ABC_errors}
\end{figure}

To evaluate the convergence of the SPARSE solution with an increase number of clouds, we perform three additional computations for each case and subdivide (or split) the single in each spatial dimension.
so that  the total number of cloud is $M_p=1$, $M_p=2^3$ and $M_p=3^3$ meaning one, two and three subdivisions in $x$, $y$ and $z$ directions respectively for the three levels of splitting considered.
To obtain a measurement of the particle averages we use the modulus of the average particle location and velocity and for a measure of the cloud's deviation in locations and velocities we use a geometrical average as follows
\begin{subequations}\label{eq: 3D_measures}
\begin{align}
|{\bf \overline{x}}_p| &= \sqrt{\overline{x}_p^2+\overline{y}_p^2+\overline{z}_p^2},  \ \ \ \ \
|{\bf \overline{u}}_p| =\sqrt{\overline{u}_p^2+\overline{v}_p^2+\overline{w}_p^2}, \label{eq: eq: 3D_measures_mean_xpup} \\
\delta_{x_p} &= \left( \sigma_{x_p} \sigma_{y_p} \sigma_{z_p} \right)^{1/3} , \ \ \ \ \
\delta_{u_p} = \left( \sigma_{u_p} \sigma_{v_p} \sigma_{w_p} \right)^{1/3} . \label{eq: eq: 3D_measures_delta_xpup}
\end{align}
\end{subequations}
The convergence of those measurements is shown in Figure~\ref{fig: ABC_errors} according to the error defined in~\eqref{eq: error} where the colors of the plots match with the clouds as depicted in Figure~\ref{fig: ABC_3Dplot}.
These trends provide evidence of the convergence of the SPARSE method when the initial condition is subdivided in macro-particles.
The erros is generally smaller for clouds with  a greater Stokes number.
This is consistent with proportionality of the right hand of the right hand side of the systems of closed SPARSE ODEs with $1/St$. 
The truncated terms are also proportional to $1/St$ and thus reduce with in an increase in  $St$.
A physics analogy that intuitively explains this error behavior, is that clouds with more inertia are more reticent to deformation according to the fluid flow and the eventual grow of high order moments (or errors) within the cloud.

\subsection{Isotropic turbulence}\label{sec: test_isoTurb}

To test the three-dimensional SPARSE formulation in a non-analytical, computed and complex velocity field, we revisit the simulation of a decaying isotropic turbulence~\cite{blaisdell1993compressibility,ristorcelli1997consistent} performed in~\cite{davis2017sparse}.
The isotropic turbulence simulation is performed in a cube with periodic boundary conditions on all sides with the validated discontinuous Galerkin code as described in \cite{klose2020assessing} and references therein, where the initial condition is adopted from \cite{jacobs2005validation}.

Computations are performed on a domain $\Omega$ spanned by coordinates $(x,y,z)$, defining a cube of size $2\pi$ so that $\Omega=[0, 2\pi]\times[0, 2\pi]\times[0, 2\pi]$.
The physical particles are initialized over a cubic domain of size $l_0=0.1$ stretching approximately 3 grid cells in each direction with $N_p=27,000$ total point-particles uniformly distributed in each direction.
The particles are released at rest and according to a one-way coupling, assuming that the flow is dilute, the particles have small but not negligible inertia with a Stokes number $St=0.5$ but the flow is not perturbed by them.
The non-dimensional particle diameter is $d_p=1.95\times10^{-3}$ and the ratio of densities $\rho_p=10^3$.
The drag and heat transfer correction factors for this case are adopted from Boiko \textit{et al.}~\cite{boiko1997shock} and Michaelides \textit{et al.}~\cite{michaelides2016multiphase} respectively and read as
\begin{subequations} \label{eq: isoTurb_f1f2}
\begin{align}
f_1&=\left(1+0.38\frac{Re_p}{24}+\frac{Re_p^{0.5}}{6}   \right)\left[ 1+\exp \left( \dfrac{-0.43}{M_p^{4.67}} \right)    \right],
\label{eq: isoTurb_f1} \\
f_2&=1+0.3Re_p^{0.5}Pr^{0.33}.
\label{eq: isoTurb_f2}
\end{align}
\end{subequations}
The computed carrier-phase velocities are used to determine the particle Reynolds as defined in~\eqref{eq: Rep_St_Pr} and the particle Mach number $M_{p}=|\boldsymbol{u}-\boldsymbol{u}_p|/\sqrt{T_{f}}$, in the forcing correction factors~\eqref{eq: isoTurb_f1f2}. 
The Prandtl number is $Pr=0.7$ and the relative heat capacity is set to unity $c_r=1$.

\begin{figure}[htbp]
	\centering
	\subfloat[]{
	    \label{fig: isoTurb_3D_1}
	    \includegraphics[width=0.45\textwidth]{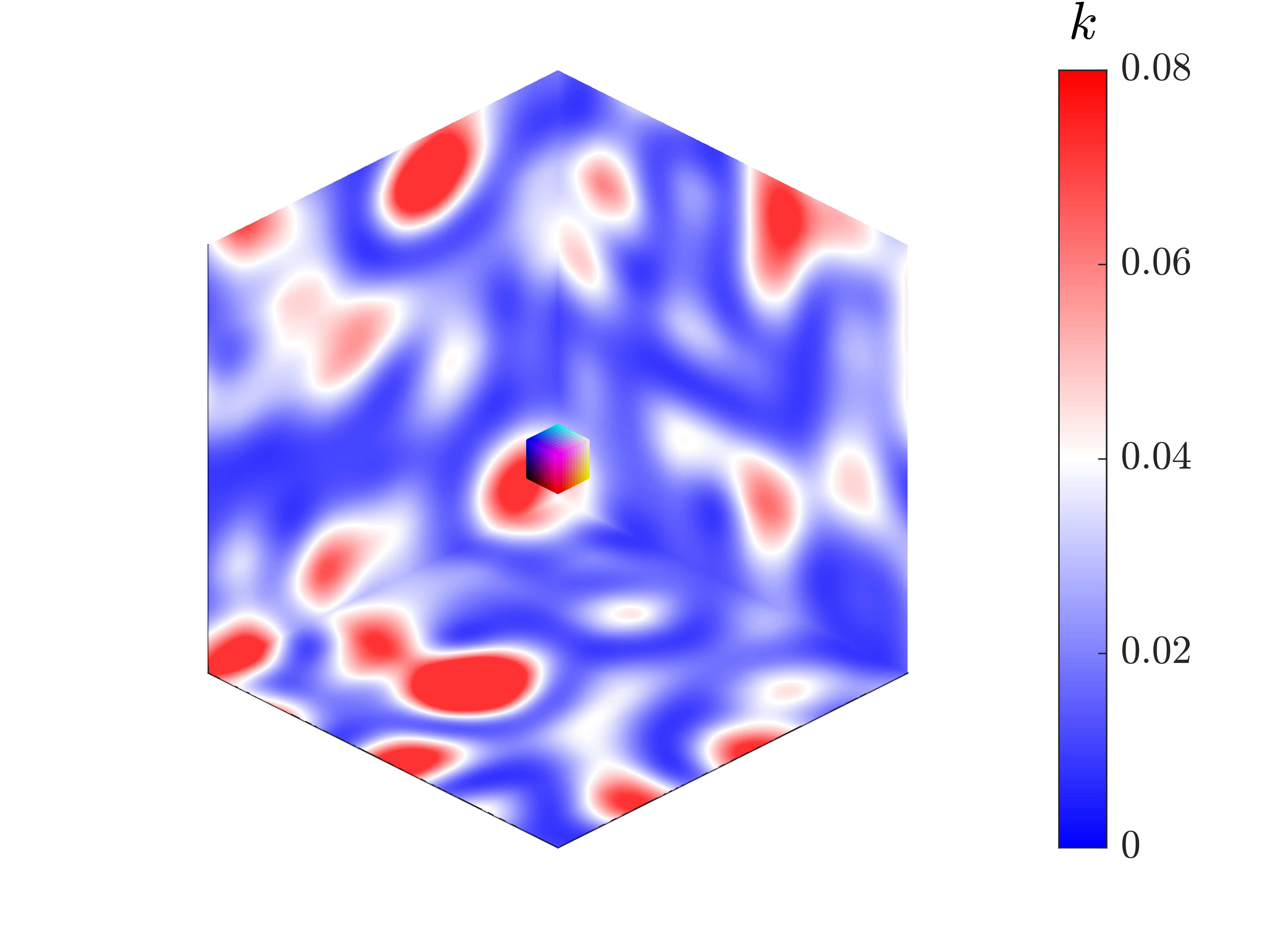}}
        \hfill	
	\subfloat[]{
		\label{fig: isoTurb_3D_2}
		\includegraphics[width=0.45\textwidth]{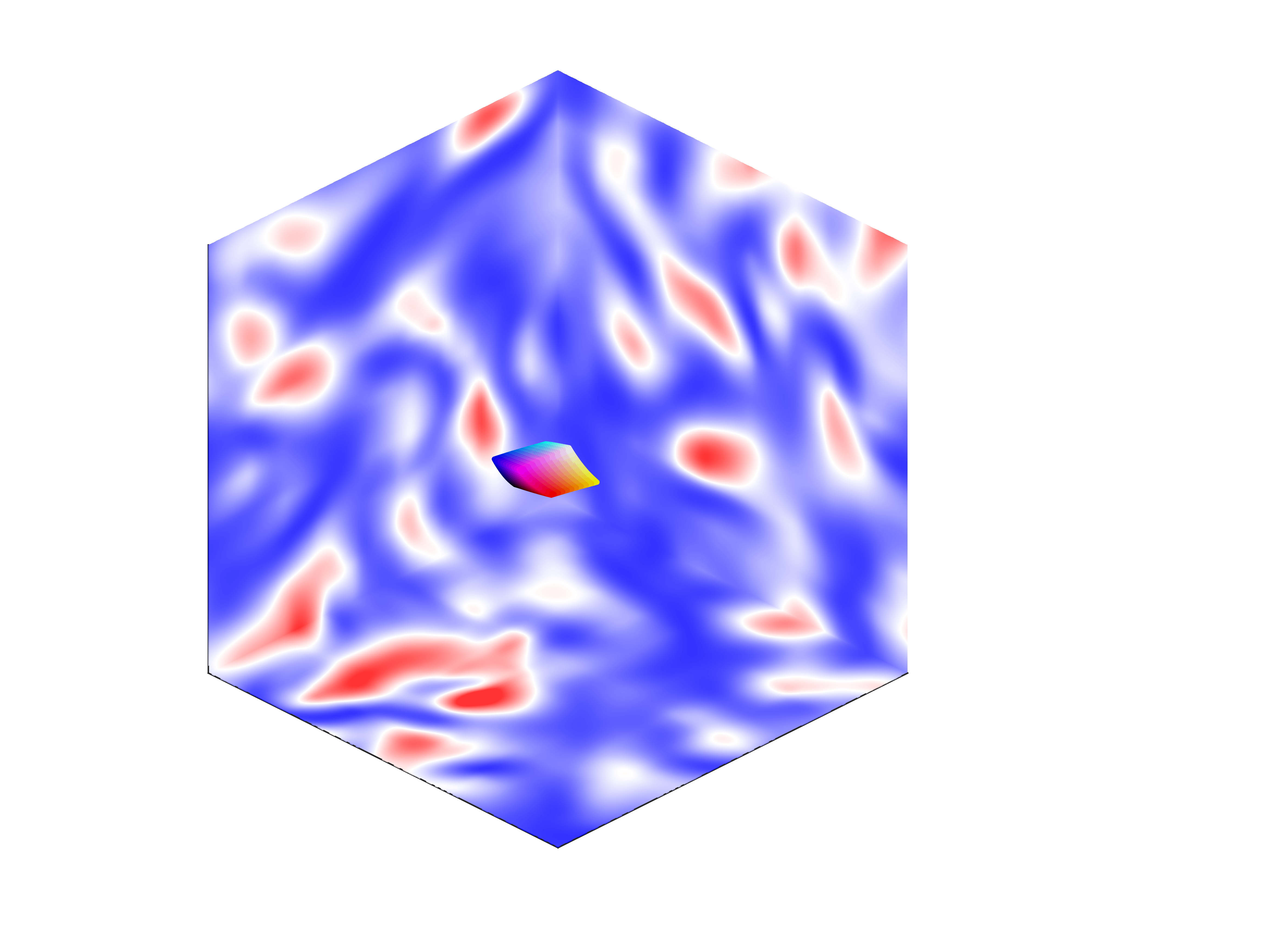}} \\
	\subfloat[]{
		\label{fig: isoTurb_3D_3}
		\includegraphics[width=0.45\textwidth]{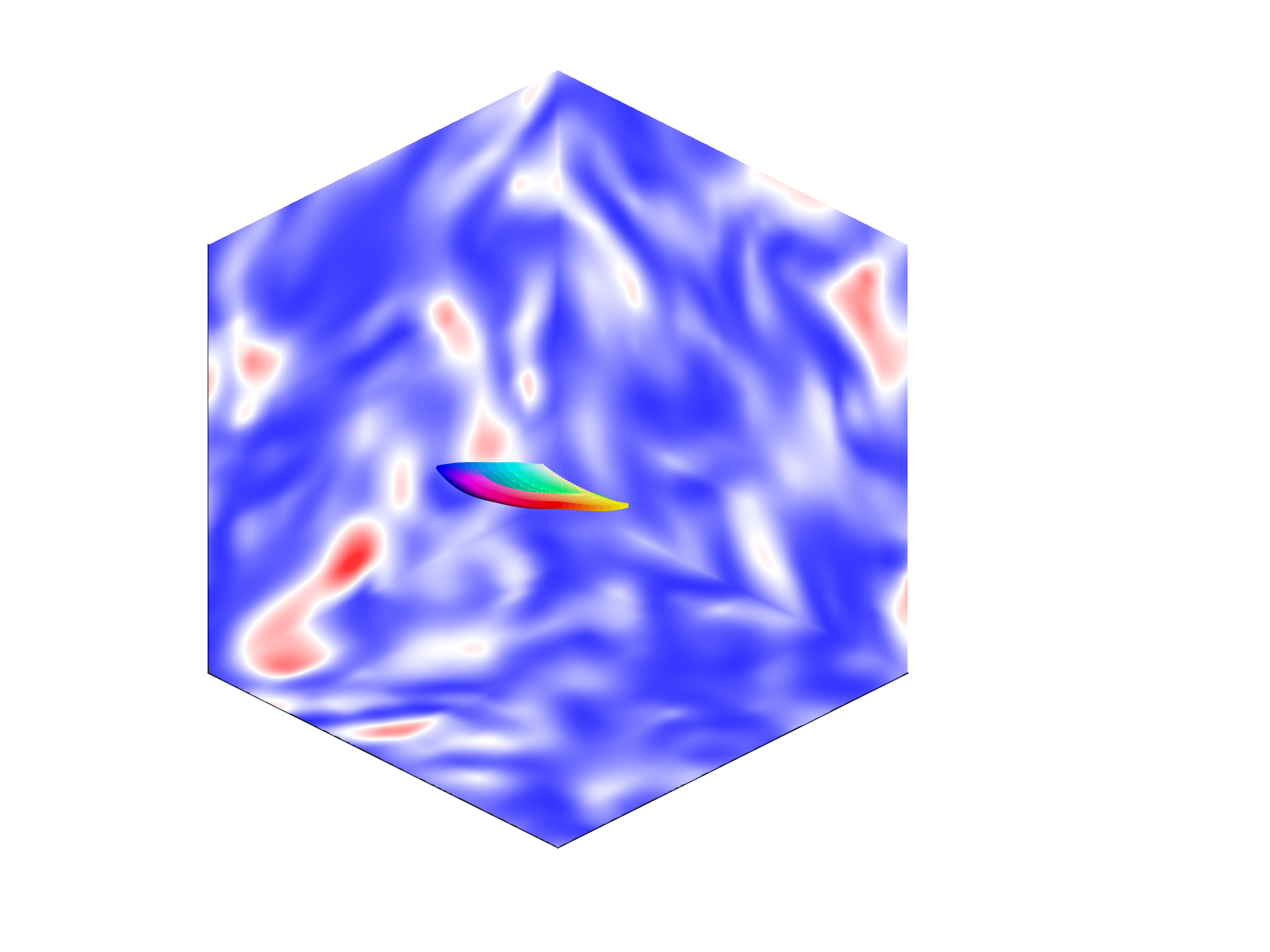}} 
		\hfill
	\subfloat[]{
		\label{fig: isoTurb_3D_4}
		\includegraphics[width=0.45\textwidth]{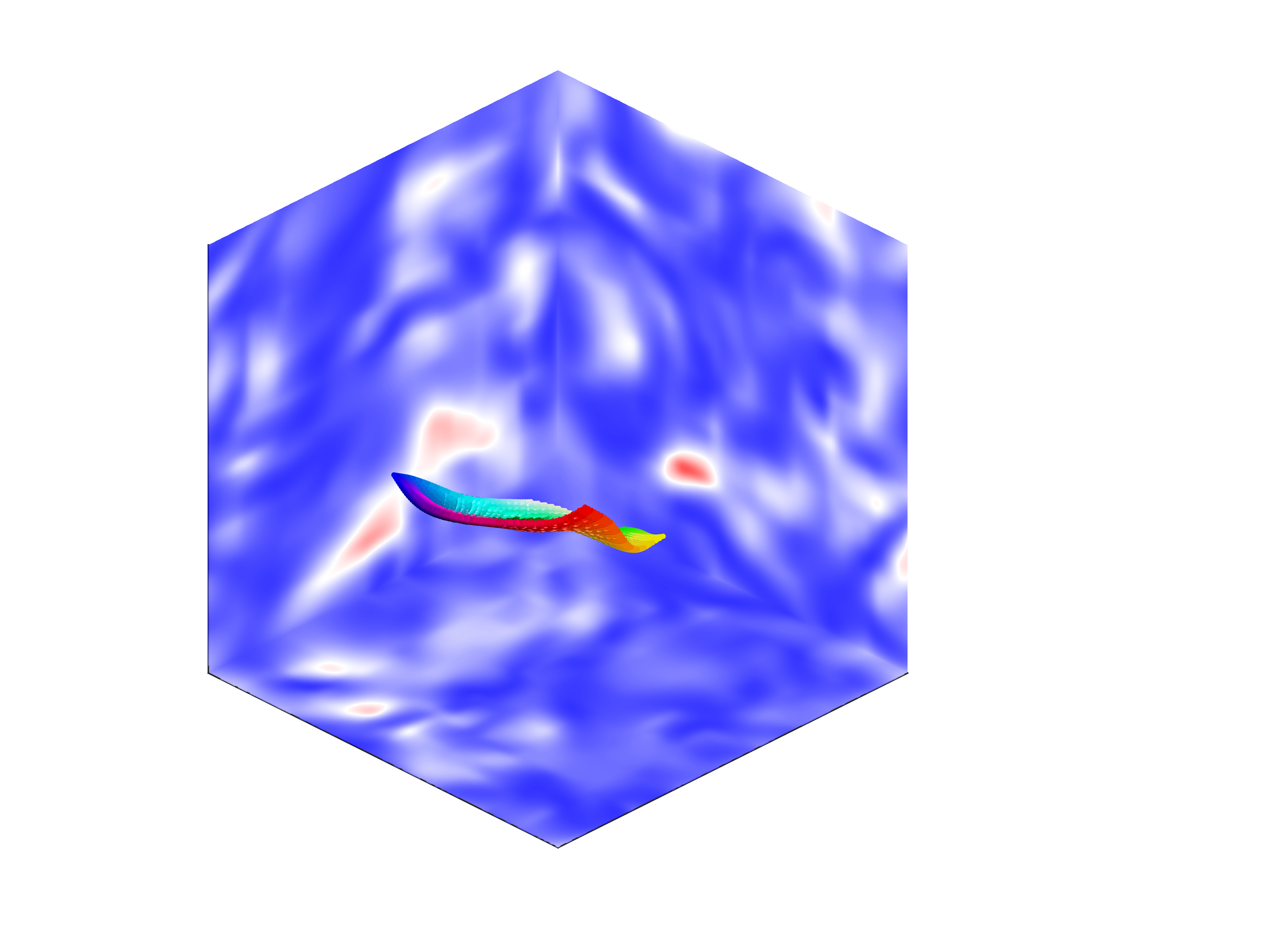}} \\
	\subfloat[]{
		\label{fig: isoTurb_3D_5}
		\includegraphics[width=0.45\textwidth]{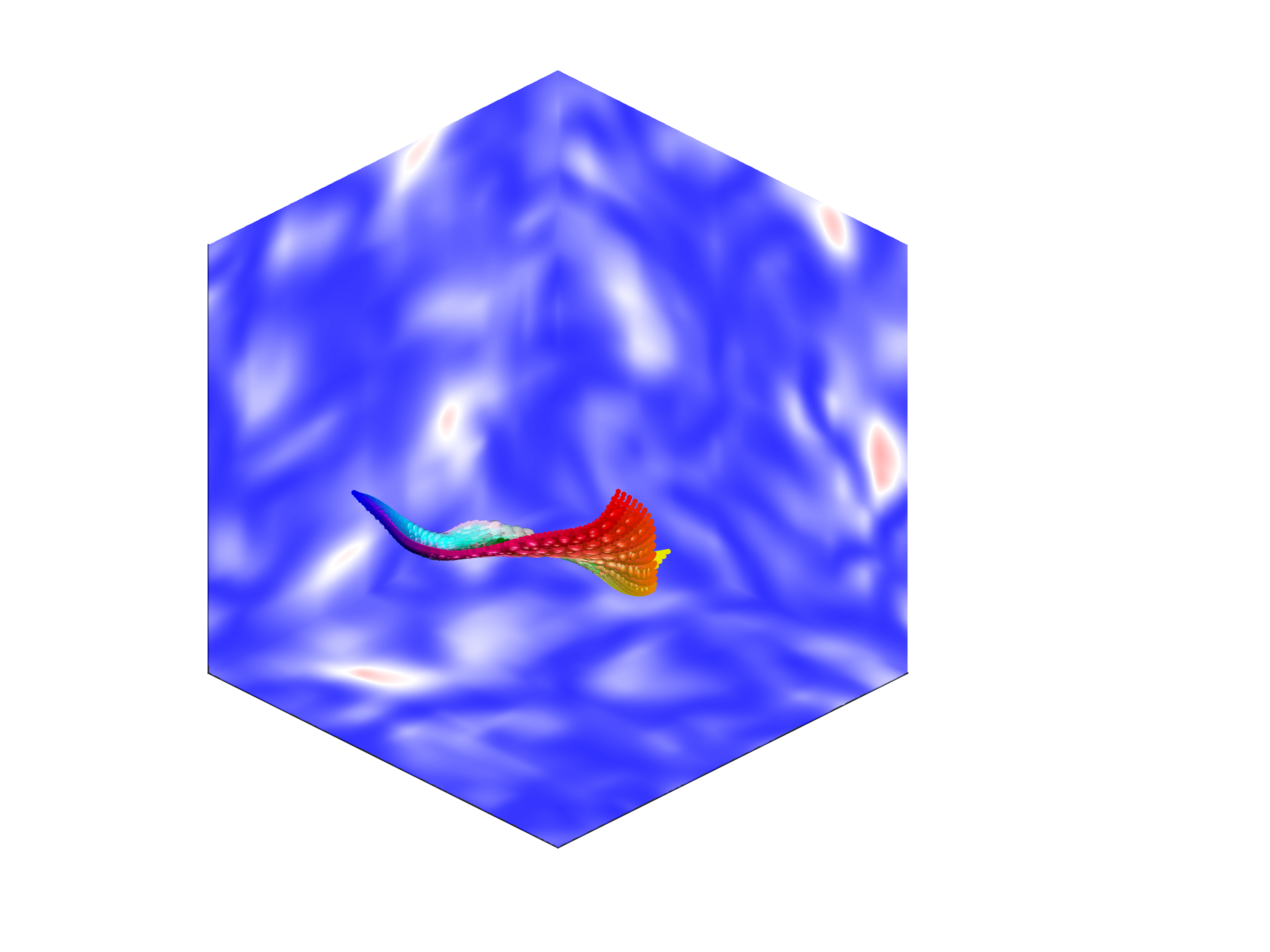}}
        \hfill	
    	\subfloat[]{
		\label{fig: isoTurb_3D_6}
		\includegraphics[width=0.45\textwidth]{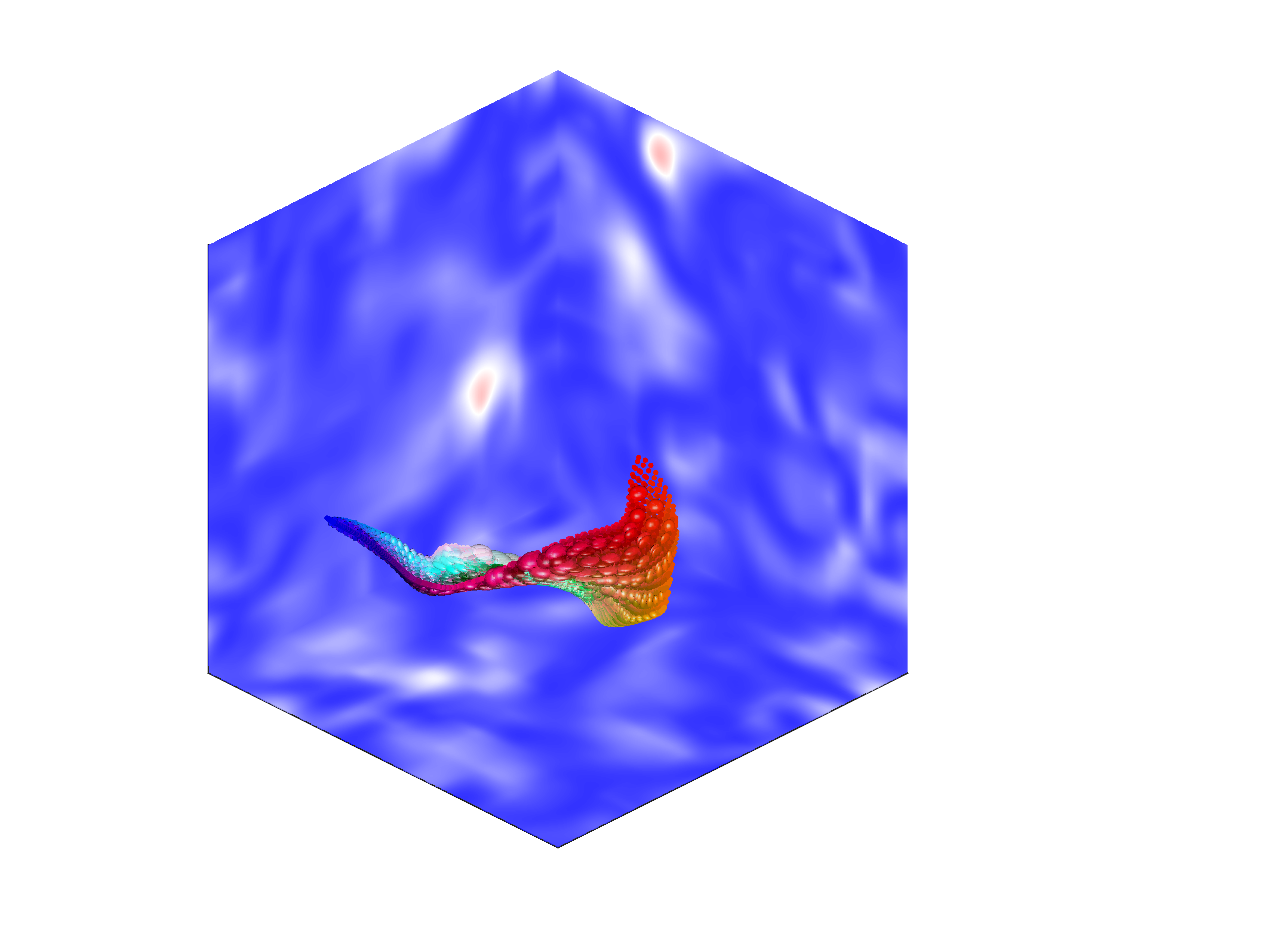}}
	\caption[]{Locations of the particles and macro-particles tracked with PSIC and SPARSE methods respectively for the three-dimensional isotropic decaying turbulence case at times (a) $t=0$, (b) $t=0.8$, (c) $t=1.6$, (d) $t=2.4$, (e) $t=3.2$ and (f) $t=4$). The contours show the turbulent kinetic energy in the boundaries of the domain.}
	\label{fig: isoTurb_3D}
\end{figure}

The result of the computations for a thousand macro-particles $M_p=10^3$, uniformly distributed in space is shown in Figure~\ref{fig: isoTurb_3D} for several instants of time where the $27,000$ point-particles computed with the PSIC method are shown as points and the SPARSE clouds as ellipsoids using the covariance matrix of the cloud location in three dimensions. 
Contours of the turbulent kinetic energy $k$ visualize the carrier-phase's turbulent structures on the boundaries of the cube.
The particles initially at rest react to the carrier flow. 
As compared to the fluid tracers (not shown), the inertial tracers have a smoother response. 
After some time, the initial cube of particles is dispersed and the macro-particles are advected and deformed according to the carrier flow.

\begin{figure}[htbp]
	\centering
	\subfloat[]{
		\label{fig: isoTurb_means_xpup_St05}
		\includegraphics[width=0.45\textwidth]{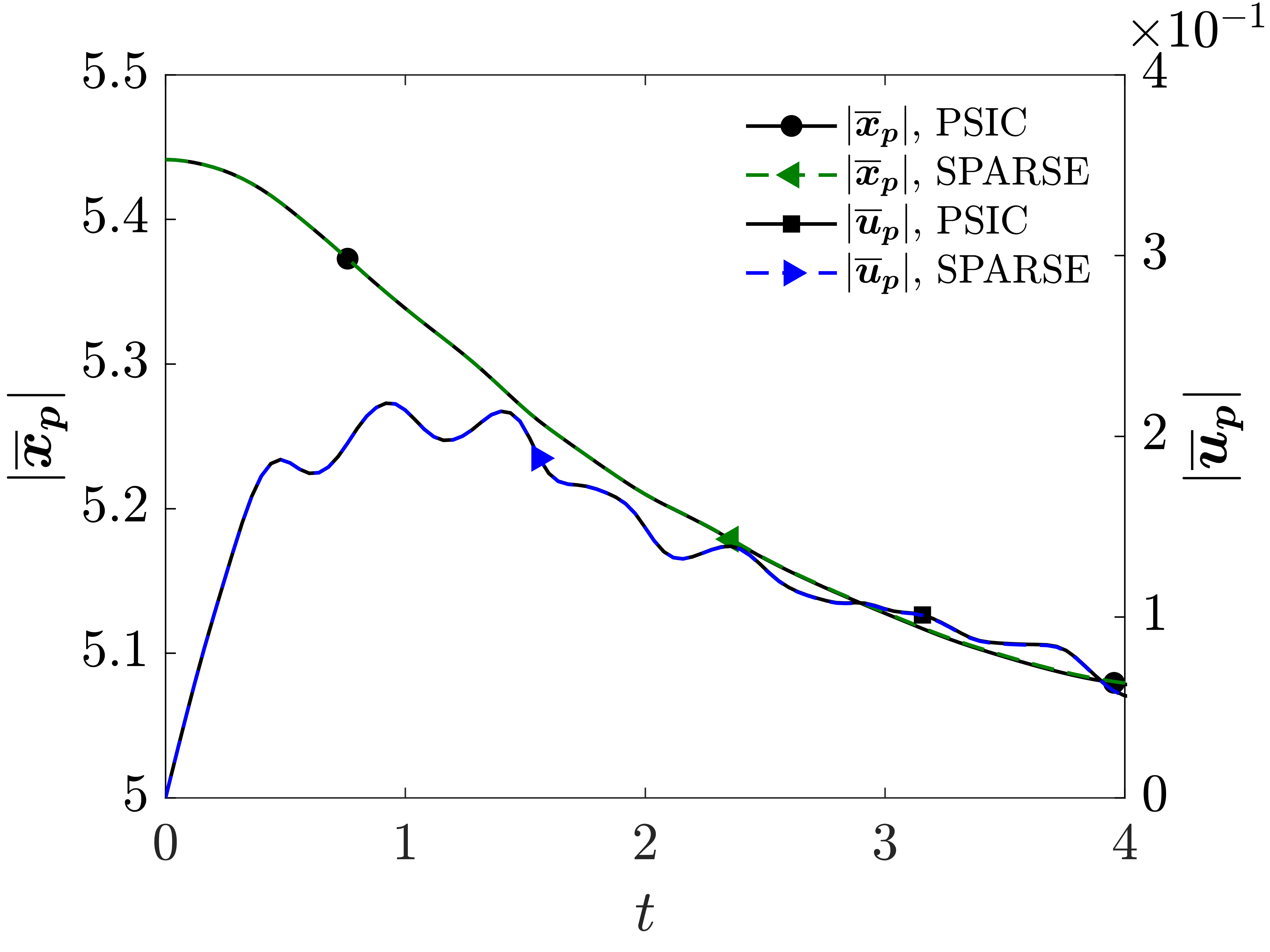}}
        \hfill	
	\subfloat[]{
		\label{fig: isoTurb_sigmaxp_St05}
		\includegraphics[width=0.45\textwidth]{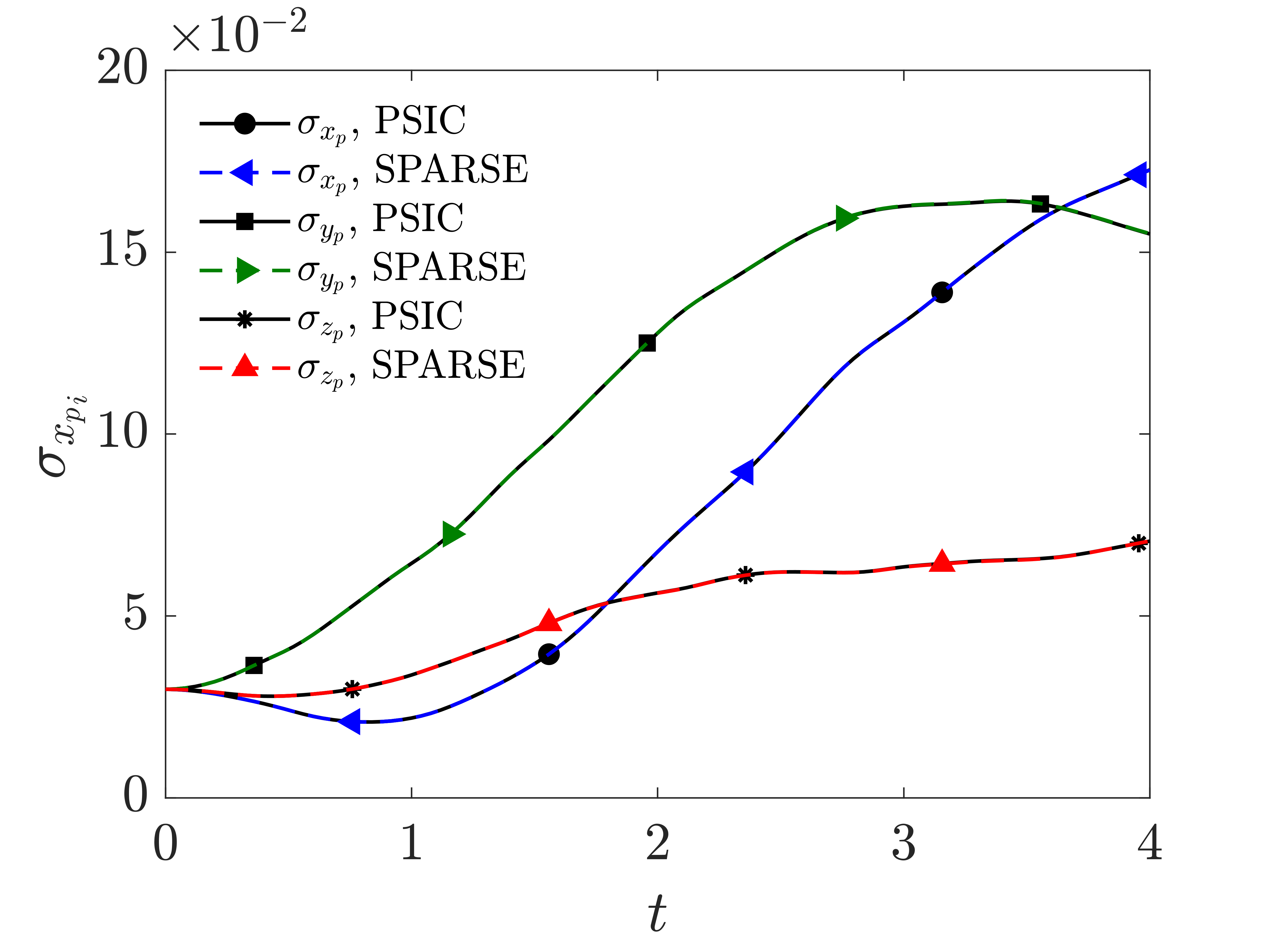}} \\
	\subfloat[]{
		\label{fig: isoTurb_meanAndSigma_Tp_St05}
		\includegraphics[width=0.45\textwidth]{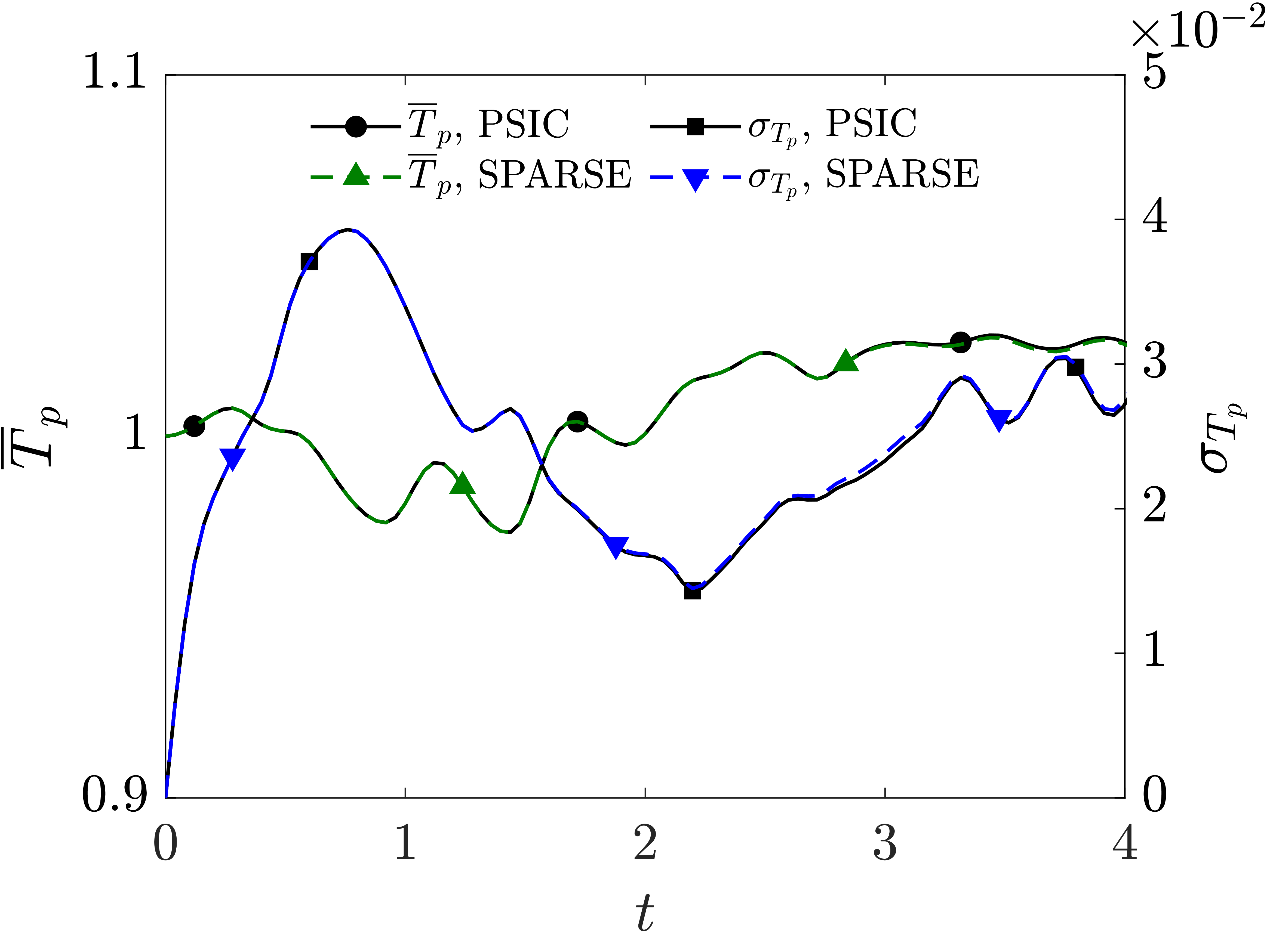}}
        \hfill	
	\subfloat[]{
		\label{fig: isoTurb_sigmaup_St05}
		\includegraphics[width=0.45\textwidth]{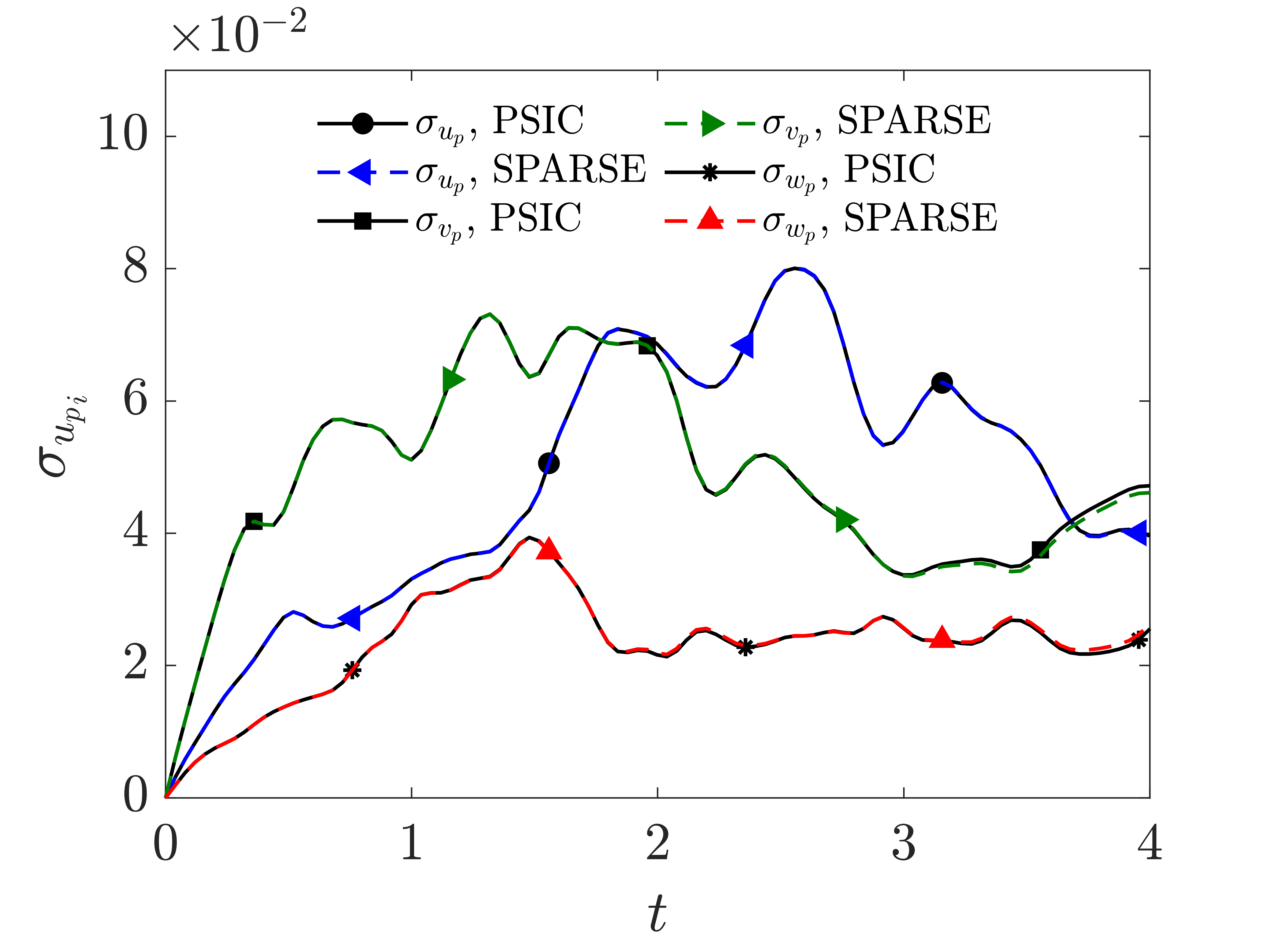}}
	\caption[]{Statistics of the inertial particle cloud in the three-dimensional decaying isotropic turbulence case computed with PSIC and SPARSE methods; (a) average module of the particle cloud location and velocity, (b) spatial deviations of the particle cloud, (c) temperature mean and deviation and (c) deviations of the particle cloud velocity.}
	\label{fig: isoTurb_mean_and_deviations}
\end{figure}

The average particle location and velocity defined in~\eqref{eq: eq: 3D_measures_mean_xpup} computed with the PSIC and SPARSE methods are compared in Figure~\ref{fig: isoTurb_means_xpup_St05}.
The plots show that the mean location and velocity initially change according to the dynamics of the eddies at the cloud location.
As the cloud spreads and the fluid velocity is sampled over a larger area, the fluid velocity through the cloud approaches zero after an initial acceleration because the turbulence is isotropic, making the average velocity in the box to evolve towards zero.
The deviations of the cloud's location is plotted versus time in Figure~\ref{fig: isoTurb_sigmaxp_St05} and show a general increase in all three dimensions consistent with turbulent diffusion mechanisms~\cite{pope2000turbulent}.
The particle standard deviations of the sub-cloud scale velocity increase from an initial rest state in which the standard deviations are zero towards a trend that correlates with the decaying carrier-phase turbulence as seen in Figure~\ref{fig: isoTurb_sigmaup_St05}.
The temperature average is almost constant and the standard deviation of the temperature small as shown in Figure~\ref{fig: isoTurb_meanAndSigma_Tp_St05} because the turbulence Mach number is low and the flow is near isothermal.
The standard deviation of the particle temperature of the cloud behaves similar to the one of the particle velocity, starting from an initial value zero according to a uniform temperature in the cloud to increase with an oscillating trend governed by the changes in the carrier phase flow temperature.
The results by PSIC and SPARSE methods are in an excellent agreement for the average magnitudes of the particle cloud as well as for its deviations.
The rest of the second moments not shown in Figure~\ref{fig: isoTurb_mean_and_deviations} are also well captured by the SPARSE method. 
The maximum discrepancy in the first two moments when comparing SPARSE with PSIC show a relative error of $3\%$ or smaller validating the closed SPARSE tracer. 
To measure the reduction of the computational cost by the use of the SPARSE tracer as compared to PSIC, the ratio of equations computed with SPARSE with respect to PSIC is $0.1852$, ensuring computational savings despite having used a thousand macro-particles for splitting the initial condition.
We also observe similar convergence rates that those shown in Figure~\ref{fig: ABC_errors} that have been omitted for brevity. 


\section{Conclusions and future work} \label{sec: conclusions}

A closed SPARSE tracer is developed that predicts the dynamics of the first two statistical moments of groups of particles and traces them as a single point. 
This cloud or macro-particle approach accounts for the effects of carrier-phase velocity distribution and the sub-cloud's second moments of the particle phase and carrier-phase.
The tracer combines a truncated Taylor expansion of the forcing correction factors around the cloud's mean relative velocity and a Reynolds decomposition of the ensemble averages of the particle variables within the cloud.
Using a Taylor expansion, averaging and truncation, the extended SPARSE formulation provides a closed set of equations for the first two moments of the particle cloud.
The closure expresses unknown combined moments of both phases in terms of those known moments of the disperse phase that are traced with the SPARSE method.
This closes the SPARSE tracer method that so far has been used with an \textit{a priori} closure.


The SPARSE method reduces the computational expense for the tracing of the first two statistical moments of a cloud as compared to a simulation with the PSIC method by reducing the required degrees of freedom. 
It improves upon the accuracy of commonly used zeroth order Cloud-in-Cell models through a second order moment correction by expanding the forcing function in the surroundings of the cloud.
The error of the SPARSE tracer is a function of the truncated terms of the Taylor expansions and the truncation of higher-order statistical moments that are heuristically shown in test cases to converge monotonically with an increasing number of subdivisions of the initial cloud.




The closed SPARSE method is verified and validated against PSIC results for analytical one-, two- and three-dimensional flows where the relative errors are either negligible or small percentages in all the test cases.

The SPARSE tracer is accurate for finite time and will require merging and join of macro-particles to adapt the number of macro-particles needed depending on the error of the model as the simulation evolves.
We intend to report on this adaptation in the near future.


\section*{Acknowledgments}
This work was supported by the Air Force Office of Scientific Research under Award No FA9550-19-1-0387 and a San Diego State University Graduate Fellowship.









\appendix

\bibliography{Bib}{}
\bibliographystyle{unsrt}

\end{document}